\documentclass[12pt,letterpaper]{article}
\usepackage{amsmath,amssymb,array,calc,rotating,epsfig,psfrag,amscd, cite}

\makeatletter
\renewcommand\section{\@startsection {section}{1}{\z@}%
                                   {-3.5ex \@plus -1ex \@minus -.2ex}
                                   {2.3ex \@plus.2ex}%
                                   {\normalfont\large\bfseries}}
\renewcommand\subsection{\@startsection{subsection}{2}{\z@}%
                                     {-3.25ex\@plus -1ex \@minus -.2ex}%
                                     {1.5ex \@plus .2ex}%
                                     {\normalfont\bfseries}}
\makeatother

\let\non\nonumber

\let\a=\alpha

\let\s=\sigma

\let\S=\Sigma

\newcommand{\bea}{\begin{eqnarray}}
\newcommand{\eea}{\end{eqnarray}}
\newcommand{\be}{\begin{equation}}
\newcommand{\ee}{\end{equation}}

\newcommand{\m}{\mu}

\newcommand{\p}{\partial}

\newcommand{\D}[1]{\ensuremath{\mathrm{D}#1}}

\newcommand{\C}[1]{$(\ref{#1})$}

\def\IZ{\relax\ifmmode\mathchoice
{\hbox{\cmss Z\kern-.4em Z}}{\hbox{\cmss Z\kern-.4em Z}}
{\lower.9pt\hbox{\cmsss Z\kern-.4em Z}} {\lower1.2pt\hbox{\cmsss
Z\kern-.4em Z}}\else{\cmss Z\kern-.4em Z}\fi}
\def\IR{\relax{\rm I\kern-.18em R}}

\def\one{{\hbox{ 1\kern-.8mm l}}}

\newlength{\bredde}
\def\slash#1{\settowidth{\bredde}{$#1$}\ifmmode\,\raisebox{.15ex}{/}
\hspace*{-\bredde} #1\else$\,\raisebox{.15ex}{/}\hspace*{-\bredde}
#1$\fi}

\newsavebox{\zzzbar}
\sbox{\zzzbar}
  {\setlength{\unitlength}{0.9em}
  \begin{picture}(0.6,0.7)
  \thinlines
  \put(0,0){\line(1,0){0.6}}
  \put(0,0.75){\line(1,0){0.575}}
  \multiput(0,0)(0.0125,0.025){30}{\rule{0.3pt}{0.3pt}}
  \multiput(0.2,0)(0.0125,0.025){30}{\rule{0.3pt}{0.3pt}}
  \put(0,0.75){\line(0,-1){0.15}}
  \put(0.015,0.75){\line(0,-1){0.1}}
  \put(0.03,0.75){\line(0,-1){0.075}}
  \put(0.045,0.75){\line(0,-1){0.05}}
  \put(0.05,0.75){\line(0,-1){0.025}}
  \put(0.6,0){\line(0,1){0.15}}
  \put(0.585,0){\line(0,1){0.1}}
  \put(0.57,0){\line(0,1){0.075}}
  \put(0.555,0){\line(0,1){0.05}}
  \put(0.55,0){\line(0,1){0.025}}
  \end{picture}}

\newcommand{\ena}{\end{eqnarray}}
\newcommand{\beqa}{\begin{eqnarray}}
\newcommand{\eeqa}{\end{eqnarray}}

\newcommand{\g}{\gamma}

\def\a{\alpha}

\def\g{\gamma}

\def\m{\mu}

\def\s{\sigma}

\def\D{\Delta}

\def\S{\Sigma}

\begin{document}
\begin{titlepage}

\begin{center}



\vskip 2 cm
{\Large \bf Eigenvalue equation for genus two modular graphs }\\
\vskip 1.25 cm { Anirban Basu\footnote{email address:
    anirbanbasu@hri.res.in} } \\
{\vskip 0.5cm  Harish--Chandra Research Institute, HBNI, Chhatnag Road, Jhusi,\\
Prayagraj 211019, India}

\end{center}

\vskip 2 cm

\begin{abstract}
\baselineskip=18pt

We obtain a second order differential equation on moduli space satisfied by certain modular graph functions at genus two, each of which has two links. This eigenvalue equation is obtained by analyzing the variations of these graphs under the variation of the Beltrami differentials. This equation involves seven distinct graphs, three of which appear in the integrand of the $D^8\mathcal{R}^4$ term in the low momentum expansion of the four graviton amplitude at genus two in type II string theory.

\end{abstract}

\end{titlepage}


\section{Introduction}

Multiloop string amplitudes provide useful insight into the structure of terms in the effective action of string theory. The effective action encodes the dynamics of the massless modes of the theory, and yields S--matrix elements which contain terms both analytic as well as non--analytic in the external momenta of the particles. The structure of these terms in the effective action can be directly read off from the low momentum expansion of the string amplitudes. While the terms analytic in the external momenta arise from the integration over the interior of the supermoduli space of the super Riemann surface, the terms non--analytic in the external momenta arise from the boundary of moduli space. 

Calculating these terms becomes progressively difficult as one considers higher genus string amplitudes\footnote{String amplitudes at every order in the genus expansion include an infinite number of $\alpha'$ corrections.}. Beyond tree level, one has to integrate over the geometric moduli of the Riemann surface which is non--trivial. At genus one, in order to calculate the analytic terms in the low momentum expansion, it is very useful to obtain eigenvalue equations which the modular invariant integrand satisfies. This helps us not only to have an understanding of the detailed structure of the integrand, but also to calculate the integral over moduli space.            

In general, at every genus if one considers the analytic terms, the integrand at a fixed order in the derivative expansion can be described diagrammatically by graphs, referred to as modular graph functions. Roughly, the vertices of the graphs are the positions of insertions of the vertex operators on the worldsheet, while the links are given by the scalar Green function connecting the vertices. These graphs depend on the moduli of the worldsheet and transform with fixed weights under $Sp(2g,\mathbb{Z})$ transformations for the genus $g$ Riemann surface, such that the integrand is $Sp(2g,\mathbb{Z})$ invariant, hence these graphs are referred to as modular. Analysis of these graphs at one loop in open and closed superstring theory has led to various insights into perturbative string theory, as well as the underlying mathematical structure~\cite{Ellis:1987dc,Abe:1988cq,Green:1999pv,Green:2008uj,Green:2013bza,Pioline:2014bra,D'Hoker:2015foa,Basu:2015ayg,DHoker:2015wxz,Basu:2016fpd,D'Hoker:2016jac,Basu:2016xrt,Basu:2016kli,Basu:2016mmk,DHoker:2016quv,Kleinschmidt:2017ege,Brown:2017qwo,Basu:2017nhs,Basu:2017zvt,Broedel:2018izr,Zerbini:2018hgs,Gerken:2018zcy,Gerken:2018jrq}.

These issues have not been as well studied beyond genus one, and our aim is to understand certain properties of some graphs at genus two. To be specific, we shall consider the low momentum expansion of the genus two four graviton amplitude in type II superstring theory~\cite{D'Hoker:2005vch,Berkovits:2005df,Berkovits:2005ng}. The integrand is simpler than other string theories, thanks to the maximal supersymmetry the type II theory enjoys. Various properties of the amplitude have been analyzed in detail upto the $D^6\mathcal{R}^4$ term, and some properties of the integrand have also been studied at higher orders in the derivative expansion~\cite{Wentworth,D'Hoker:2005jhf,Jong,D'Hoker:2013eea,D'Hoker:2014gfa,Pioline:2015qha,Basu:2015dqa,Pioline:2015nfa,DHoker:2017pvk,DHoker:2018mys}. For the $1/4$ BPS $D^4\mathcal{R}^4$ term, the integral boils down to simply the integral over the volume element of moduli space~\cite{D'Hoker:2005jhf}. For the $1/8$ BPS $D^6\mathcal{R}^4$ term, the integrand involves the Kawazumi--Zhang invariant~\cite{Kawazumi,Zhang} which satisfies an eigenvalue equation on moduli space, which allows the integral to be evaluated~\cite{D'Hoker:2014gfa}. 

Our aim is to look at some genus two modular graphs beyond the $D^6\mathcal{R}^4$ term in the low momentum expansion of the type II amplitude, which lead to non--BPS terms in the effective action. In particular, we shall consider the graphs that arise in the integrand of the $D^8\mathcal{R}^4$ term in the derivative expansion. Like other modular graphs studied at genus one and two, do these graphs satisfy some eigenvalue equation(s) on moduli space? The answer to this question generalizes in several ways the structure of the eigenvalue equations obtained in other cases, as we now summarize.

For the $D^8\mathcal{R}^4$ term, there are three topologically distinct skeleton graphs which contribute to the amplitude. While these skeleton graphs only involve scalar Green functions connecting vertices, to specify a modular graph completely one also has to specify what we call the ``dressing factors'' of the skeleton graphs. To define them, let $z_i$ and $z_j$ be two vertices in the graph (not necessarily distinct), and let $\Omega$ be the period matrix of the genus two Riemann surface defined in the standard way by the integral of the abelian differentials over the homology cycles. Then the dressing factor is defined by
\be ({\rm Im} \Omega)^{-1}_{IJ} \omega_I (z_i) \overline{\omega_J (z_j)},\ee
where $\omega_I (z ) = \omega_I (z) dz$ is the Abelian differential.  
The structure of these dressing factors is uniquely determined for the three skeleton graphs for the $D^8\mathcal{R}^4$ term given the string amplitude. Thus we see that genus two (and higher) modular graphs are specified by the choice of both the skeleton graphs as well as the dressing factors\footnote{Note that independent dressing factors given the various integrated vertices start only from genus two, because $({\rm Im} \Omega)^{-1}_{IJ} \omega_I (z_i) \overline{\omega_J (z_j)} = ({\rm Im}\tau)^{-1}$ at genus one for all $z_i,{z_j}$, where $\tau$ is the complex structure of the torus.}. 

For the three modular graphs that arise for the $D^8\mathcal{R}^4$ term, we show that a specific linear combination of them comes close to satisfying a simple second order eigenvalue differential equation on moduli space, apart from certain problematic terms. Hence we consider three other modular graphs having the same skeleton graphs that arise for the $D^8\mathcal{R}^4$ term, but having different dressing factors. We show that a specific linear combination of these three graphs also comes close to satisfying an eigenvalue equation, apart from certain problematic terms which are exactly the same as those obtained earlier for the other three graphs. Thus the difference of these two equations yields the desired eigenvalue equation. Apart from involving these six modular graphs, this eigenvalue equation also involves a seventh modular graph as a source term. The seventh modular graph also has the same skeleton graph as one of the three skeleton graphs, but with a different dressing factor compared to the other skeleton graphs. 

Note that the structure of the differential equation involves seven modular graphs of which four are not in the list of graphs for the $D^8\mathcal{R}^4$ term in the type II theory. Hence it is interesting to ask if they arise in the low momentum expansion of other amplitudes, for example, higher point amplitudes in the same theory. Also the fact that several of these graphs are involved in the same eigenvalue equation leads us to the question of asking if there are several equations these set of graphs satisfy, leading to relations among them. These issues naturally generalize analysis at genus one, and are worthwhile to study. 

We would like to mention that the structure of the eigenvalue equation we obtain seems natural, as they all involve modular graphs having the same set of skeleton graphs with different dressing factors. This is perhaps expected as no set of graphs is preferred over the others. Our analysis suggests that there should be such eigenvalue equations involving graphs at higher orders in the derivative expansion as well.          

We begin with a review of the genus two four graviton amplitude in the type II theory, followed by a discussion of the modular graphs upto the $D^8\mathcal{R}^4$ term.  To obtain the eigenvalue equation satisfied by the various modular graphs, we first analyze their holomorpic variations, followed by the anti--holomorphic variations of the holomorphic variations when the Beltrami differential is varied.  We then proceed with the similar calculation for the three more modular graphs that are relevant in order to obtain a simple eigenvalue equation. Finally we obtain the eigenvalue equation involving the seven modular graphs. In our analysis, a crucial role is played by auxiliary graphs which we introduce at various stages. These auxiliary graphs reduce to combinations of graphs we want to evaluate. However, they can be evaluated differently to give useful relations, which leads to simplifications, finally leading to the desired equation. Several technical details are given in the appendices.                                      

\section{The genus two four graviton amplitude in type II string theory}

The genus two four graviton amplitude is the same in type IIA and IIB string theory~\cite{Green:1999pu}. It is given by~\cite{D'Hoker:2005vch,Berkovits:2005df,Berkovits:2005ng}
\be \mathcal{A} = \frac{\pi}{64} \kappa_{10}^2 e^{2\phi} \mathcal{R}^4\int_{\mathcal{M}_2} \frac{\vert d^3\Omega\vert^2}{({\rm det} Y)^3} \mathcal{B} (s,t,u; \Omega,\bar\Omega),\ee
where $2\kappa_{10}^2 = (2\pi)^7 \alpha'^4$, the period matrix is given by $\Omega = X+iY$, where $X, Y$ are matrices with real entries, and the measure is 
\be \vert d^3\Omega\vert^2 = \prod_{I\leq J} i d\Omega_{IJ} \wedge d\bar\Omega_{IJ}.\ee
The integral is over $\mathcal{M}_2$, the fundamental domain of $Sp(4,\mathbb{Z})$. Also the dynamics is contained in
\be \label{def}\mathcal{B} (s,t,u; \Omega,\bar\Omega) = \int_{\S^4} \frac{\vert \mathcal{Y} \vert^2}{({\rm det} Y)^2} e^{-\alpha'\sum_{i< j}k_i \cdot k_j G(z_i,z_j)/2},\ee
where each factor of $\S$ represents an integral over the genus two worldsheet. In \C{def}, the string Green function is given by
\be G(z,w) = -{\rm ln} \vert E(z,w)\vert^2 + 2\pi Y^{-1}_{IJ} \Big({\rm Im} \int_z^w \omega_I \Big)\Big({\rm Im} \int_z^w \omega_J\Big),\ee
where $Y^{-1}_{IJ} = (Y^{-1})_{IJ}$, $E(z,w)$ is the prime form and $\omega_I$ $(I=1,2)$ are the abelian differential one forms. Also in \C{def} we have that
\be 3\mathcal{Y} = (t-u) \Delta(1,2) \wedge \Delta(3,4) + (s-t) \Delta(1,3) \wedge \Delta(4,2) +(u-s) \Delta(1,4)\wedge \Delta(2,3).\ee
where
\be \Delta(i,j) = \epsilon_{IJ} \omega_I (z_i) \wedge \omega_J (z_j) .\ee
The Mandelstam variables are given by $s= -\alpha' (k_1 + k_2)^2/4, t= -\alpha'(k_1 + k_4)^2/4, u= -\alpha' (k_1 +k_3)^2/4$. We have that $\sum_i k_i =0$ and $k_i^2 =0$.

Now \C{def} is conformally invariant as it is invariant under 
\be \label{shift}G(z,w) \rightarrow G(z,w) + c(z) + c(w)\ee
using $s+t+u=0$, even though the string Green function $G(z,w)$ is not.

While considering analytic terms in the low momentum expansion, we define
\be  \label{exp}\mathcal{B} (s,t,u; \Omega,\bar\Omega) =  \sum_{p,q=0}^\infty\mathcal{B}^{(p,q)} (\Omega,\bar\Omega) \frac{\s_2^p \s_3^q}{p!q!}\ee
where
\be \s_n =  s^n + t^n + u^n.\ee

Thus we need to analyze $\mathcal{B}^{(2,0)}$ in order to study the $D^8\mathcal{R}^4$ term. The aim it to obtain $Sp(4,\mathbb{Z})$ invariant eigenvalue equation(s) satisfied by the various modular graph functions in $\mathcal{B}^{(2,0)} (\Omega,\bar\Omega)$, along with their generalizations if needed. 

In the low momentum expansion \C{exp}, $\mathcal{B}^{(p,q)}(\Omega, \bar\Omega)$ is a sum of various graphs with distinct topologies as we shall discuss below. From \C{def} we see that each of them involves factors of $G(z,w)$ in the integrand and hence is not generically conformally invariant, even though it is modular invariant. Of course, the total contribution from all the graphs is conformally invariant. While one can consider eigenvalue equations satisfied by these graphs\footnote{The field theory limit of these graphs directly in the worldline formalism has been considered in~\cite{Basu:2018eep}.}, we shall instead consider contributions coming from graphs each of which is conformally as well as modular invariant. This is obtained by replacing \C{def} with          
\be \label{Def}\mathcal{B} (s,t,u; \Omega,\bar\Omega) = \int_{\S^4} \frac{\vert \mathcal{Y} \vert^2}{({\rm det} Y)^2} e^{-\alpha'\sum_{i< j}k_i \cdot k_j \mathcal{G}(z_i,z_j)/2}\ee
and performing the low energy expansion, where $\mathcal{G} (z,w)$ is the conformally invariant Arakelov Green function given by\footnote{See~\cite{D'Hoker:2013eea,DHoker:2017pvk} for a detailed discussion on these issues. }
\be \label{arakelov} \mathcal{G} (z,w) = G (z,w) - \g(z) - \g(w) +\g_1 \ee
which is invariant under \C{shift}. 
To define the various expressions in \C{arakelov}, consider the Kahler form
\be \kappa = \frac{1}{4} Y^{-1}_{IJ} \omega_I  \wedge \overline{\omega_J},\ee 
which satisfies
\be \int _{\S} \kappa = 1, \ee 
on using the Riemann bilinear relation\footnote{We use the convention that the volume element is given by
\be dz \wedge d \bar{z} = 2 d({\rm Re}z) \wedge d({\rm Im}z) \equiv d^2 z. \ee
Thus \C{riemann} yields
\be \int_{\S} d^2 z \omega_I (z) \overline{\omega_J (z)} = 2 Y_{IJ}.\ee 
Also $2\delta^2 (z) \equiv \delta({\rm Re}z) \delta({\rm Im}z) $, leading to
\be \int_{\S} dz \wedge d\bar{z} \delta^2 (z) = \int_{\S} d^2 z \delta^2 (z)=1.\ee}
\be \label{riemann}\int_{\S} \omega_I \wedge \overline{\omega_J} = 2 Y_{IJ}.\ee
Now in \C{arakelov}, we have that\footnote{An integral over $\S_w$ means integrating over the $w$ coordinate on the worldsheet. We shall write it as $\S$ when there is no scope for confusion.}
\be \g (z) = \int_{\S_w} \kappa(w) G(z,w)  = \frac{1}{4}Y^{-1}_{IJ}\int_{\S} d^2 w  \omega_I (w) \overline{\omega_J (w)}G(z,w), \ee
and
\be \g_1 = \int_{\S} \kappa(z) \g (z) = \frac{1}{16} Y^{-1}_{IJ} Y^{-1}_{KL}\int_{\S^2} d^2 z d^2 w  \omega_I (z) \overline{\omega_J (z)}  \omega_K (w) \overline{\omega_L (w)}G(z,w).\ee
Thus we obtain the useful relation
\be \label{vanish} \int_{\S_z} \kappa(z) \mathcal{G} (z,w)=0.\ee
Note that \C{def} and \C{Def} are exactly the same using momentum conservation, though the low momentum expansion of the later always yields conformally invariant modular graphs, hence \C{Def} is natural in string theory. Also we shall see that the low momentum expansion of \C{Def} is algebraically simpler to manipulate than that obtained from \C{def}, given the extra terms in \C{arakelov}.     

\section{The various modular graph functions}

We now consider the various modular graph functions that arise in the low momentum expansion of the four graviton amplitude, keeping terms upto the $D^8\mathcal{R}^4$ term. These are obtained from the low momentum expansion of \C{Def}.

\subsection{The $D^4\mathcal{R}^4$ term}
The $D^4\mathcal{R}^4$ term is simply given by~\cite{D'Hoker:2005jhf}

\be \mathcal{B}^{(1,0)} (\Omega,\bar\Omega) = \frac{1}{2} \int_{\S^4} \frac{\vert \Delta(1,2) \wedge \Delta(3,4)\vert^2}{({\rm det} Y)^2} = 32.\ee

In obtaining this and the various relations below, we use the relations\cite{D'Hoker:2013eea}
\bea \int_{\S_i} \Delta(i,j) \wedge \overline{\Delta(i,k)} &=& -2 {\rm det }Y ~Y^{-1}_{IJ} \omega_I (j) \wedge \overline{\omega_J (k)}, \non \\ \int_{\S_j,\S_k} \Delta(i,j) \wedge \overline{\Delta(j,k)} \wedge \Delta(k,l) &=& 4 {\rm det} Y \Delta(i,l),\eea
which follow from \C{riemann}.

\subsection{The $D^6\mathcal{R}^4$ term}

At the next order in the low momentum expansion, the $D^6\mathcal{R}^4$ term is given by~\cite{D'Hoker:2013eea,D'Hoker:2014gfa}
\bea \label{01}\mathcal{B}^{(0,1)} (\Omega,\bar\Omega) &=& -\frac{1}{3} \int_{\S^4} \frac{\vert \Delta(1,2) \wedge \Delta(3,4) - \Delta(1,4) \wedge \Delta (2,3)\vert^2}{({\rm det} Y)^2} \non \\ && \times \Big(\mathcal{G}(z_1,z_2) + \mathcal{G}(z_3,z_4) - \mathcal{G}(z_1,z_3) - \mathcal{G}(z_2,z_4)\Big) \non \\ &=& -8 \int_{\S^2} \prod_{i=1,2} d^2 z_i \mathcal{G}(z_1,z_2) \hat{P}(z_1, z_2),\eea
where
\be \hat{P}(z_1,z_2) = \Big(Y^{-1}_{IJ} Y^{-1}_{KL} - 2 Y^{-1}_{IL} Y^{-1}_{JK}\Big) \omega_I (z_1) \overline{\omega_J (z_1)} \omega_K (z_2) \overline{\omega_L(z_2)}. \ee
Now using the identity
\be \int_{\S_z}d^2 z \hat{P}(z,w)=0,\ee
from \C{01} we also have that
\be \mathcal{B}^{(0,1)} (\Omega,\bar\Omega) = -8\int_{\S^2} \prod_{i=1}^2 d^2 z_i G(z_1,z_2) \hat{P}(z_1, z_2).\ee
Alternatively on using \C{vanish}, we have that
\be \label{D6R4}\mathcal{B}^{(0,1)} (\Omega,\bar\Omega) = 16\int_{\S^2} \prod_{i=1}^2 d^2 z_i \mathcal{G}(z_1,z_2)P(z_1,z_2),\ee
where
\be \label{defP}P(z_1,z_2) = Y^{-1}_{IL} Y^{-1}_{JK}\omega_I (z_1) \overline{\omega_J (z_1)} \omega_K (z_2) \overline{\omega_L(z_2)}.\ee
The eigenvalue equation satisfied by \C{D6R4} is analyzed in appendix A.

The skeleton graph for the $D^6 \mathcal{R}^4$ term, which involves a single power of the Arakelov Green function, is depicted by figure 1. 

\begin{figure}[ht]
\begin{center}
\[
\mbox{\begin{picture}(110,30)(0,0)
\includegraphics[scale=.55]{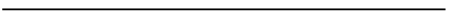}
\end{picture}}
\]
\caption{Skeleton graph for the $D^6\mathcal{R}^4$ term}
\end{center}
\end{figure}

\subsection{The $D^8\mathcal{R}^4$ term} 

Our primary interest lies in the graphs that arise in the integrand of the $D^8\mathcal{R}^4$ term. For this term, we have that~\cite{D'Hoker:2013eea,DHoker:2017pvk}
\be \mathcal{B}^{(2,0)} (\Omega,\bar\Omega) =  \frac{1}{4} \int_{\S^4} \frac{\vert \Delta(1,2) \wedge \Delta(3,4)\vert^2}{({\rm det} Y)^2}\Big(\mathcal{G}(z_1,z_4) + \mathcal{G}(z_2,z_3) -\mathcal{G}(z_1,z_3) - \mathcal{G}(z_2,z_4) \Big)^2.\ee

Thus there are modular graph functions of three distinct topologies involving two factors of the Arakelov Green function, with skeleton graphs depicted by figure 2.

\begin{figure}[ht]
\begin{center}
\[
\mbox{\begin{picture}(200,120)(0,0)
\includegraphics[scale=.65]{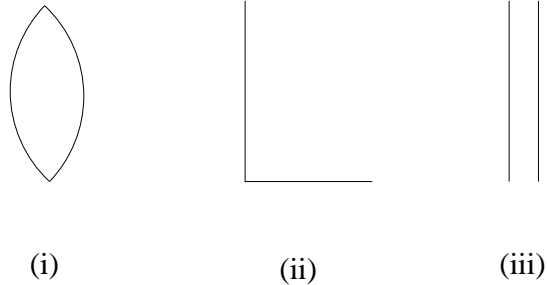}
\end{picture}}
\]
\caption{Skeleton graphs for the $D^8\mathcal{R}^4$ term}
\end{center}
\end{figure}

For the $D^8\mathcal{R}^4$ term, we denote
\be   \mathcal{B}^{(2,0)}  (\Omega,\bar\Omega)= \sum_{i=1}^3 \mathcal{B}_i^{(2,0)} (\Omega,\bar\Omega) \ee 
where each contribution arises from a separate skeleton graph, as defined below.

\subsubsection{Defining $\mathcal{B}_1^{(2,0)} $}

For the skeleton graph depicted by figure 2 (i), we have that
\bea \label{O}\mathcal{B}_1^{(2,0)} (\Omega,\bar\Omega) &=&  \int_{\S^4} \frac{\vert \Delta(1,2) \wedge \Delta(3,4)\vert^2}{({\rm det} Y)^2}\mathcal{G}(z_1,z_4)^2  \non \\ &=& 4 \int_{\S^2} \prod_{i=1}^2 d^2 z_i \mathcal{G}(z_1,z_2)^2 Q_1 (z_1,z_2),\eea
where
\be Q_1 (z_1,z_2) = Y^{-1}_{IJ} Y^{-1}_{KL} \omega_I (z_1) \overline{\omega_J (z_1)} \omega_K (z_2) \overline{\omega_L (z_2)} = \m(z_1) \m(z_2), \ee
where we have defined\footnote{Thus
\be \kappa = \frac{1}{4} \mu(z) dz \wedge d\overline{z},\ee
and \C{vanish} gives
\be \int_{\S_z} d^2 z \mu(z) \mathcal{G} (z,w)=0.\ee}
\be \m(z) = Y^{-1}_{IJ} \omega_I (z) \overline{\omega_J (z)}.\ee
Thus $P(z_1,z_2)$ (defined by \C{defP}) and $Q_1 (z_1,z_2)$ are two independent modular invariants involving the two points $z_i$ which are constructed using the abelian differentials and the imaginary part of the period matrix\footnote{As mentioned in the introduction, we refer to any such invariant as a dressing factor. To specify a modular graph, this data is needed along with the skeleton graph.}. While $Q_1(z_1,z_2)$ involves separate modular invariants at each $z_i$, $P(z_1,z_2)$ is twisted as $z_1$ and $z_2$ are intertwined. 

Thus we have the useful relation
\be \int_{\S_z} d^2 z P(z,w) = 2\m(w)\ee  
which we use at various places in our analysis. 

\subsubsection{Defining $\mathcal{B}_2^{(2,0)} $}

For the skeleton graph depicted by figure 2 (ii), we have that
\bea \label{T}\mathcal{B}_2^{(2,0)} (\Omega,\bar\Omega) &=&  -2\int_{\S^4} \frac{\vert \Delta(1,2) \wedge \Delta(3,4)\vert^2}{({\rm det} Y)^2}\mathcal{G}(z_1,z_4) \mathcal{G}(z_1,z_3)  \non \\ &=& -4 \int_{\S^3} \prod_{i=1}^3 d^2 z_i  \mathcal{G}(z_1,z_2) \mathcal{G}(z_1,z_3) Q_2 (z_1;z_2,z_3),\eea
where
\bea Q_2 (z_1;z_2,z_3) &=& Y^{-1}_{IJ} \Big(Y^{-1}_{KL} Y^{-1}_{MN} - Y^{-1}_{KN} Y^{-1}_{LM}\Big) \omega_I (z_1) \overline{\omega_J (z_1)} \omega_K (z_2) \overline{\omega_L (z_2)} \omega_M (z_3) \overline{\omega_N (z_3)}\non \\ &=& \m(z_1) \Big(\m(z_2)\m(z_3)- P(z_2,z_3)\Big).\eea
For later purposes, it is useful to note that
\be \int_{\S} d^2 z_2 Q_2 (z_1;z_2,z_3)= 2Q_1 (z_1,z_3), \quad \int_{\S} d^2 z_3 Q_2 (z_1;z_2,z_3)= 2Q_1 (z_1,z_2).\ee
Using \C{vanish}, we have that
\be \label{Two}\mathcal{B}_2^{(2,0)} (\Omega,\bar\Omega) =4 \int_{\S^3} \prod_{i=1}^3 d^2 z_i  \mathcal{G}(z_1,z_2) \mathcal{G}(z_1,z_3) \m (z_1) P(z_2,z_3),\ee
which is the expression we shall use in our analysis.

\subsubsection{Defining $\mathcal{B}_3^{(2,0)} $}

Finally for the skeleton graph depicted by figure 2 (iii), we have that
\bea \label{Th}\mathcal{B}_3^{(2,0)} (\Omega,\bar\Omega) &=&  \int_{\S^4} \frac{\vert \Delta(1,2) \wedge \Delta(3,4)\vert^2}{({\rm det} Y)^2}\mathcal{G}(z_1,z_4) \mathcal{G}(z_2,z_3) \non \\ &=&  \int_{\S^4} \prod_{i=1}^4 d^2 z_i \mathcal{G}(z_1,z_4) \mathcal{G}(z_2,z_3) Q_3 (z_1,z_2,z_3,z_4),\eea
where we have that
\bea Q_3 (z_1,z_2,z_3,z_4) &=& \Big(Y^{-1}_{IJ} Y^{-1}_{KL} - Y^{-1}_{IL} Y^{-1}_{JK}\Big)\Big(Y^{-1}_{MN} Y^{-1}_{PQ} - Y^{-1}_{MQ} Y^{-1}_{NP}\Big) \non \\&& \times \omega_I (z_1) \overline{\omega_J (z_1)} \omega_K (z_2) \overline{\omega_L (z_2)} \omega_M (z_3) \overline{\omega_N (z_3)} \omega_P (z_4) \overline{\omega_Q (z_4)} \non \\ &=& \Big(\m(z_1)\m(z_2) - P(z_1,z_2)\Big)\Big(\m(z_3)\m(z_4) - P(z_3,z_4)\Big).\eea
Again for later purpose, it is useful to note that
\be \int_{\S}d^2 z_1 Q_3 (z_1,z_2,z_3,z_4) = 2 Q_2 (z_2;z_3,z_4).\ee
Thus using \C{vanish}, we have that
\be \label{Three}\mathcal{B}_3^{(2,0)} (\Omega,\bar\Omega) = \int_{\S^4} \prod_{i=1}^4 d^2 z_i \mathcal{G}(z_1,z_4) \mathcal{G}(z_2,z_3) P(z_1,z_2) P(z_3,z_4),\ee
which is the expression we shall use in our analysis. 

Once again it is useful to note that
\bea &&\int_{\S} d^2 z \mathcal{G} (z,w) Q_1 (z,w) = \int_{\S} d^2 w \mathcal{G} (z,w) Q_1 (z,w) =0, \non \\ &&\int_{\S} d^2 z \mathcal{G} (z,w) Q_2 (z;u,v)=0.\eea

Thus we see that the dressing factors for the various skeleton graphs are fixed given the structure of the string amplitude leading to the expressions for the various modular graphs. 

\section{Varying the Beltrami differentials}

We shall obtain the eigenvalue equation by first performing holomorphic and then anti--holomorphic variations with respect to the Beltrami differentials of each modular graph. Here we briefly summarize results that are relevant for our purposes for calculating these variations. 

From now onwards, we shall use the notation
\be Y^{-1}_{IJ} \omega_I (z_1) \overline{\omega_J (z_2)} = (z_1,\overline{z_2})\ee
for the dressing factors for brevity\footnote{Thus $\m (z) = (z,\overline{z})$ and $P(z_1,z_2) = (z_1,\overline{z_2}) (z_2,\overline{z_1})$.}.

The single--valued string Green function satisfies
\bea \label{eigen}\overline\p_{w}\p_zG(z,w) &=& 2\pi\delta^2 (z-w) - \pi (z,\overline{w}), \non \\ \overline\p_{z}\p_zG(z,w) &=& -2\pi\delta^2 (z-w) + \pi \m(z),
\eea
which leads to the equations satisfied by the Arakelov Green function 
\bea \label{Eigen}\overline\p_{w}\p_z\mathcal{G}(z,w) &=& 2\pi\delta^2 (z-w) - \pi (z,\overline{w}), \non \\ \overline\p_{z}\p_z \mathcal{G}(z,w) &=& -2\pi\delta^2 (z-w) + \frac{\pi}{2} \m(z)
\eea
on using \C{arakelov}. 

The holomorphic deformation with respect to the Beltrami differential $\m$ is given by
\be \delta_{\m} \phi = \frac{1}{2\pi}\int_{\S} d^2 w \mu_{\bar{w}}^{~w} \delta_{ww}\phi,\ee
where the basic variations of the Abelian differentials, period matrix and prime form are given by~\cite{DHoker:1988pdl,Verlinde:1986kw}
\bea \label{beltvar}\delta_{ww} \omega_I (z) &=& \omega_I (w) \p_z\p_w {\rm ln} E(z,w), \non \\ \delta_{ww} \Omega_{IJ} &=& 2\pi i\omega_I (w) \omega_J (w), \non \\ \delta_{ww} {\rm ln} E(u,v) &=& -\frac{1}{2} \Big( \p_w {\rm ln} E(w,u) - \p_w {\rm ln} E(w,v)\Big)^2.\eea

This leads to the useful formula~\cite{D'Hoker:2014gfa} 
\be \delta_{ww} \Big( Y^{-1}_{IJ} \omega_J (z)\Big) = - Y^{-1}_{IJ} \omega_J (w) \p_z \p_w G(w,z),\ee
as well as
\be \delta_{ww} G(u,v) = \frac{1}{2} \Big( \p_w G(w,u) - \p_w G(w,v)\Big)^2.\ee

\section{Calculating the holomorphic variations}

Using the above results, we now calculate the holomorphic variations for the three modular graphs in $\mathcal{B}^{(2,0)}$.

\subsection{Equation involving $\mathcal{B}^{(2,0)}_1$}

First let us consider the variation resulting from varying $\mathcal{B}^{(2,0)}_1$ in \C{O}.
We have that
\bea \label{wQ1}\delta_{ww} Q_1 (z_1,z_2) = - \Big(\p_w \p_{z_1} \mathcal{G} (w,z_1)(w, \overline{z_1}) \m (z_2)+\p_w \p_{z_2} \mathcal{G} (w,z_2) (w,\overline{z_2}) \m(z_1)\Big).\eea
Both the terms in \C{wQ1} contribute equally to the integral in \C{O}. 
We next consider $\delta_{ww} \mathcal{G} (z_1,z_2)$. In the intermediate steps here as well as later, it is useful to note that the string Green function is single valued and hence we can freely integrate by parts such terms in the Arakelov Green function and discard total derivatives appropriately. This gives us that 
\bea \label{varG}\delta_{ww} \mathcal{G} (z_1,z_2) &=& - \p_w \mathcal{G}(w,z_1) \p_w \mathcal{G}(w,z_2) \non \\ &&- \frac{1}{4} \int_{\S} d^2 u (w,\overline{u}) \p_w \mathcal{G} (w,u) \p_u \Big( \mathcal{G} (u,z_1) + \mathcal{G} (u,z_2)\Big)\eea
leading to manifestly conformally covariant variations involving the Arakelov Green function.

We see that the second term in \C{varG} (given in brackets, involving a sum of two terms neither of which depends on both $z_1$ and $z_2$) does not contribute to the variation given by $\int_{\S^2} \prod_{i=1,2} d^2 z_i \mathcal{G}(z_1,z_2) Q_1 (z_1,z_2) \delta_{ww}\mathcal{G}(z_1,z_2)$.
Thus we get that
\bea \label{one}\frac{\delta_{ww} \mathcal{B}^{(2,0)}_1}{8} &=&  -\int_{\S^2} \prod_{i=1}^2 d^2 z_i \mathcal{G}(z_1,z_2) Q_1 (z_1,z_2)\p_w \mathcal{G}(w,z_1) \p_w \mathcal{G}(w,z_2)\non \\ &&+ 2   \int_{\S^2}\prod_{i=1}^2 d^2 z_i \mathcal{G}(z_1,z_2) \p_{z_1} \mathcal{G}(z_1,z_2) \p_w \mathcal{G}(w,z_1) (w,\overline{z_1}) \m(z_2) .\eea

\subsection{Equation involving $\mathcal{B}^{(2,0)}_2$}

Next we consider the variation resulting from varying $\mathcal{B}^{(2,0)}_2$ in \C{Two}. Proceeding as before, we have that
\bea \label{wQ2}&&\delta_{ww} \Big(\m(z_1)P(z_2,z_3)\Big) = - \p_w \p_{z_1} \mathcal{G}(w,z_1) (w,\overline{z_1}) P(z_2,z_3) \non \\  &&- \m(z_1)\Big(\p_w \p_{z_2} \mathcal{G}(w,z_2)(w,\overline{z_3}) (z_3,\overline{z_2})+ \p_w \p_{z_3} \mathcal{G}(w,z_3)(z_2,\overline{z_3}) (w,\overline{z_2})\Big)  .  \eea
The last two terms in \C{wQ2} contribute equally to the integral in \C{Two}. 
As before, the  second term in \C{varG} does not contribute to $ \int_{\S^3} \prod_{i=1}^3 d^2 z_i \mathcal{G}(z_1,z_3) \m (z_1) P(z_2,z_3) \delta_{ww} \mathcal{G}(z_1,z_2)$.
This leads to
\bea \label{two}-\frac{\delta_{ww} \mathcal{B}^{(2,0)}_2}{8} &=&  \int_{\S^3} \prod_{i=1}^3 d^2 z_i \mathcal{G}(z_1,z_3) \m (z_1)P(z_2,z_3) \p_w \mathcal{G}(w,z_1) \p_w \mathcal{G}(w,z_2) \non \\ &&- \int_{\S^3}  \prod_{i=1}^3 d^2 z_i\mathcal{G}(z_1,z_3) \Big[ \p_{z_1}\mathcal{G}(z_1,z_2)  \p_w \mathcal{G}(w,z_1)   (w,\overline{z_1}) P(z_2,z_3)\non \\ &&+  \m(z_1)\p_{z_2} \mathcal{G}(z_1,z_2) \p_w \mathcal{G}(w,z_2) (z_3,\overline{z_2})  (w,\overline{z_3})\Big].  \eea

\subsection{Equation involving $\mathcal{B}^{(2,0)}_3$}

Finally let us consider the variation resulting from varying $\mathcal{B}^{(2,0)}_3$ in \C{Three}.
Proceeding as before, we have that
\bea \label{wQ3}&&\delta_{ww} \Big(P (z_1,z_2)P(z_3,z_4)\Big) = - \Big[\Big( \p_w \p_{z_1} \mathcal{G}(w,z_1) (w,\overline{z_2})(z_2,\overline{z_1})\non \\ &&+ \p_w \p_{z_2} \mathcal{G}(w,z_2)  (z_1,\overline{z_2})(w,\overline{z_1})\Big) P(z_3,z_4)+ \Big(\p_w \p_{z_3} \mathcal{G}(w,z_3) (w,\overline{z_4})(z_4,\overline{z_3})\non \\ &&+ \p_w \p_{z_4} \mathcal{G}(w,z_4)(z_3,\overline{z_4})(w,\overline{z_3})\Big) P(z_1,z_2)\Big].\eea

All the terms in \C{wQ3} contribute equally to the integral in \C{Three}. 
Once again, the  second term in \C{varG} does not contribute to $ \int_{\S^4} \prod_{i=1}^4 d^2 z_i \mathcal{G}(z_2,z_3) P (z_1,z_2)P(z_3,z_4) \delta_{ww} \mathcal{G}(z_1,z_4)$.
Thus we obtain
\bea \label{three}&&\frac{\delta_{ww} \mathcal{B}^{(2,0)}_3}{2} =  - \int_{\S^4} \prod_{i=1}^4 d^2 z_i \mathcal{G}(z_2,z_3) P(z_1,z_2)P(z_3,z_4)  \p_w \mathcal{G}(w,z_1) \p_w \mathcal{G}(w,z_4)\non \\ &&+ 2 \int_{\S^4} \prod_{i=1}^4 d^2 z_i \mathcal{G}(z_2,z_3) \p_{z_1} \mathcal{G}(z_1,z_4) \p_w \mathcal{G}(w,z_1)  (w,\overline{z_2})  (z_2,\overline{z_1})  P(z_3,z_4). \eea
For each of the above contributions, we have that
\be \overline\p_w (\delta_{ww} \mathcal{B}^{(2,0)}_i) =0.\ee
Hence the variations are holomorphic. Regarding this issue of holomorphicity, it is interesting at this stage to contrast with the equations obtained if one starts with non--conformal graphs instead. This is discussed in appendix B.

\section{Calculating the anti--holomorphic variations of the holomorphic variations}

Given the expressions above, to calculate the mixed variation $\overline\delta_{uu} \delta_{ww} \mathcal{B}^{(2,0)}_i$, we also use~\cite{DHoker:1988pdl}
\be \label{cxstr}\overline\delta_{uu} \p_z = 2\pi\delta^2 (z-u) \bar\p_z,\ee
which leads to~\cite{D'Hoker:2014gfa}
\be \overline\delta_{uu} \p_w G(w,z)  = -\pi \overline\p_{{w}}\delta^2(w-u) + \pi (w,\overline{u}) \Big(\overline\p_{{u}} G(u,z) - \overline\p_{{u}} G(u,w)\Big).\ee

This gives us the useful relation\footnote{This can also be obtained from the complex conjugate of \C{varG}, and \C{cxstr}.}
\bea \label{morevar}\overline\delta_{uu} \p_w \mathcal{G}(w,z)  =  \pi  (w,\overline{u}) \Big(\overline\p_u \mathcal{G}(u,z) - \frac{1}{2}\overline\p_{u} \mathcal{G}(u,w)\Big)  +\frac{\pi}{4}  \int_{\S} d^2 x (x,\overline{u}) (w,\overline{x}) \overline\p_u \mathcal{G} (u,x)\eea
thus yielding a manifestly conformally covariant expression.

We now evaluate the antiholomorphic variations of \C{one}, \C{two} and \C{three}. 

\subsection{Equation involving $\mathcal{B}^{(2,0)}_1$}

From varying \C{one}, we get that
\be \label{onevary}\frac{1}{8} \overline\delta_{uu}\delta_{ww} \mathcal{B}^{(2,0)}_1 = \sum_{\a = A}^F \Phi_{1,\a} \ee
where we now mention the various contributions. In variations of the type $\overline\delta_{uu} \p_x \mathcal{G} (x,y)$ in \C{one}, only the very first of the three terms in \C{morevar} contributes (this is the only term which depends on $y$ in $\overline\delta_{uu} \p_x \mathcal{G} (x,y)$). 

In \C{onevary}, $\Phi_{1,A}$ contains a term having four derivatives, two of which involve $\p_w$ while the remaining two involve $\overline\p_u$ derivatives, and is given by 
\be \label{1A}\Phi_{1,A} = \int_{\S^2} \prod_{i=1,2} d^2 z_i Q_1 (z_1,z_2) \p_w \mathcal{G} (w,z_1) \p_w \mathcal{G} (w,z_2) \overline\p_u \mathcal{G} (u,z_1) \overline\p_u \mathcal{G}(u,z_2).\ee

Next $\Phi_{1,B}$ is a sum of two terms, each of which has four derivatives. One of the terms has two $\p_w$ derivatives and one $\overline\p_u$ derivative, while the other term has two $\overline\p_u$ derivatives and one $\p_w$ derivative. We have that 
\bea \label{1B}\Phi_{1,B} &=& -2 \int_{\S^2} \prod_{i=1,2} d^2 z_i \m(z_2) \p_w \mathcal{G} (w,z_1)\overline\p_u \mathcal{G} (u,z_1)\Big[ (w,\overline{z_1})  \overline\p_u \mathcal{G} (u,z_2)  \p_{z_1} \mathcal{G}(z_1,z_2) \non \\ &&+  (z_1,\overline{u})\p_w\mathcal{G} (w,z_2)  \overline\p_{z_1} \mathcal{G} (z_1,z_2)\Big].\eea

We next consider several terms each having four derivatives, of which there is only one $\p_w$ derivative and one $\overline\p_u$ derivative. They are given by
\bea \label{1CDE}&&\Phi_{1,C} = 2  \int_{\S^2} \prod_{i=1,2} d^2 z_i (w,\overline{z_1})(z_2,\overline{u})\p_w \mathcal{G} (w,z_1) \overline\p_u \mathcal{G} (u,z_2)  \p_{z_1} \mathcal{G} (z_1,z_2) \overline\p_{z_2} \mathcal{G}(z_1,z_2), \non \eea
\bea
 &&\Phi_{1,D} = 2  (w,\overline{u})   \int_{\S^2} \prod_{i=1,2} d^2 z_i  \m(z_2)\p_w \mathcal{G} (w,z_1) \overline\p_u \mathcal{G} (u,z_1) \p_{z_1} \mathcal{G} (z_1,z_2) \overline\p_{z_1} \mathcal{G}(z_1,z_2), \non \\ &&\Phi_{1,E} = -\frac{1}{2}  \int_{\S^2} \prod_{i=1,2,3} d^2 z_i \m(z_3)(w,\overline{z_1})(z_2,\overline{u})\p_w \mathcal{G} (w,z_1) \overline\p_u \mathcal{G} (u,z_2) \p_{z_1} \mathcal{G} (z_1,z_3) \overline\p_{z_2} \mathcal{G} (z_2,z_3).\non \\ \eea
In this analysis, as well as the ones for the graphs below, it shall be clear later why we treat these contributions having the same derivative structure as a sum of distinct terms.

Finally, $\Phi_{1,F}$ contains $\p_w \mathcal{G} (w,z_1) \overline\p_u \mathcal{G} (u,z_2)$ as its only contribution involving derivatives, and is given by
\bea \label{1F}\Phi_{1,F} &=& 2\pi Y^{-1}_{IJ} Y^{-1}_{KL} Y^{-1}_{MN} \int_{\S^2} \prod_{i=1,2} d^2 z_i \mathcal{G} (z_1,z_2) \p_w \mathcal{G} (w,z_1) \overline\p_u \mathcal{G} (u,z_2)\omega_I (z_2) \overline{\omega_L (z_1)}  \non \\ && \times \Big(\omega_K (w) \omega_M (z_1) - \omega_K (z_1)\omega_M (w)\Big)\Big( \overline{\omega_J (z_2)} \overline{\omega_N (u)} - \overline{\omega_J (u)} \overline{\omega_N (z_2)}\Big).\eea

In our analysis of $\mathcal{B}^{(2,0)}_1$, we always ignore contact term contributions of the form $\mathcal{G} (z_1,z_2) \delta^2 (z_1-z_2)$. These are contributions from the boundary of moduli space arising from colliding vertex operators, and do not contribute to local higher derivative interactions\footnote{We shall have more to say about such ignored contributions later.}.    

Each $\Phi_{1,\alpha}$ $(\alpha =A, \ldots, F)$ is invariant under $w \leftrightarrow \bar{u}$, and hence the total variation is hermitian. 

Also neglecting contact term contributions, we have that 
\be \overline\p_w (\overline{\delta}_{uu} \delta_{ww} \mathcal{B}^{(2,0)}_1)=0,\ee 
and $\p_u (\overline{\delta}_{uu} \delta_{ww} \mathcal{B}^{(2,0)}_1)=0$, the later following from hermiticity. Hence holomorphy in $w$ and anti--holomorphy in $u$
 is maintained for the variation.

We now perform the same analysis for the holomorphic variations of the other two modular graphs. 

\subsection{Equation involving $\mathcal{B}^{(2,0)}_2$}

From varying \C{two}, we get that
\be -\frac{1}{8} \overline\delta_{uu}\delta_{ww} \mathcal{B}^{(2,0)}_2 = \sum_{\a= A}^G \Phi_{2,\a} \ee
as given below. 
In variations of the type $\overline\delta_{uu} \p_x \mathcal{G} (x,y)$ in \C{two}, as above only the very first of the three terms in \C{morevar} contributes. We classify the various contributions based on the derivative structure as we have done above. 

Thus $\Phi_{2,A}$ is the term containing four derivatives, involving $\p_w^2$ and $\overline\p_u^2$ given by
\be \label{2A}\Phi_{2,A} = -\int_{\S^3} \prod_{i=1,2,3}d^2 z_i \m(z_1) P(z_2,z_3) \p_w \mathcal{G} (w,z_1) \p_w \mathcal{G} (w,z_2) \overline{\p}_u \mathcal{G} (u,z_1) \overline{\p}_u \mathcal{G} (u,z_3).\ee 

Next $\Phi_{2,B}$ involves terms each having four derivatives. Each term has either two $\p_w$ and one $\overline\p_u$, or two $\overline\p_u$ and one $\p_w$. This gives us that
\bea \label{2B}\Phi_{2,B} &=& \int_{\S^3} \prod_{i=1,2,3}d^2 z_i \overline{\p}_u \mathcal{G} (u,z_1)\overline{\p}_u \mathcal{G} (u,z_2) \Big[ P(z_2,z_3) (w,\overline{z_1}) \p_{z_1} \mathcal{G} (z_1,z_3) \p_w \mathcal{G} (w,z_1) \non \\ &&+ \m(z_1) (w,\overline{z_2}) (z_2,\overline{z_3}) \p_{z_3} \mathcal{G} (z_1,z_3) \p_w \mathcal{G} (w,z_3)\Big] \non \\&& + \int_{\S^3} \prod_{i=1,2,3}d^2 z_i \p_w \mathcal{G} (w,z_1) \p_w \mathcal{G} (w,z_2) \Big[ P(z_2,z_3) (z_1,\overline{u}) \overline\p_{z_1} \mathcal{G} (z_1,z_3) \overline\p_u \mathcal{G} (u,z_1)\non \\ && + \m(z_1) (z_2,\overline{u}) (z_3,\overline{z_2}) \overline\p_{z_3} \mathcal{G} (z_1,z_3) \overline\p_u \mathcal{G} (u,z_3)\Big]. \eea

Next consider terms that contain four derivatives, of which there is one $\p_w$ and one $\overline\p_u$ derivative. They are given by
\bea \label{2CDE}&&\Phi_{2,C} = -\int_{\S^3} \prod_{i=1,2,3}d^2 z_i \Big[ (w,\overline{z_1}) (z_2,\overline{u}) (z_3,\overline{z_2})\p_w  \mathcal{G} (w,z_1) \overline\p_u  \mathcal{G} (u,z_3) \p_{z_1}  \mathcal{G} (z_1,z_2) \overline\p_{z_3}  \mathcal{G} (z_1,z_3)\non \\ &&+(z_1,\overline{u}) (w,\overline{z_2}) (z_2,\overline{z_3}) \overline\p_u  \mathcal{G} (u,z_1) \p_w  \mathcal{G} (w,z_3) \p_{z_3}  \mathcal{G} (z_1,z_3) \overline\p_{z_1}  \mathcal{G} (z_1,z_2)\Big], \non \\&& \Phi_{2,D} = -(w,\overline{u})\int_{\S^3} \prod_{i=1,2,3}d^2 z_i P(z_2,z_3) \p_w  \mathcal{G} (w,z_1) \overline\p_u  \mathcal{G} (u,z_1) \p_{z_1}  \mathcal{G} (z_1,z_2) \overline\p_{z_1}  \mathcal{G} (z_1,z_3) \non \\ &&- (w,\overline{u})\int_{\S^3} \prod_{i=1,2,3}d^2 z_i \m(z_1) (z_3,\overline{z_2}) \p_w  \mathcal{G} (w,z_2) \overline\p_u  \mathcal{G} (u,z_3) \p_{z_2}  \mathcal{G} (z_1,z_2) \overline\p_{z_3}  \mathcal{G} (z_1,z_3), \non \\&& \Phi_{2,E} = \frac{1}{2}\int_{\S^3} \prod_{i=1,2,3}d^2 z_i \m(z_1) (w,\overline{z_2}) (z_3,\overline{u}) \p_w  \mathcal{G} (w,z_2)  \overline\p_u  \mathcal{G} (u,z_3)  \p_{z_2}  \mathcal{G} (z_1,z_2) \overline\p_{z_3}  \mathcal{G} (z_1,z_3) \non \\ &&+\frac{1}{4} \int_{\S^4} \prod_{i=1,2,3,4}d^2 z_i P(z_2,z_3) (w,\overline{z_1}) (z_4,\overline{u}) \p_w  \mathcal{G} (w,z_1) \overline\p_u  \mathcal{G} (u,z_4) \p_{z_1}  \mathcal{G} (z_1,z_2) \overline\p_{z_4}  \mathcal{G} (z_3,z_4) .\non \\ \eea

Also $\Phi_{2,F}$ contains two derivatives only involving $\p_w$ and $\overline\p_u$, and is given by
\bea \label{2F}\Phi_{2,F} &=& 2\pi Y^{-1}_{IJ} Y^{-1}_{KL} Y^{-1}_{MN} \int_{\S^2} \prod_{i=1,2}d^2 z_i  \mathcal{G} (z_1,z_2) \p_w  \mathcal{G} (w,z_1) \overline\p_u  \mathcal{G} (u,z_1) w_K (z_2) \overline{\omega_N (z_2)} \non \\ &&\times \Big( \omega_I (z_1) \omega_M (w)- \omega_I (w)\omega_M (z_1) \Big)\Big( \overline{\omega_J (z_1)} \overline{\omega_L (u)}-\overline{\omega_J (u)} \overline{\omega_L (z_1)}\Big) \non \eea
\bea &&+ \pi Y^{-1}_{IJ} Y^{-1}_{KL} Y^{-1}_{MN} \int_{\S^3} \prod_{i=1,2,3}d^2 z_i \p_w  \mathcal{G} (w,z_1) \overline\p_u  \mathcal{G} (u,z_2) \non \\ &&\times \Big(\omega_M (z_1) \omega_K (w)- \omega_M (w)\omega_K (z_1)\Big) \Big(\overline{\omega_J (z_2)} \overline{\omega_L (u)}-\overline{\omega_J (u)} \overline{\omega_L (z_2)}\Big)\non \\ &&\times \Big[ (z_3,\overline{z_1}) \omega_I (z_2) \overline{\omega_N (z_3)}\mathcal{G}(z_2,z_3) + (z_2,\overline{z_3}) \omega_I (z_3) \overline{\omega_N (z_1)} \mathcal{G} (z_1,z_3)\Big]. \eea

Finally $\Phi_{2,G}$ contains terms with no derivatives, giving us
\bea \label{2G}\Phi_{2,G} &=& \frac{\pi^2}{2} Y^{-1}_{IJ} Y^{-1}_{KL}\int_{\S^3} \prod_{i=1,2,3}d^2 z_i \mathcal{G} (z_1,z_2) \mathcal{G} (z_1,z_3) \m(z_1) (w,\overline{z_3}) (z_2,\overline{u}) \non \\ && \times \Big(\omega_I (w) \omega_K (z_3)- \omega_I (z_3)\omega_K (w)\Big)\Big(\overline{\omega_J (z_2)} \overline{\omega_L (u)}-\overline{\omega_J (u)} \overline{\omega_L (z_2)}\Big).\eea

Now each $\Phi_{2,\alpha}$ ($\alpha= A, \ldots, G$) is invariant under $w \leftrightarrow \bar{u}$, leading to a hermitian variation\footnote{In fact, the two terms in $\Phi_{2,D}$, $\Phi_{2,E}$ and $\Phi_{2,F}$ are all individually hermitian.}. Also we have that
\be \overline\p_w \Big(\overline\delta_{uu}\delta_{ww} \mathcal{B}^{(2,0)}_2\Big)=0,\ee
and hence $\p_u (\overline\delta_{uu}\delta_{ww} \mathcal{B}^{(2,0)}_2)=0$.

\subsection{Equation involving $\mathcal{B}^{(2,0)}_3$}

Finally from varying \C{three}, we get that
\be \frac{1}{2} \overline\delta_{uu}\delta_{ww} \mathcal{B}^{(2,0)}_3 = \sum_{\a =A}^G\Phi_{3,\a} \ee
as given below. 
In variations of the type $\overline\delta_{uu} \p_x \mathcal{G} (x,y)$ in \C{two}, once again only the very first term in \C{morevar} contributes, and we classify the various contributions based on the derivatives they contain as before. 

To start with, $\Phi_{3,A}$ contains $\p_w^2$ and $\overline\p_u^2$ and is given by
\be \label{3A}\Phi_{3,A} = \int_{\S^4} \prod_{i=1,2,3,4}d^2 z_i P(z_1,z_2) P(z_3,z_4) \p_w  \mathcal{G} (w,z_1) \p_w  \mathcal{G} (w,z_4) \overline\p_u  \mathcal{G} (u,z_2) \overline\p_u  \mathcal{G} (u,z_3).\ee

Next $\Phi_{3,B}$ contains terms having either $\p_w$ and $\overline\p_u^2$, or $\overline\p_u$ and $\p_w^2$. This gives us
\bea \label{3B}&&\Phi_{3,B} = -2\int_{\S^4} \prod_{i=1,2,3,4}d^2 z_i P(z_3,z_4) (w,\overline{z_2}) (z_2,\overline{z_1})\overline\p_u  \mathcal{G} (u,z_2) \overline\p_u  \mathcal{G} (u,z_3) \p_w  \mathcal{G} (w,z_1) \p_{z_1} \mathcal{G} (z_1,z_4) \non \\ &&- 2 \int_{\S^4} \prod_{i=1,2,3,4}d^2 z_iP(z_3,z_4) (z_2,\overline{u}) (z_1,\overline{z_2}) \p_w  \mathcal{G} (w,z_2) \p_w  \mathcal{G} (w,z_3) \overline\p_u \mathcal{G} (u,z_1)\overline\p_{z_1} \mathcal{G} (z_1,z_4).\eea

We next consider terms containing four derivatives, only one being $\p_w$ and one being $\overline\p_u$. They are 
\bea \label{3CDE}\Phi_{3,C} &=& 2\int_{\S^4} \prod_{i=1,2,3,4}d^2 z_i \Psi(w,u,z_1,z_2,z_3,z_4)  (w,\overline{z_3})(z_3,\overline{z_1}) (z_2,\overline{z_4}) (z_4,\overline{u}), \non \\ \Phi_{3,D} &=& 2(w,\overline{u})\int_{\S^4} \prod_{i=1,2,3,4}d^2 z_i \Psi(w,u,z_1,z_2,z_3,z_4)  (z_2,\overline{z_1}) P(z_3,z_4), \non \\ \Phi_{3,E} &=& -\int_{\S^4} \prod_{i=1,2,3,4}d^2 z_i  \Psi(w,u,z_1,z_2,z_3,z_4) (w,\overline{z_1}) (z_2,\overline{u}) P(z_3,z_4),\eea
where
\be \Psi(w,u,z_1,z_2,z_3,z_4) \equiv \p_w \mathcal{G} (w,z_1) \overline\p_u \mathcal{G} (u,z_2) \p_{z_1} \mathcal{G} (z_1,z_4) \overline\p_{z_2} \mathcal{G} (z_2,z_3) .\ee

Also $\Phi_{3,F}$ contains only two derivatives $\p_w$ and $\overline\p_u$ and is given by
\bea \label{3F}&&\Phi_{3,F} = \frac{\pi}{2} Y^{-1}_{IJ} Y^{-1}_{KL} Y^{-1}_{MN} Y^{-1}_{PQ}\int_{\S^3} \prod_{i=1,2,3}d^2 z_i \mathcal{G} (z_2,z_3) \p_w \mathcal{G}(w,z_1) \overline\p_u \mathcal{G} (u,z_1)\non \\ &&\times \Big(\omega_I (z_2)\omega_K (z_3)-\omega_I (z_3)\omega_K (z_2)\Big)\Big(\omega_M (w)\omega_P (z_1)-\omega_M (z_1)\omega_P (w)\Big) \non \\ &&\times \Big(\overline{\omega_J (z_1)} \overline{\omega_L (u)} - \overline{\omega_J (u)} \overline{\omega_L (z_1)}\Big)\Big(\overline{\omega_N (z_2)} \overline{\omega_Q (z_3)} - \overline{\omega_N (z_3)} \overline{\omega_Q (z_2)}\Big)\non \\ &&+ 2\pi Y^{-1}_{IJ} Y^{-1}_{KL} Y^{-1}_{MN}\int_{\S^4} \prod_{i=1,2,3,4}d^2 z_i\mathcal{G} (z_2,z_3) \p_w \mathcal{G}(w,z_1) \overline\p_u \mathcal{G} (u,z_4)\omega_K (z_2) \overline{\omega_N (z_3)} \non \\ &&\times (z_4,\overline{z_2}) (z_3,\overline{z_1})\Big(\omega_M (z_1)\omega_I (w)-\omega_M (w)\omega_I (z_1) \Big)\Big(\overline{\omega_J (z_4)} \overline{\omega_L (u)} - \overline{\omega_J (u)} \overline{\omega_L (z_4)}\Big). \non \\ \eea

Finally, $\Phi_{3,G}$ contains terms without derivatives, and is given by
\bea \label{3G}\Phi_{3,G} = \pi^2 Y^{-1}_{IJ} Y^{-1}_{KL}\int_{\S^4} \prod_{i=1,2,3,4}d^2 z_i P(z_3,z_4) (w,\overline{z_2}) (z_1,\overline{u}) \mathcal{G} (z_1,z_4) \mathcal{G} (z_2,z_3)\non \\ \times \Big( \omega_I (z_2) \omega_K (w) - \omega_I (w) \omega_K (z_2)\Big) \Big(\overline{\omega_J (z_1)} \overline{\omega_L (u)} - \overline{\omega_J (u)} \overline{\omega_L (z_1)}\Big).\eea

Each $\Phi_{3,\alpha}$ ($\alpha= A, \ldots, G$) is invariant under $w \leftrightarrow \bar{u}$, leading to a hermitian variation\footnote{In fact, the two terms in $\Phi_{3,F}$ are individually hermitian.}. Also
\be\overline\p_w \Big(\overline\delta_{uu}\delta_{ww} \mathcal{B}^{(2,0)}_3\Big) =0,\ee
and hence $\p_u (\overline\delta_{uu}\delta_{ww} \mathcal{B}^{(2,0)}_3)=0$.

\section{Simplifying the structure of variations of the modular graph functions}

From the above analysis, we see that the variations $\overline\delta_{uu}\delta_{ww}$ of the modular graph functions $\mathcal{B}^{(2,0)}_i$ yield complicated expressions. We now show that though each of them is complicated by itself, we can manipulate these expressions such that certain specific combinations of these graphs yield variations which can be drastically simplified.

For this purpose, it is very useful to depict various contributions to the variations of the modular graphs schematically by skeleton graphs, where only the structure of the Green functions are depicted. In these graphs, we follow the convention that for the link from $z_i$ to $z_j$ that involves a derivative of the Arakelov Green function $\p_{z_i} \mathcal{G} (z_i,z_j)$ or $\bar\p_{z_i} \mathcal{G} (z_i,z_j)$, the tip of the arrow points towards the vertex $z_i$. The vertices depicted by $w$ and $u$ are not integrated over, while all the other vertices are integrated over. In the graphs, $\p$ is depicted by $\delta$, while $\bar\p$ is depicted by $\bar\delta$.    

First let us consider the variations that yield terms of the form $O(\p_w^2 \bar\p_u^2)$ for the various modular graphs, given by $\Phi_{1,A}$ in \C{1A}, $\Phi_{2,A}$ in \C{2A} and $\Phi_{3,A}$ in \C{3A}. These are schematically depicted by figure 3.   

\begin{figure}[ht]
\begin{center}
\[
\mbox{\begin{picture}(420,150)(0,0)
\includegraphics[scale=.7]{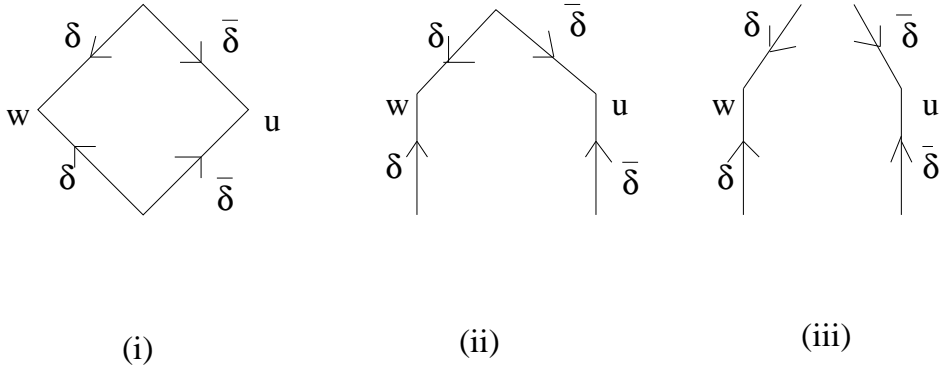}
\end{picture}}
\]
\caption{Skeleton graphs for (i) $\Phi_{1,A}$, (ii) $\Phi_{2,A}$ and (iii) $\Phi_{3,A}$}
\end{center}
\end{figure}

Next let us consider the variations that yield terms of the form $O(\p_w \bar\p_u^2)$ and its hermitian conjugate for the various modular graphs, given by \C{1B}, \C{2B} and \C{3B}. In the graphs, we only denote that $O(\p_w \bar\p_u^2)$ part, while the other can be obtained simply by hermitian conjugation. Figure 4 depicts the graphs for $\Phi_{1,B}$ and $\Phi_{3,B}$, while figure 5 depicts the graph for $\Phi_{2,B}$.  

\begin{figure}[ht]
\begin{center}
\[
\mbox{\begin{picture}(280,140)(0,0)
\includegraphics[scale=.8]{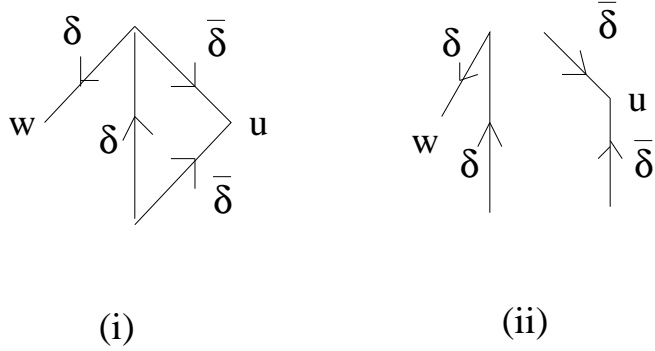}
\end{picture}}
\]
\caption{Skeleton graphs for (i) $\Phi_{1,B}$ and (ii) $\Phi_{3,B}$}
\end{center}
\end{figure}

\begin{figure}[ht]
\begin{center}
\[
\mbox{\begin{picture}(290,110)(0,0)
\includegraphics[scale=.8]{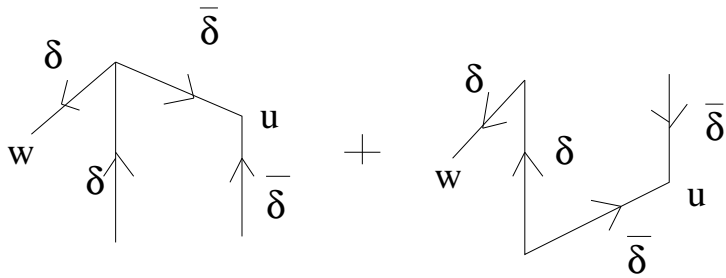}
\end{picture}}
\]
\caption{Skeleton graphs for $\Phi_{2,B}$}
\end{center}
\end{figure}

Next consider the variations having four derivatives, of which only one is $\p_w$ and one is $\bar\p_u$. Figure 6 depicts the graphs for $\Phi_{1,C}$, $\Phi_{1,D}$ and $\Phi_{1,E}$ resulting from \C{1CDE}. Figures 7, 8 and 9 depict the graphs for $\Phi_{2,C}$, $\Phi_{2,D}$ and $\Phi_{2,E}$ respectively, which result from \C{2CDE}\footnote{The second graph in figure 8 for $\Phi_{2,D}$ has the same structure as the one in figure 6 for $\Phi_{1,E}$, and the first graph in figure 9 for $\Phi_{2,E}$. We club the various contributions as we have done since it is convenient for our purposes.}. Figure 10 depicts the graphs for $\Phi_{3,C}$, $\Phi_{3,D}$ and $\Phi_{3,E}$ which result from \C{3CDE} as they have the same factor of $\Psi(w,u,z_1,z_2,z_3,z_4)$.

\begin{figure}[ht]
\begin{center}
\[
\mbox{\begin{picture}(380,130)(0,0)
\includegraphics[scale=.65]{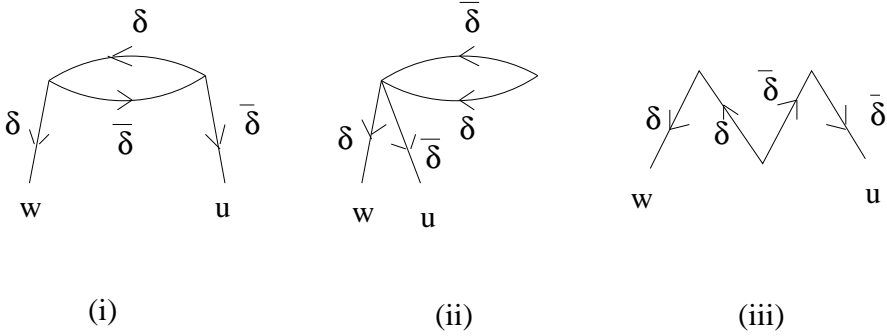}
\end{picture}}
\]
\caption{Skeleton graphs for (i) $\Phi_{1,C}$, (ii) $\Phi_{1,D}$ and (iii) $\Phi_{1,E}$}
\end{center}
\end{figure}

\begin{figure}[ht]
\begin{center}
\[
\mbox{\begin{picture}(230,110)(0,0)
\includegraphics[scale=.7]{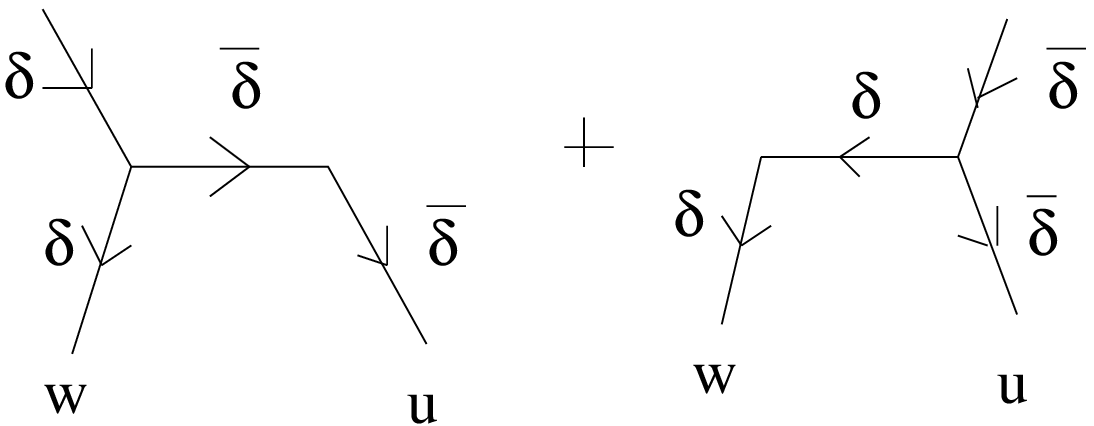}
\end{picture}}
\]
\caption{Skeleton graphs for $\Phi_{2,C}$}
\end{center}
\end{figure}

\begin{figure}[ht]
\begin{center}
\[
\mbox{\begin{picture}(230,110)(0,0)
\includegraphics[scale=.7]{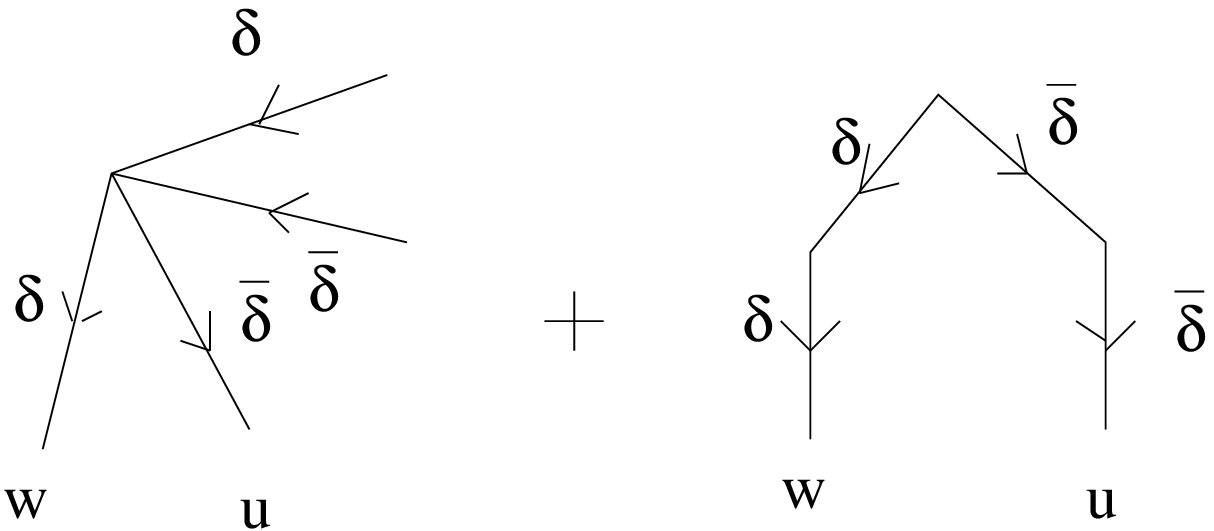}
\end{picture}}
\]
\caption{Skeleton graphs for $\Phi_{2,D}$}
\end{center}
\end{figure}

\begin{figure}[ht]
\begin{center}
\[
\mbox{\begin{picture}(360,100)(0,0)
\includegraphics[scale=.75]{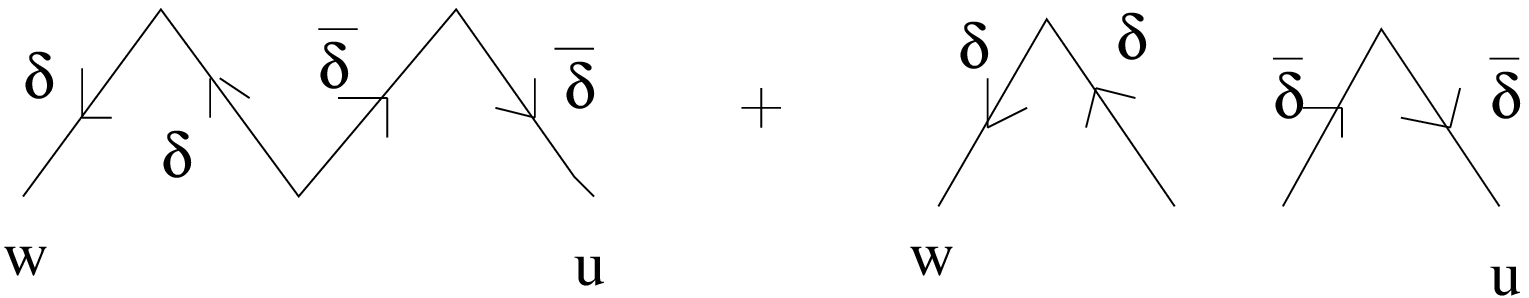}
\end{picture}}
\]
\caption{Skeleton graphs for $\Phi_{2,E}$}
\end{center}
\end{figure}

\begin{figure}[ht]
\begin{center}
\[
\mbox{\begin{picture}(170,110)(0,0)
\includegraphics[scale=.7]{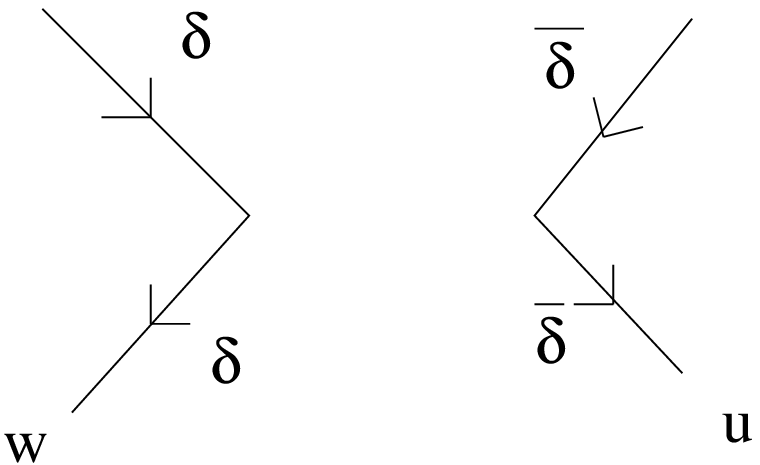}
\end{picture}}
\]
\caption{Skeleton graphs for $\Phi_{3,C}$, $\Phi_{3,D}$ and $\Phi_{3,E}$}
\end{center}
\end{figure}

All other variations involve either two derivatives (one $\p_w$ and one $\bar\p_u$) or none. We shall treat them separately in our analysis below.

The variations depicted by figures 3 to 10 all involve four derivatives and are individually quite complicated, and we now proceed to analyze these variations  in detail.

First let us focus  only on those graphs that arise in the variations of $\mathcal{B}^{(2,0)}_1$ and $\mathcal{B}^{(2,0)}_2$.   

\subsection{Relating variations of $\mathcal{B}^{(2,0)}_1$ and $\mathcal{B}^{(2,0)}_2$}

For these variations, rather than considering $\Phi_{1,\alpha}$  and $\Phi_{2,\alpha}$  $(\alpha =A,\ldots, E)$ individually, it is very useful to consider the graphs $\Phi_{1,\alpha}$ and $\Phi_{2,\alpha}$ together for every $\a$.

\begin{figure}[ht]
\begin{center}
\[
\mbox{\begin{picture}(170,110)(0,0)
\includegraphics[scale=.7]{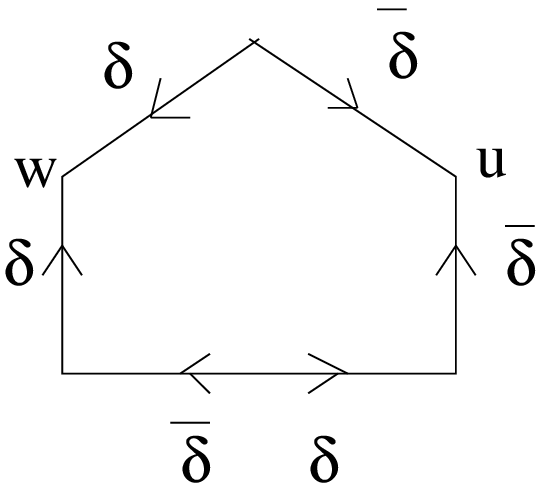}
\end{picture}}
\]
\caption{Skeleton graph for $\Phi_{12,A}$}
\end{center}
\end{figure}

This is because we can start from an auxiliary graph\footnote{Auxiliary graphs introduced in~\cite{Basu:2016xrt} for genus one graphs proved very useful in obtaining eigenvalue equations. Here we see the crucial role they play at genus two.} for each $\alpha$ such that it reduces to a linear combination of $\Phi_{1,\a}$ and $\Phi_{2,\a}$. On the other hand, these auxiliary graphs are constructed such that each of them can be manipulated to yield graphs involving only two derivatives. Hence performing such a construction for each $\alpha$, we shall see that a certain linear combination of graphs involving $\Phi_{1,\a}$ and $\Phi_{2,\a}$ for every $\a$ contains graphs with only two derivatives (one $\p$ and one $\overline\p$). Moreover, we shall see that the linear combination is precisely the same for all $\a$. Hence it is only for a specific linear combination of the variations resulting from $\mathcal{B}^{(2,0)}_1$ and $\mathcal{B}^{(2,0)}_2$ that we can get eliminate all graphs with four derivatives in terms of graphs with only two derivatives, leading to a drastic simplification.           

To start with, consider the skeleton graph depicted by figure 11. We define $\Phi_{12,A}$ as
\be \label{12A}\Phi_{12,A} = \int_{\S^3} \prod_{i=1,2,3}d^2 z_i \p_w \mathcal{G} (w,z_1) \p_w \mathcal{G} (w,z_2) \overline\p_u \mathcal{G} (u,z_1) \overline\p_u \mathcal{G} (u,z_3) \m (z_1) (z_2,\overline{z_3}) \overline\p_{z_2} \p_{z_3} \mathcal{G} (z_2,z_3).\ee
Using the first equation in \C{Eigen} for the link that has both the $\p$ and $\overline\p$ derivatives in this graph, we trivially get that   
\be \Phi_{12,A} = \pi(2\Phi_{1,A} +\Phi_{2,A}).\ee

\begin{figure}[ht]
\begin{center}
\[
\mbox{\begin{picture}(270,110)(0,0)
\includegraphics[scale=.7]{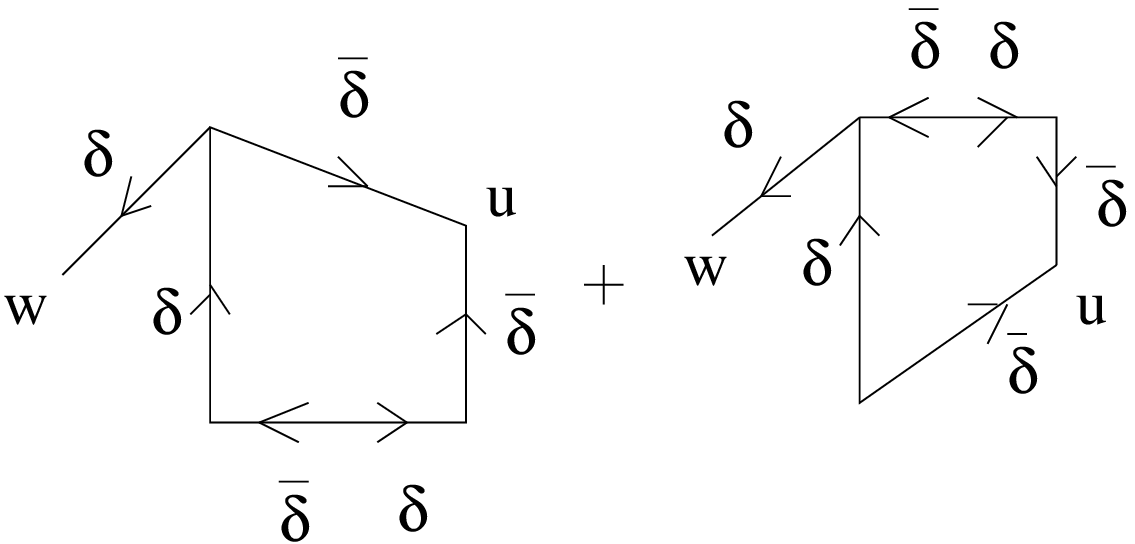}
\end{picture}}
\]
\caption{Skeleton graphs for $\Phi_{12,B}$}
\end{center}
\end{figure}

We now proceed along similar lines for the other graphs. Using the skeleton graphs depicted by figure 12 (the complete graph is given by adding the hermitian conjugate of what is given in figure 12), we define $\Phi_{12,B}$ by   
\bea \label{12B}\Phi_{12,B} &=& \int_{\S^3} \prod_{i=1,2,3}d^2 z_i  \overline\p_u \mathcal{G} (u,z_1) \overline\p_u \mathcal{G} (u,z_2) \Big[ \p_w \mathcal{G} (w,z_1)   \p_{z_1} \mathcal{G} (z_1,z_3)(z_3,\overline{z_2})(w,\overline{z_1})\non \\&& + \p_w \mathcal{G} (w,z_3) \p_{z_3} \mathcal{G} (z_1,z_3)\m(z_1) (w,\overline{z_2})\Big] \overline\p_{z_3} \p_{z_2} \mathcal{G} (z_2,z_3)\non \eea
\bea&&+\int_{\S^3} \prod_{i=1,2,3}d^2 z_i  \p_w \mathcal{G} (w,z_1)  \p_w \mathcal{G} (w,z_2) \Big[ \overline\p_u \mathcal{G} (u,z_1) \overline\p_{z_1} \mathcal{G} (z_1,z_3) (z_2,\overline{z_3}) (z_1, \overline{u}) \non \\ &&+\overline\p_u \mathcal{G} (u,z_3) \overline\p_{z_3} \mathcal{G} (z_1,z_3) \m(z_1) (z_2, \overline{u}) \Big]\overline\p_{z_2} \p_{z_3} \mathcal{G} (z_2,z_3),\eea
and using \C{Eigen}, we trivially get that
\be \Phi_{12,B} = -\pi(2\Phi_{1,B} +\Phi_{2,B}).\ee

\begin{figure}[ht]
\begin{center}
\[
\mbox{\begin{picture}(270,110)(0,0)
\includegraphics[scale=.65]{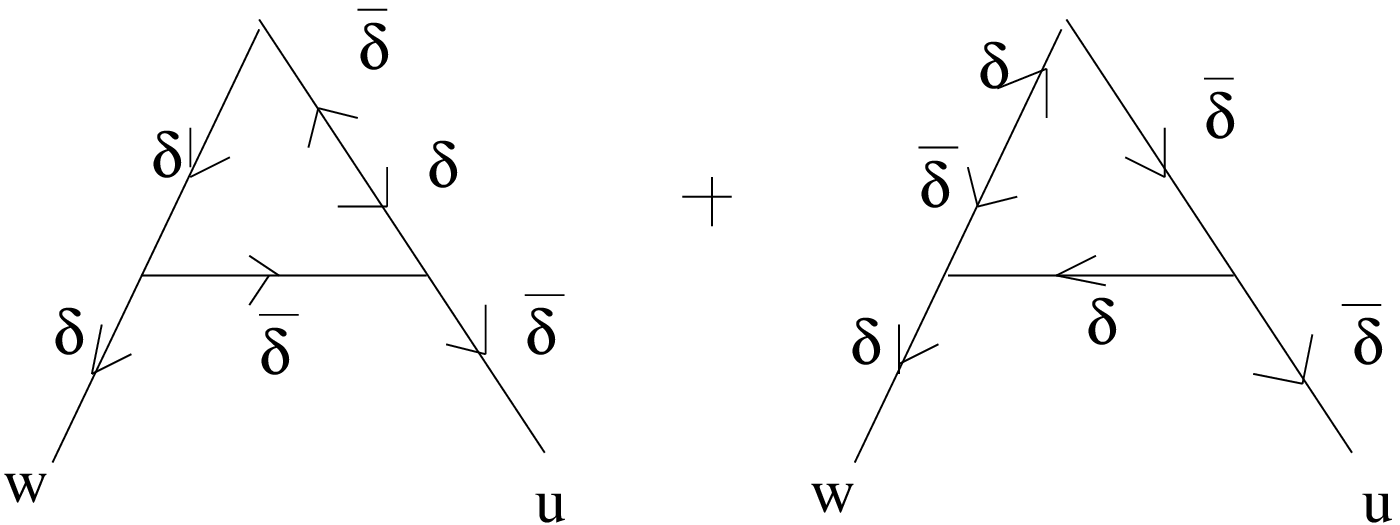}
\end{picture}}
\]
\caption{Skeleton graphs for $\Phi_{12,C}$}
\end{center}
\end{figure}

Next using the skeleton graphs depicted by figure 13, we define   
\bea \label{12C}\Phi_{12,C} &=&\int_{\S^3} \prod_{i=1,2,3}d^2 z_i  \Big[ \overline\p_{z_2} \mathcal{G} (z_1,z_2) \p_{z_1} \mathcal{G} (z_1,z_3) (w,\overline{z_1}) (z_3,\overline{u})\overline\p_{z_3} \p_{z_2} \mathcal{G} (z_2,z_3)\non \\ &&+\overline\p_{z_2} \mathcal{G} (z_2,z_3) \p_{z_1} \mathcal{G} (z_1,z_2) (w,\overline{z_3}) (z_2,\overline{u}) \overline\p_{z_1} \p_{z_3} \mathcal{G} (z_1,z_3)\Big]\p_w \mathcal{G} (w,z_1) \overline\p_u \mathcal{G} (u,z_2), \non \\ \eea
and using \C{Eigen}, we trivially get that
\be \Phi_{12,C} = \pi(2\Phi_{1,C} +\Phi_{2,C}).\ee

\begin{figure}[ht]
\begin{center}
\[
\mbox{\begin{picture}(270,130)(0,0)
\includegraphics[scale=.8]{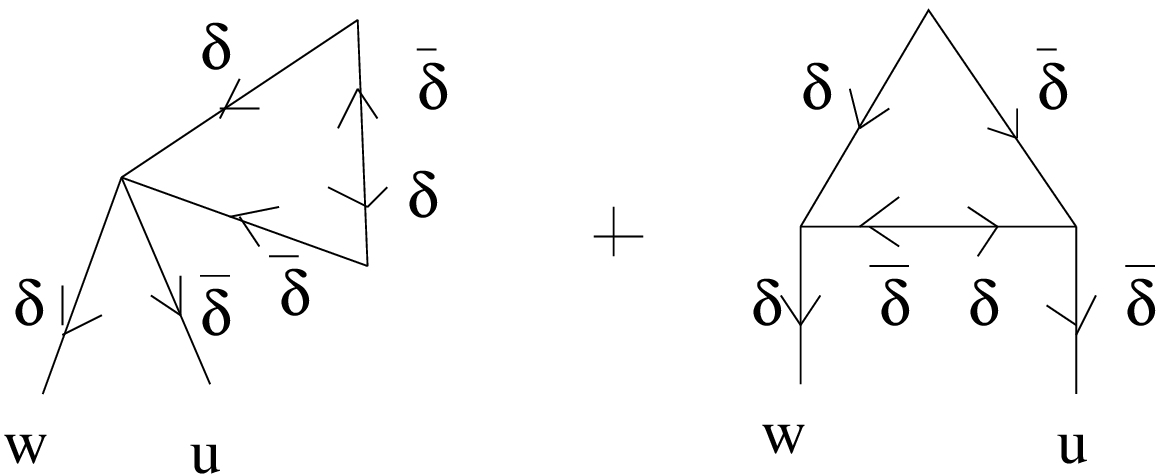}
\end{picture}}
\]
\caption{Skeleton graphs for $\Phi_{12,D}$}
\end{center}
\end{figure}

Similarly using the skeleton graphs depicted by figure 14, we define 
\bea \label{12D}\Phi_{12,D} &=& (w,\overline{u}) \int_{\S^3} \prod_{i=1,2,3}d^2 z_i  \Big[ \overline\p_u \mathcal{G} (u,z_1) \overline\p_{z_1} \mathcal{G} (z_1,z_3) (z_2,\overline{z_3}) \overline\p_{z_2} \p_{z_3} \mathcal{G} (z_2,z_3)\non \\ &&+ \overline\p_u \mathcal{G} (u,z_3) \overline\p_{z_3} \mathcal{G} (z_2,z_3) \m(z_2)\overline\p_{z_1} \p_{z_3} \mathcal{G} (z_1,z_3)\Big] \p_w \mathcal{G} (w,z_1) \p_{z_1} \mathcal{G} (z_1,z_2),\eea
and using \C{Eigen}, we trivially get that
\be \Phi_{12,D} = \pi(2\Phi_{1,D} +\Phi_{2,D}).\ee

\begin{figure}[ht]
\begin{center}
\[
\mbox{\begin{picture}(230,100)(0,0)
\includegraphics[scale=.85]{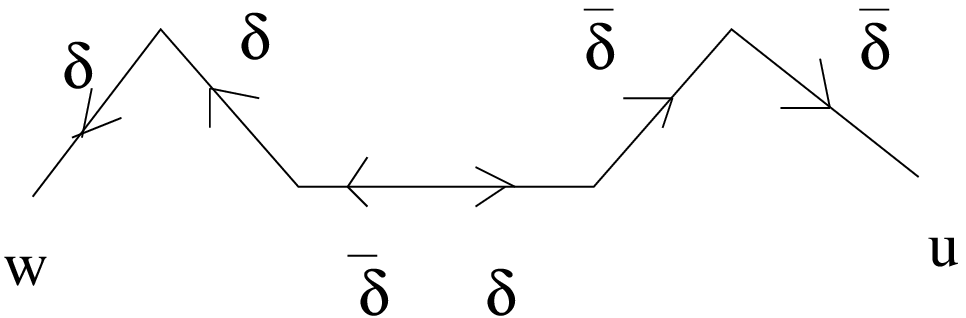}
\end{picture}}
\]
\caption{Skeleton graph for $\Phi_{12,E}$}
\end{center}
\end{figure}

Finally using the skeleton graph depicted by figure 15, we define
\bea \Phi_{12,E} &=&  \int_{\S^4} \prod_{i=1,2,3,4}d^2 z_i \p_w\mathcal{G} (w,z_1) \overline\p_u\mathcal{G} (u,z_2) \p_{z_1} \mathcal{G} (z_1,z_4) \overline\p_{z_2} \mathcal{G} (z_2,z_3) \non \\ && \times (z_2,\overline{u}) (w,\overline{z_1}) (z_4,\overline{z_3}) \overline\p_{z_4} \p_{z_3} \mathcal{G} (z_3,z_4),\eea
and using \C{Eigen}, we trivially get that
\be \Phi_{12,E} = -4\pi(2\Phi_{1,E} +\Phi_{2,E}).\ee

Adding all these contributions, we get that
\bea \label{onetwo}\frac{1}{4} \overline\delta_{uu} \delta_{ww} \Big(\mathcal{B}^{(2,0)}_1 -\frac{1}{2} \mathcal{B}^{(2,0)}_2 \Big) &=& 2 \Phi_{1,F}  + (\Phi_{2,F} +\Phi_{2,G}) \non \\ &&+\frac{1}{\pi} \Big(\Phi_{12,A} -\Phi_{12,B} +\Phi_{12,C} + \Phi_{12,D} - \frac{1}{4} \Phi_{12,E} \Big).\eea
Thus the terms in the last line of \C{onetwo} involve the various auxiliary diagrams given above. Note that we obtain the structure in \C{onetwo} because only a specific linear combination of $\Phi_{1,\a}$ and $\Phi_{2,\a}$ arises in our analysis on considering the auxiliary graphs. We expect such techniques to generalize to modular graphs at higher orders in the low momentum expansion involving more factors of the Arakelov Green function.     

Now we used the first equation in \C{Eigen} for the link containing both the $\p$ and $\overline\p$ derivatives for each of the auxiliary modular graphs to obtain the linear combination $2\Phi_{1,\alpha} +\Phi_{2,\a}$ for each $\a$, leading to \C{onetwo}. Of course, the resulting graphs have four derivatives we started with. 

However, each of these auxiliary graphs can be evaluated in a separate manner by integrating by parts the $\p$ and $\overline\p$ derivatives that are on the same link and moving them to the neighboring links. In doing so, for each graph each $\p$ encounters a $\overline\p$ (and each $\overline\p$ encounters a $\p$) which can be simplified again using \C{Eigen}. Now we are left with graphs involving only two derivatives (one $\p$ and one $\overline\p$) resulting in simpler expressions. Though we are very far from obtaining any eigenvalue equation, it is encouraging that the right hand side of \C{onetwo} only contains graphs with at most two derivatives.          

As an example, proceeding this way we have that
\bea &&\Phi_{12,A} = 4\pi^2 (w,\overline{u}) \mathcal{G}(w,u) \int_{\S} d^2 z \m(z) \p_w \mathcal{G} (w,z) \overline\p_u \mathcal{G} (u,z) \non \\ &&-  2\pi^2 \int_{\S^2} \prod_{i=1,2}d^2 z_i \m(z_1) (w,\overline{z_2}) (z_2,\overline{u})  \p_w \mathcal{G} (w,z_1) \overline\p_u \mathcal{G} (u,z_1) \Big[\mathcal{G} (w,z_2)+\mathcal{G} (u,z_2)\Big]\non \\ &&+\pi^2 \int_{\S^3} \prod_{i=1,2,3}d^2 z_i \m(z_1)  (w,\overline{z_2}) (z_2,\overline{z_3}) (z_3,\overline{u})  \mathcal{G} (z_2,z_3)\p_w \mathcal{G} (w,z_1) \overline\p_u \mathcal{G} (u,z_1).\eea

While there are many terms in the resulting expressions that are obtained on simplifying the various auxiliary graphs along the lines described above, there are also many cancellations among various terms (in particular, those involving graphs having loops), and we are left with only a few graphs having skeleton diagrams of specific topologies, which we now describe. 

\subsection{Simplifying contributions to $\mathcal{B}^{(2,0)}_1 -\mathcal{B}^{(2,0)}_2/2$}

To begin with, given \C{onetwo} let us consider the graphs that arise in
\bea \label{firstone}\Phi_{12,A} - \Phi_{12,B} +\Phi_{12,C} +\Phi_{12,D}. \eea
We write down the various contributions that contain upto two derivatives which are obtained as explained above.   

There is a potential contribution (with topology in figure 16 (i)) given by
\bea &&4\pi^2 \int_{\S^2} \prod_{i=1}^2 d^2 z_i \mathcal{G}(z_1,z_2) \p_w\mathcal{G} (w,z_1) \overline\p_u \mathcal{G} (u,z_1) \Big[ (w,\overline{z_1}) (z_1,\overline{z_2}) (z_2,\overline{u}) \non \\ &&+ (w,\overline{z_2})(z_2,\overline{z_1})(z_1,\overline{u}) - \m(z_1) (w,\overline{z_2})(z_2,\overline{u}) - (w,\overline{u})P(z_1,z_2)\Big]\non \\ &&=- \frac{4\pi^2}{{\rm det}Y} \int_{\S^2}\prod_{i=1}^2 d^2 z_i \mathcal{G}(z_1,z_2) \m(z_2)\p_w\mathcal{G} (w,z_1) \overline\p_u \mathcal{G} (u,z_1)\Delta (w,z_1) \overline{\Delta (u,z_1)}\non \\ &&= 0\eea  
leading to a vanishing contribution\footnote{$\Delta(z_i,z_j)$ is defined in \C{holbiform}.}. 

Including all the terms, the non--vanishing contributions to \C{firstone} lead to
\be \label{add12}\Phi_{12,A} - \Phi_{12,B} +\Phi_{12,C} +\Phi_{12,D} = \sum_{i=1}^5\Psi_i ,\ee
where the various graphs on the right hand side are topologically distinct. 

First, we have that $\Psi_1$ is given by 
\bea \label{Psi1}&&\Psi_{1}=\frac{\pi^2}{{\rm det}Y} \int_{\S^3} \prod_{i=1}^3 d^2 z_i (w,\overline{z_2})(w,\overline{z_3}) \mathcal{G} (z_2,z_3) \p_{z_3} \mathcal{G} (z_1,z_3) \overline\p_u \mathcal{G} (u,z_1) \Delta (z_1,z_2) \overline{\Delta (u,z_1)}\non \\ &&+\frac{\pi^2}{{\rm det}Y} \int_{\S^3} \prod_{i=1}^3 d^2 z_i (z_2,\overline{u})(z_3,\overline{u}) \mathcal{G} (z_2,z_3) \overline\p_{z_3} \mathcal{G} (z_1,z_3) \p_w \mathcal{G} (w,z_1) \overline{\Delta (z_1,z_2)} \Delta (w,z_1),\non \\\eea
as depicted by figure 16 (ii) and its hermitian conjugate. Note that it involves graphs with two derivatives one of which does not involve $\p_w$ or $\overline\p_u$. These are the only graphs of this kind.

\begin{figure}[ht]
\begin{center}
\[
\mbox{\begin{picture}(330,230)(0,0)
\includegraphics[scale=.7]{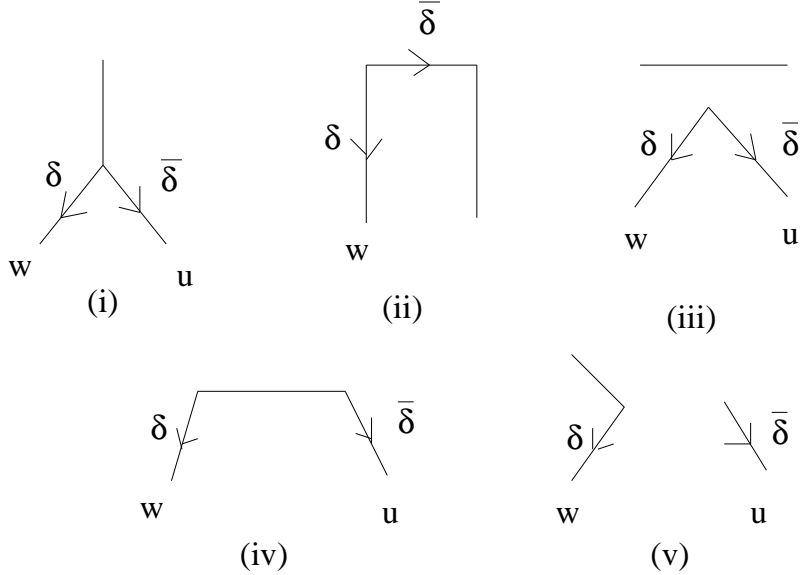}
\end{picture}}
\]
\caption{Skeleton graphs for various contributions}
\end{center}
\end{figure}

Next, $\Psi_2$ is given by
\bea \Psi_2 = \frac{\pi^2}{{\rm det}Y} \int_{\S^3} \prod_{i=1}^3 d^2 z_i P(z_2,z_3)\mathcal{G} (z_2,z_3) \p_w\mathcal{G} (w,z_1) \overline\p_u \mathcal{G} (u,z_1)\Delta (w,z_1) \overline{\Delta (u,z_1)},\eea
as depicted by figure 16 (iii), while $\Psi_3$ is given by
\bea \Psi_3 &=& \pi^2 \int_{\S^2} \prod_{i=1}^2 d^2 z_i \mathcal{G} (z_1,z_2) \p_w\mathcal{G} (w,z_1) \overline\p_u \mathcal{G} (u,z_2) \Big[ \m(z_2) (w,\overline{z_1})(z_1,\overline{u})\non \\ &&+ \m(z_1) (w,\overline{z_2})(z_2,\overline{u})- \m(z_1)\m(z_2)(w,\overline{u})\Big],\eea
as depicted by figure 16 (iv). We also have $\Psi_4$ given by
\bea \Psi_4 &=& -\frac{\pi^2}{2} \int_{\S^3} \prod_{i=1}^3 d^2 z_i \mathcal{G} (z_1,z_3) \p_w\mathcal{G} (w,z_1) \overline\p_u \mathcal{G} (u,z_2) \m(z_1) (w,\overline{z_3})(z_3,\overline{z_2}) (z_2,\overline{u}) \non \\ &&- \frac{\pi^2}{2} \int_{\S^3} \prod_{i=1}^3 d^2 z_i \mathcal{G} (z_2,z_3) \p_w\mathcal{G} (w,z_1) \overline\p_u \mathcal{G} (u,z_2)\m(z_2) (w,\overline{z_1})(z_1,\overline{z_3})(z_3,\overline{u}),\non \\ \eea
as depicted by figure 16 (v) and its hermitian conjugate. 
Finally we have $\Psi_5$, the contribution without derivatives given by
\bea \Psi_5 &=& 2\pi^3 (w,\overline{u}) \int_{\S^2} \prod_{i=1}^2 d^2 z_i \mathcal{G} (z_1,z_2)^2 \m(z_2) (w,\overline{z_1})(z_1,\overline{u}) \non \\ &&-\pi^3 (w,\overline{u}) \int_{\S^3} \prod_{i=1}^3 d^2 z_i \m(z_2) \mathcal{G} (z_1,z_2) \mathcal{G} (z_2,z_3) (w,\overline{z_1})(z_1,\overline{z_3})(z_3,\overline{u}).\eea

We can also perform a similar analysis for $\Phi_{12,E}$, which we shall discuss later. The expressions for $\Phi_{1,F}$, $\Phi_{2,F}$ and $\Phi_{2,G}$ are given in appendix C. Including them, we obtain all the terms on the right hand side of \C{onetwo}. Note that while this structure has simplified, we would like to obtain an expression without any derivatives, for reasons to be explained later. 

We now perform a similar analysis for the graphs that arise in the variations of $\mathcal{B}^{(2,0)}_2$ and $\mathcal{B}^{(2,0)}_3$. 

\subsection{Relating variations of $\mathcal{B}^{(2,0)}_2$ and $\mathcal{B}^{(2,0)}_3$}

The strategy behind the analysis is the same as what we have done above, and so we only mention the relevant auxiliary graphs.

\begin{figure}[ht]
\begin{center}
\[
\mbox{\begin{picture}(230,100)(0,0)
\includegraphics[scale=.8]{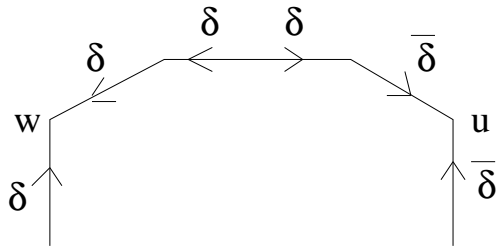}
\end{picture}}
\]
\caption{Skeleton graph for $\Phi_{23,A}$}
\end{center}
\end{figure}

Using the skeleton graph depicted by figure 17, we define 
\bea \label{23A}\Phi_{23,A} &=&  \int_{\S^4} \prod_{i=1,2,3,4}d^2 z_i \p_w \mathcal{G} (w,z_1) \p_w \mathcal{G} (w,z_2) \overline\p_u  \mathcal{G} (u,z_3) \overline\p_u \mathcal{G} (u,z_4) \non \\ && \times (z_1, \overline{z_4}) P(z_2,z_3) \overline\p_{z_1} \p_{z_4} \mathcal{G} (z_1,z_4),   \eea
which leads to
\bea \Phi_{23,A} = -\pi(2\Phi_{2,A} + \Phi_{3,A}).\eea

Next using the skeleton graphs depicted by figure 18 (the complete graph is given by adding the hermitian conjugate of what is given in figure 18), we define
\bea \label{23B} \Phi_{23,B} &=& \int_{\S^4} \prod_{i=1,2,3,4}d^2 z_i  \p_w \mathcal{G} (w,z_1) \p_{z_1} \mathcal{G} (z_1,z_4) \overline\p_u \mathcal{G} (u,z_2) \bar\p_u \mathcal{G} (u,z_3)(w,\overline{z_2}) \non \\ &&\times \Big[ P(z_3,z_4)\overline\p_{z_1} \p_{z_2} \mathcal{G} (z_1,z_2) 
+(z_4,\overline{z_3}) (z_2,\overline{z_1})\overline\p_{z_4} \p_{z_3} \mathcal{G} (z_3,z_4)\Big] \non \\ &&+ \int_{\S^4} \prod_{i=1,2,3,4}d^2 z_i  \overline\p_u \mathcal{G} (u,z_1) \overline\p_{z_1} \mathcal{G} (z_1,z_4) \p_w \mathcal{G} (w,z_2) \p_w \mathcal{G} (w,z_3)(z_2,\overline{u}) \non \\ &&\times \Big[ P(z_3,z_4)\overline\p_{z_2} \p_{z_1} \mathcal{G} (z_1,z_2) 
+(z_3,\overline{z_4}) (z_1,\overline{z_2})\overline\p_{z_3} \p_{z_4} \mathcal{G} (z_3,z_4)\Big],\eea
which gives us
\bea \Phi_{23,B} = \pi(2\Phi_{2,B} + \Phi_{3,B}).\eea

\begin{figure}[ht]
\begin{center}
\[
\mbox{\begin{picture}(360,130)(0,0)
\includegraphics[scale=.7]{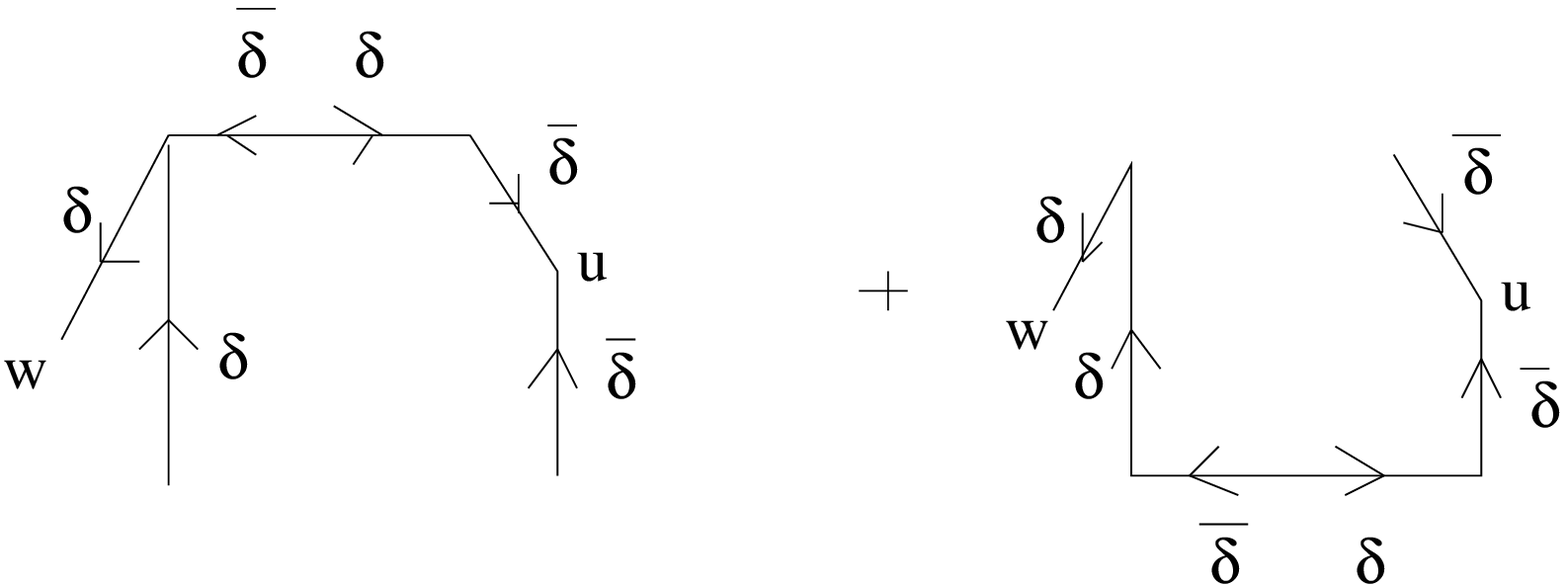}
\end{picture}}
\]
\caption{Skeleton graphs for $\Phi_{23,B}$}
\end{center}
\end{figure}

Using the skeleton graphs depicted by figure 19, we define 
\bea \label{23C}\Phi_{23,C} &=& \int_{\S^4} \prod_{i=1,2,3,4}d^2 z_i  \p_w \mathcal{G} (w,z_1) \p_{z_1} \mathcal{G} (z_1,z_4)  \overline\p_u \mathcal{G} (u,z_2) \overline\p_{z_2} \mathcal{G} (z_2,z_3) (w,\overline{z_3})(z_4,\overline{u}) \non \\&& \times \Big[(z_2,\overline{z_4}) \overline\p_{z_1} \p_{z_3} \mathcal{G} (z_1,z_3) + (z_3,\overline{z_1}) \overline\p_{z_4} \p_{z_2} \mathcal{G} (z_2,z_4)\Big],\eea
and get that
\bea \Phi_{23,C} = -\pi(2\Phi_{2,C} + \Phi_{3,C}).\eea

\begin{figure}[ht]
\begin{center}
\[
\mbox{\begin{picture}(360,130)(0,0)
\includegraphics[scale=.7]{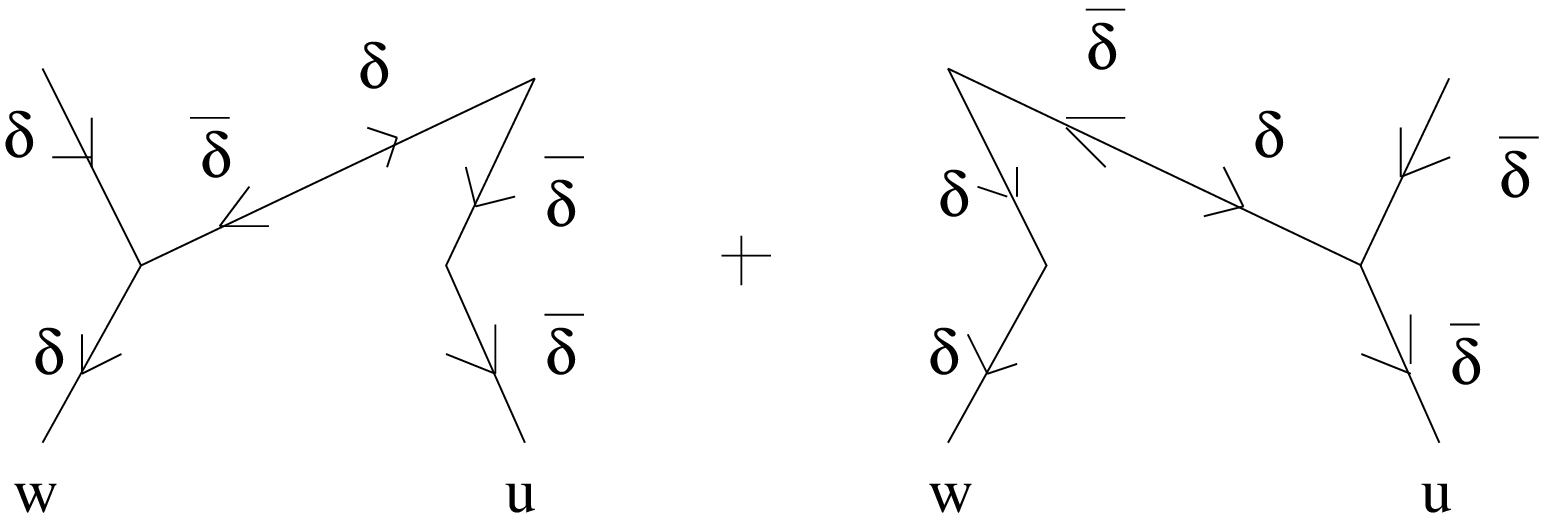}
\end{picture}}
\]
\caption{Skeleton graphs for $\Phi_{23,C}$}
\end{center}
\end{figure}

Using the skeleton graphs depicted by figure 20, we define 
\bea \label{23D}\Phi_{23,D} = (w,\overline{u})\int_{\S^4} \prod_{i=1,2,3,4}d^2 z_i \p_w \mathcal{G} (w,z_1) \p_{z_1} \mathcal{G} (z_1,z_4)\overline\p_u \mathcal{G} (u,z_2) \overline\p_{z_2} \mathcal{G} (z_2,z_3)\non \\  \times \Big[ P(z_3,z_4) \overline\p_{z_1} \p_{z_2} \mathcal{G} (z_1,z_2)+ (z_2,\overline{z_1}) (z_4,\overline{z_3}) \overline\p_{z_4} \p_{z_3} \mathcal{G} (z_3,z_4)\Big],\eea
giving us
\bea \Phi_{23,D} = -\pi(2\Phi_{2,D} + \Phi_{3,D}).\eea

Finally, simply using
\bea \Phi_{23,E} = \Phi_{12,E},\eea
we get that
\bea \Phi_{23,E} = 2\pi(2\Phi_{2,E} + \Phi_{3,E}),\eea
where we have used \C{Eigen} in each case as in the previous analysis.

\begin{figure}[ht]
\begin{center}
\[
\mbox{\begin{picture}(330,120)(0,0)
\includegraphics[scale=.7]{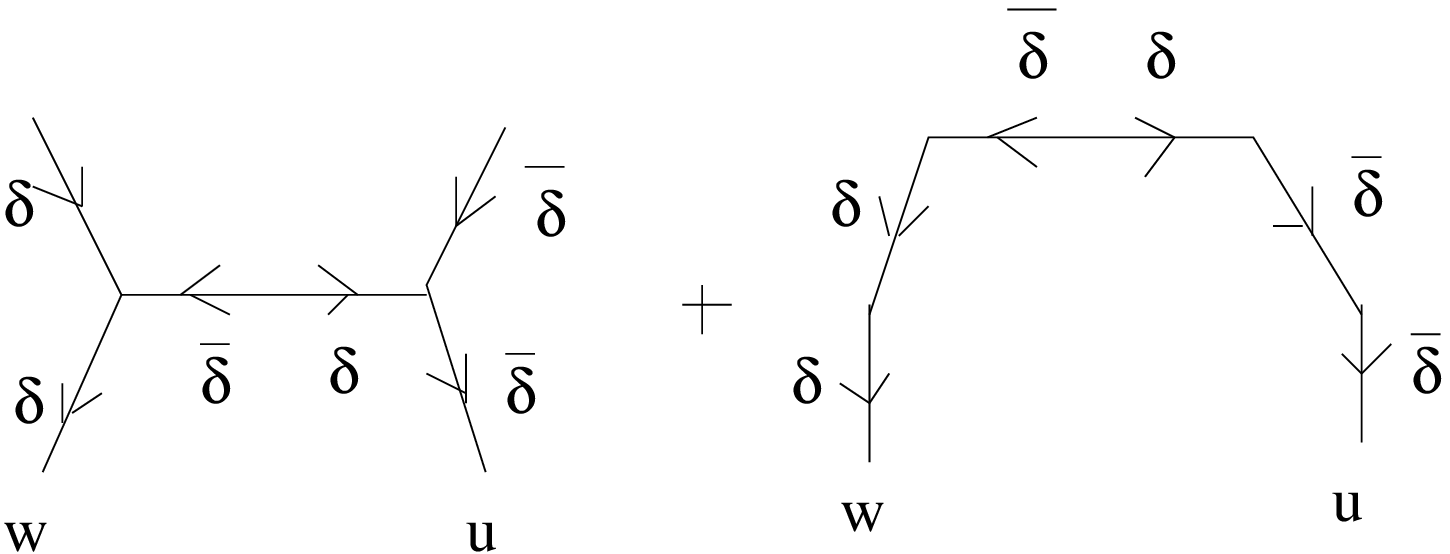}
\end{picture}}
\]
\caption{Skeleton graphs for $\Phi_{23,D}$}
\end{center}
\end{figure}

Thus adding up the various contributions, we get that
\bea \label{twothree}\frac{1}{2} \overline\delta_{uu} \delta_{ww} \Big(\mathcal{B}^{(2,0)}_3 -\frac{1}{2} \mathcal{B}^{(2,0)}_2 \Big) &=& 2 (\Phi_{2,F} +\Phi_{2,G}) + (\Phi_{3,F} +\Phi_{3,G}) \non \\ &&-\frac{1}{\pi} \Big(\Phi_{23,A} -\Phi_{23,B} +\Phi_{23,C} + \Phi_{23,D} - \frac{1}{2} \Phi_{23,E} \Big),\non \\ \eea
where the terms in the last line involving auxiliary graphs can be expressed in terms of graphs having only two derivatives. 

\subsection{Simplifying contributions to $\mathcal{B}^{(2,0)}_3 - \mathcal{B}^{(2,0)}_2/2$}

Given the right hand side of \C{twothree} and proceeding as we have done earlier, we consider the terms that arise in
\be \label{more}\Phi_{23,A} - \Phi_{23,B} +\Phi_{23,C} +\Phi_{23,D},\ee
postponing the analysis for $\Phi_{23,E}$ later. The expressions for $\Phi_{2,F}$, $\Phi_{2,G}$, $\Phi_{3,F}$ and $\Phi_{3,G}$ are given in appendix C.

\begin{figure}[ht]
\begin{center}
\[
\mbox{\begin{picture}(100,100)(0,0)
\includegraphics[scale=.7]{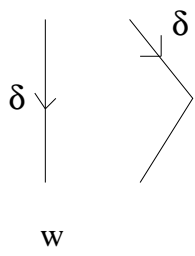}
\end{picture}}
\]
\caption{Another contribution}
\end{center}
\end{figure}

There is a contribution to \C{more} given by
\bea \label{above}&&\frac{\pi^2}{{\rm det}Y} \int_{\S^4} \prod_{i=1}^4 d^2 z_i (w,\overline{z_1})(w,\overline{z_2}) (z_3,\overline{z_4})\mathcal{G}(z_1,z_2) \p_{z_1} \mathcal{G} (z_1,z_4) \overline\p_u \mathcal{G} (u,z_3) \Delta(z_4,z_2) \overline{\Delta(u,z_3)} \non \\ &&+ \frac{\pi^2}{{\rm det}Y} \int_{\S^4} \prod_{i=1}^4 d^2 z_i (z_1,\overline{u})(z_2,\overline{u}) (z_4,\overline{z_3})\mathcal{G}(z_1,z_2) \overline\p_{z_1} \mathcal{G} (z_1,z_4) \p_w \mathcal{G} (w,z_3) \overline{\Delta(z_4,z_2)} \Delta(w,z_3) \non \\ \eea
as depicted by figure 21 and its hermitian conjugate, which is distinct from the graphs $\Psi_i$ $(i=1,\ldots,5)$ that are depicted by the skeleton graphs in figure 16. However, using
\be \label{repeat}(z_i,\overline{z_j}) = 2\delta^2 (z_i - z_j) -\frac{1}{\pi} \overline\p_{z_j} \p_{z_i} \mathcal{G} (z_i,z_j)\ee   
which follows from \C{Eigen} for $(z_3,\overline{z_4})$ and $(z_4,\overline{z_3})$ in the first and second lines of \C{above} respectively and integrating by parts, we get that \C{above} equals  
\bea \label{sim}&&\frac{2\pi^3}{{\rm det}Y} \int_{\S^3} \prod_{i=1}^3 d^2 z_i \mathcal{G} (z_1,z_2) \mathcal{G} (z_1,z_3) \Big[ (w,\overline{z_1})(w,\overline{z_2}) (z_3,\overline{u}) \Delta(z_1,z_2) \overline{\Delta(u,z_3)}\non \\ &&+ (z_1,\overline{u})(z_2,\overline{u}) (w,\overline{z_3}) \overline{\Delta(z_1,z_2)}\Delta(w,z_3)\Big]\non \\ &&-\frac{\pi^3}{{\rm det}Y} \int_{\S^4} \prod_{i=1}^4 d^2 z_i \mathcal{G} (z_1,z_2) \mathcal{G} (z_3,z_4) \Big[ (w,\overline{z_1})(w,\overline{z_2}) (z_3,\overline{u})(z_1,\overline{z_4})\Delta(z_4,z_2) \overline{\Delta(u,z_3)}\non\\ &&+ (z_1,\overline{u})(z_2,\overline{u}) (w,\overline{z_3}) (z_4,\overline{z_1}) \overline{\Delta(z_4,z_2)}\Delta(w,z_3)\Big] +2\Psi_1,\eea
where $\Psi_1$ is given in \C{Psi1}.

There is also a potential contribution with the topology of figure 16 (v) given by (along with its hermitian conjugate)
\bea &&2\pi^2 \int_{\S^3} \prod_{i=1}^3 d^2 z_i \mathcal{G} (z_1,z_3) \p_w\mathcal{G} (w,z_1) \overline\p_u \mathcal{G} (u,z_2) \Big[ (w,\overline{z_3})(z_3,\overline{z_2})(z_2,\overline{z_1})(z_1,\overline{u}) \non \\ &&- (w,\overline{u})(z_1,\overline{z_3})(z_3,\overline{z_2})(z_2,\overline{z_1}) + (w,\overline{z_2})(z_2,\overline{z_1})(z_1,\overline{z_3})(z_3,\overline{u}) -P(z_1,z_2)(w,\overline{z_3})(z_3,\overline{u})\Big]\non \\ &&= -\frac{2\pi^2}{{\rm det}Y}\int_{\S^3}\prod_{i=1}^3 d^2 z_i \mathcal{G} (z_1,z_3) \p_w\mathcal{G} (w,z_1) \overline\p_u \mathcal{G} (u,z_2)\m(z_3) (z_2,\overline{z_1})\Delta(w,z_1)\overline{\Delta(u,z_2)} \non \\ &&=0,\eea
which vanishes.

Thus adding the various contributions, we finally get that
\be \label{add23}\Phi_{23,A} - \Phi_{23,B} +\Phi_{23,C} +\Phi_{23,D} = 2\sum_{i=1}^4\Psi_i + \sum_{i=6}^8\Psi_i,\ee
where $\Psi_i$ $(i=1,\ldots,4)$ have already been defined before, and we need to define $\Psi_6$, $\Psi_7$ and $\Psi_8$.

We have that $\Psi_6$ is given by
\bea &&\Psi_{6}=-\frac{2\pi^2}{{\rm det}Y} \int_{\S^3} \prod_{i=1}^3 d^2 z_i (z_1,\overline{z_2})(w,\overline{z_3}) \mathcal{G} (z_2,z_3) \p_{z_3} \mathcal{G} (z_1,z_3) \overline\p_u \mathcal{G} (u,z_1) \Delta (w,z_2) \overline{\Delta (u,z_1)}\non \\ &&-\frac{2\pi^2}{{\rm det}Y} \int_{\S^3} \prod_{i=1}^3 d^2 z_i (z_2,\overline{z_1})(z_3,\overline{u}) \mathcal{G} (z_2,z_3) \overline\p_{z_3} \mathcal{G} (z_1,z_3) \p_w \mathcal{G} (w,z_1) \overline{\Delta (u,z_2)} \Delta (w,z_1),\non \\\eea
which has two derivatives, and has a skeleton graph depicted by figure 16 (ii).

Also $\Psi_7$ has two derivatives, and is given
by
\bea \Psi_7= \frac{8\pi^2}{{\rm det}Y} \int_{\S^2} \prod_{i=1}^2 d^2 z_i \mathcal{G}(z_1,z_2)\p_w\mathcal{G} (w,z_1) \overline\p_u\mathcal{G}(u,z_2)(z_2,\overline{z_1})\Delta (w,z_1)\overline{\Delta(u,z_2)} ,\non \\ \eea
and has a skeleton graph depicted by figure 16 (iv). 

Finally, $\Psi_8$ involves terms without any derivatives and is given by
\bea &&\Psi_8 = 2\pi^3 (w,\overline{u})  \int_{\S^3} \prod_{i=1}^3 d^2 z_i (w,\overline{z_1})(z_1,\overline{u})P(z_2,z_3)\mathcal{G} (z_1,z_2) \mathcal{G} (z_1,z_3)\non \\ &&-\pi^3 (w,\overline{u}) \int_{\S^4} \prod_{i=1}^4 d^2 z_i P(z_3,z_4) (w,\overline{z_1})(z_1,\overline{z_2})(z_2,\overline{u}) \mathcal{G}(z_1,z_4)\mathcal{G}(z_2,z_3)\non \\&&+ \frac{2\pi^3}{{\rm det}Y} \int_{\S^3} \prod_{i=1}^3 d^2 z_i \mathcal{G} (z_1,z_2) \mathcal{G} (z_1,z_3) \Big[ (w,\overline{z_1})(w,\overline{z_2}) (z_3,\overline{u}) \Delta(z_1,z_2) \overline{\Delta(u,z_3)}\non \\ &&+ (z_1,\overline{u})(z_2,\overline{u}) (w,\overline{z_3}) \overline{\Delta(z_1,z_2)}\Delta(w,z_3)\Big] \non \\ 
&&-\frac{\pi^3}{{\rm det}Y} \int_{\S^4} \prod_{i=1}^4 d^2 z_i \mathcal{G} (z_1,z_2) \mathcal{G} (z_3,z_4) \Big[ (w,\overline{z_1})(w,\overline{z_2}) (z_3,\overline{u})(z_1,\overline{z_4})\Delta(z_4,z_2) \overline{\Delta(u,z_3)}\non\\ &&+ (z_1,\overline{u})(z_2,\overline{u}) (w,\overline{z_3}) (z_4,\overline{z_1}) \overline{\Delta(z_4,z_2)}\Delta(w,z_3)\Big]\non \\ &&-\frac{\pi^3}{{\rm det}Y}  \int_{\S^4} \prod_{i=1}^4 d^2 z_i P(z_3,z_4) (w,\overline{z_1})(z_2,\overline{u})\mathcal{G} (z_1,z_2) \mathcal{G}(z_3,z_4) \Delta(w,z_1)\overline{\Delta(u,z_2)}.\eea

\subsection{Relating variations of $\mathcal{B}^{(2,0)}_1$, $\mathcal{B}^{(2,0)}_2$ and $\mathcal{B}^{(2,0)}_3$}

From \C{onetwo} and \C{twothree}, we see that the variations of the combinations of the graphs $\mathcal{B}^{(2,0)}_1 -\mathcal{B}^{(2,0)}_2/2$ and $\mathcal{B}^{(2,0)}_3 -\mathcal{B}^{(2,0)}_2/2$ are special, in the sense that the right hand side of these equations can be expressed in terms of graphs having at most two derivatives. However, we would like the right hand side of the eigenvalue equation to only have graphs with no derivatives, for reasons to be explained later. Hence it is not clear to us how to proceed and get a simple eigenvalue equation using either \C{onetwo} or \C{twothree}.  

Hence we look for an equation involving all the three modular graphs which might lead to some futher simplification compared to either \C{onetwo} or \C{twothree}.  In order to see such simplifications, we now obtain relations between the auxiliary graphs $\Phi_{12,\a}$ and $\Phi_{23,\a}$ for every $\a$. Note that $\Phi_{12,E} = \Phi_{23,E}$ and hence we are interested in $\alpha = A,B,C$ and $D$ only.          

First let us consider $\Phi_{23,A}$ in \C{23A}. Writing $P(z_2,z_3) = (z_2,\overline{z_3})(z_3,\overline{z_2})$, and using \C{repeat} for $(z_3,\overline{z_2})$, we get that
\bea \label{difA}\Phi_{23,A}  - 2 \Phi_{12,A} = -\frac{1}{\pi} \Big(\int_{\S^2} \prod_{i=1,2} d^2 z_i (z_1,\overline{z_2})\p_w \mathcal{G} (w,z_1) \overline\p_u \mathcal{G} (u,z_2) \overline\p_{z_1} \p_{z_2} \mathcal{G} (z_1,z_2)\Big)^2\eea
on using \C{12A}. Thus the left hand side of \C{difA} can be obtained from a single auxiliary graph whose skeleton graph is depicted by figure 22.  

\begin{figure}[ht]
\begin{center}
\[
\mbox{\begin{picture}(180,150)(0,0)
\includegraphics[scale=.65]{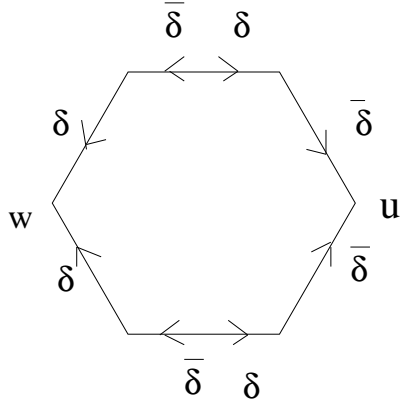}
\end{picture}}
\]
\caption{Skeleton graph for $\Phi_{23,A} - 2\Phi_{12,A}$}
\end{center}
\end{figure}

Next let us consider $\Phi_{23,B}$ in \C{23B}. Using \C{repeat} for $(z_3,\overline{z_4})$, $(z_2,\overline{z_1})$, $(z_4,\overline{z_3})$ and $(z_1,\overline{z_2})$ in the first, second, third and fourth terms in \C{23B} respectively, we get that

\bea \label{difB} \Phi_{23,B}  - 2 \Phi_{12,B} &=& - \frac{2}{\pi} \int_{\S^4} \prod_{i=1,2,3,4} d^2 z_i  \p_w \mathcal{G} (w,z_1) \overline\p_u \mathcal{G} (u,z_2) \overline\p_u \mathcal{G} (u,z_3) \p_{z_1} \mathcal{G} (z_1,z_4) \non \\ && \times \overline\p_{z_1} \p_{z_2} \mathcal{G} (z_1,z_2) \overline\p_{z_4} \p_{z_3} \mathcal{G} (z_3,z_4) (w,\overline{z_2}) (z_4,\overline{z_3}) \non \\ && - \frac{2}{\pi} \int_{\S^4} \prod_{i=1,2,3,4} d^2 z_i  \overline\p_u \mathcal{G} (u,z_1)\p_w \mathcal{G} (w,z_2) \p_w \mathcal{G} (w,z_3) \overline\p_{z_1} \mathcal{G} (z_1,z_4) \non \\ && \times \overline\p_{z_2} \p_{z_1} \mathcal{G} (z_1,z_2) \overline\p_{z_3} \p_{z_4} \mathcal{G} (z_3,z_4) (z_2,\overline{u}) (z_3,\overline{z_4})\eea
on using \C{12B}. Again, the left hand side of \C{difB} can be obtained from a single auxiliary graph whose skeleton graph is depicted by figure 23 and its hermitian conjugate. 

\begin{figure}[ht]
\begin{center}
\[
\mbox{\begin{picture}(190,120)(0,0)
\includegraphics[scale=.7]{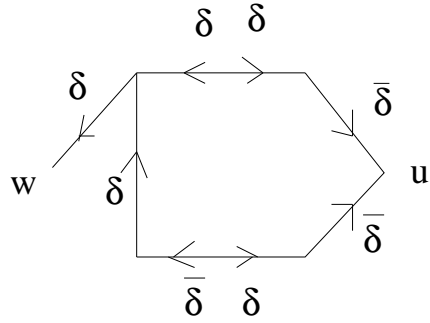}
\end{picture}}
\]
\caption{Skeleton graph for $\Phi_{23,B} - 2\Phi_{12,B}$}
\end{center}
\end{figure}

We next consider $\Phi_{23,C}$ in \C{23C}. Using \C{repeat} for $(z_2,\overline{z_4})$ and $(z_3,\overline{z_1})$ in the first and second terms in \C{23C} respectively, we get that
\bea \label{difC}\Phi_{23,C}  - 2 \Phi_{12,C} &=&-\frac{2}{\pi} \int_{\S^4} \prod_{i=1,2,3,4} d^2 z_i  \p_w \mathcal{G} (w,z_1) \overline\p_u \mathcal{G} (u,z_2) \p_{z_1} \mathcal{G} (z_1,z_4) \overline\p_{z_2} \mathcal{G} (z_2,z_3) \non \\ && \times \overline\p_{z_1} \p_{z_3} \mathcal{G} (z_1,z_3) \overline\p_{z_4} \p_{z_2} \mathcal{G} (z_2,z_4)(w,\overline{z_3}) (z_4,\overline{u})\eea
on using \C{12C}. As above, the left hand side of \C{difC} can be obtained from a single auxiliary graph whose skeleton graph is depicted by figure 24. 

\begin{figure}[ht]
\begin{center}
\[
\mbox{\begin{picture}(190,140)(0,0)
\includegraphics[scale=.7]{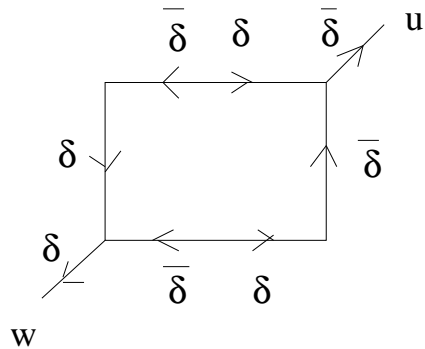}
\end{picture}}
\]
\caption{Skeleton graph for $\Phi_{23,C} - 2\Phi_{12,C}$}
\end{center}
\end{figure}

Finally let us consider $\Phi_{23,D}$ in \C{23D}. Using \C{repeat} for $(z_3,\overline{z_4})$ and $(z_2,\overline{z_1})$ in the first and second terms in \C{23D} respectively, we get that
\bea \label{difD} \Phi_{23,D}  - 2 \Phi_{12,D} &=& - \frac{2}{\pi} (w,\overline{u}) \int_{\S^4} \prod_{i=1,2,3,4} d^2 z_i  \p_w \mathcal{G} (w,z_1) \overline\p_u \mathcal{G} (u,z_2)  \p_{z_1} \mathcal{G} (z_1,z_4) \overline\p_{z_2} \mathcal{G} (z_2,z_3)\non \\ &&\times \overline\p_{z_4} \p_{z_3} \mathcal{G} (z_3,z_4) \overline\p_{z_1} \p_{z_2} \mathcal{G} (z_1,z_2) (z_4,\overline{z_3})\eea
on using \C{12D}. Once again, the left hand side of \C{difD} can be obtained from a single auxiliary graph whose skeleton graph is depicted by figure 25.

\begin{figure}[ht]
\begin{center}
\[
\mbox{\begin{picture}(170,160)(0,0)
\includegraphics[scale=.7]{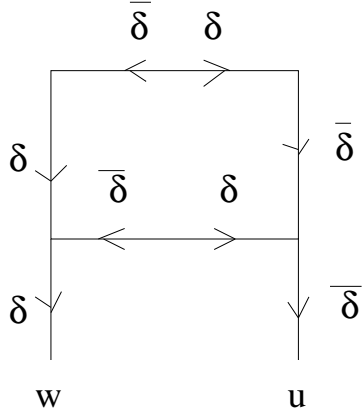}
\end{picture}}
\]
\caption{Skeleton graph for $\Phi_{23,D} - 2\Phi_{12,D}$}
\end{center}
\end{figure}

Thus we see that the combination $\Phi_{23,\a} -2 \Phi_{12,\a}$ $(\a = A,\ldots, D)$ is special, and can be derived naturally from a single auxiliary graph for each $\a$. 

Motivated by this observation, from \C{onetwo} and \C{twothree}, we have that
\bea \label{mainone}&&\frac{1}{4} \overline{\delta}_{uu} \delta_{ww} \Big(\mathcal{B}^{(2,0)}_1 - \mathcal{B}^{(2,0)}_2 +\mathcal{B}^{(2,0)}_3\Big) = 2(\Phi_{1,F} +\Phi_{1,G} + \Phi_{2,F} +\Phi_{2,G}) +\frac{1}{2} \Big( \Phi_{3,F} +\Phi_{3,G}\Big)\non \\ &&+\frac{1}{2\pi} \Big[ \Big(2\Phi_{12,A} - \Phi_{23,A}\Big) - \Big(2\Phi_{12,B} - \Phi_{23,B}\Big) + \Big(2\Phi_{12,C} - \Phi_{23,C}\Big) + \Big(2\Phi_{12,D} - \Phi_{23,D}\Big)\Big].\non \\ \eea
Note that the contributions from $\Phi_{12,E}$ and $\Phi_{23,E}$ cancel in \C{mainone} using $\Phi_{12,E} = \Phi_{23,E}$, hence we need not analyze them. Obviously, we have chosen the specific combination of \C{onetwo} and \C{twothree} in obtaining \C{mainone} such that its second line  involves $\Phi_{23,\a} -2 \Phi_{12,\a}$ $(\a = A,\ldots, D)$. 
  
\subsection{Simplifying contributions to $\mathcal{B}^{(2,0)}_1 - \mathcal{B}^{(2,0)}_2 +\mathcal{B}^{(2,0)}_3$}

Adding the various contributions involving $\Phi_{12,\a}$ and $\Phi_{23,\a}$ ($\a=1,\ldots,D$) from \C{add12} and \C{add23} and using \C{addmore}, we get that 
\bea \label{simple1}\frac{1}{4} \overline{\delta}_{uu} \delta_{ww} \Big(\mathcal{B}^{(2,0)}_1 - \mathcal{B}^{(2,0)}_2 +\mathcal{B}^{(2,0)}_3\Big) = \frac{1}{\pi}\Big[\Psi_2 - \frac{1}{2}\Big(\Psi_6 +\Psi_7\Big)\Big]+\frac{1}{2\pi}(2\Psi_5 - \Psi_8) +\widetilde{\Phi}_0,\eea
where $\widetilde{\Phi}_0$ is defined in \C{tilde}.

We see that \C{simple1} has drastically simplified compared to the earlier expressions. However, we are still left with terms involving derivatives on the right hand side of \C{simple1}, and hence it is not clear to us how to get a simple eigenvalue equation out of it. Hence we proceed differently.   

\section{Adding more modular graph functions}

One natural way of remedying the problem in \C{simple1} is to simply add more  modular graph functions to the list of $\mathcal{B}^{(2,0)}_1$, $\mathcal{B}^{(2,0)}_2$ and $\mathcal{B}^{(2,0)}_3$. This can be easily obtained by looking at the skeleton graphs in figure 2. We can simply keep the structure of the Green functions same as in these graphs but obtain different modular invariants by contracting the dressing factors $(z_i,\overline{z_j})$ differently using the locations of the integrated vertex operators. Then we can go through the same analysis we did for the previous graphs to obtain the analogue of \C{mainone} for these graphs as well, and see if further simplifications can be achieved by combining these equations. Though these new graphs we now consider do not arise in the integrand of the $D^8\mathcal{R}^4$ term, we still label them by $\mathcal{B}^{(2,0)}_i$ for the sake of uniformity as they have two factors of the Arakelov Green function.

Thus we now consider the modular graph 
\be \label{graph4}\mathcal{B}_4^{(2,0)} = 4\int_{\S^2}\prod_{i=1}^2 d^2 z_i \mathcal{G} (z_1,z_2)^2 P(z_1,z_2)\ee
which has the same skeleton graph as figure 2 (i), but different dressing factors compared to $\mathcal{B}_1^{(2,0)}$,
\be \label{graph5}\mathcal{B}_5^{(2,0)} = 4\int_{\S^3}\prod_{i=1}^3 d^2 z_i \mathcal{G} (z_1,z_2) \mathcal{G} (z_1,z_3) (z_1,\overline{z_2}) (z_2,\overline{z_3}) (z_3,\overline{z_1})\ee
which has the same skeleton graph as figure 2 (ii), but different dressing factors compared to $\mathcal{B}_2^{(2,0)}$, and
\be \label{graph6}\mathcal{B}_6^{(2,0)} =  \int_{\S^4}\prod_{i=1}^4 d^2 z_i \mathcal{G} (z_1,z_4) \mathcal{G}(z_2,z_3)(z_1,\overline{z_4}) (z_4,\overline{z_3}) (z_3,\overline{z_2})(z_2,\overline{z_1}) \ee
which has the same skeleton graph as figure 2 (iii), but different dressing factors compared to $\mathcal{B}_3^{(2,0)}$.

Another obvious graph is 
\be \mathcal{B}_7^{(2,0)} =  \int_{\S^4}\prod_{i=1}^4 d^2 z_i  \mathcal{G} (z_1,z_4) \mathcal{G}(z_2,z_3)(z_1,\overline{z_2}) (z_2,\overline{z_4}) (z_4,\overline{z_3})(z_3,\overline{z_1}),\ee
which has the same skeleton graph as figure 2 (iii), but different dressing factors compared to $\mathcal{B}_3^{(2,0)}$ and $\mathcal{B}_6^{(2,0)}$. This shall arise only at the end of  our analysis. 

Finally, one can also consider the graph $(\mathcal{B}^{(0,1)})^2$ involving the square of the graph for the $D^6\mathcal{R}^4$ term. However, this will not be needed in our analysis.

We now proceed with our analysis of the modular graphs \C{graph4}, \C{graph5} and \C{graph6}. Our strategy will be the same as what we did for the three graphs earlier, and so we shall skip some details for the sake of brevity.

\section{Calculating the holomorphic variations}

We first calculate the holomorphic variations of the graphs given by \C{graph4}, \C{graph5} and \C{graph6}. As we have done before, this is obtained by using the relevant formulae in section 5. We simply give the final expressions.

\subsection{Equation involving $\mathcal{B}_4^{(2,0)}$}

From \C{graph4}, we have that

\bea \label{4hol}\frac{\delta_{ww}\mathcal{B}_4^{(2,0)}}{8} &=& -\int_{\S^2}\prod_{i=1}^2 d^2 z_i \mathcal{G}(z_1,z_2) P(z_1,z_2) \p_w\mathcal{G} (w,z_1) \p_w\mathcal{G} (w,z_2) \non \\ &&+2 \int_{\S^2}\prod_{i=1}^2 d^2 z_i \mathcal{G}(z_1,z_2) (w,\overline{z_2}) (z_2,\overline{z_1}) \p_{z_1}\mathcal{G} (z_1,z_2) \p_w \mathcal{G} (w,z_1) \non \\ &&- \frac{1}{2} \int_{\S^3}\prod_{i=1}^3 d^2 z_i \mathcal{G} (z_1,z_2) P(z_1,z_2) (w,\overline{z_3}) \p_{z_3}\mathcal{G} (z_1,z_3)\p_w \mathcal{G} (w,z_3). \eea

\subsection{Equation involving $\mathcal{B}_5^{(2,0)}$}

From \C{graph5}, we have that
\bea \label{5hol}&&-\frac{\delta_{ww}\mathcal{B}_5^{(2,0)}}{4} = \int_{\S^3}\prod_{i=1}^3 d^2 z_i \mathcal{G} (z_1,z_3) \p_w \mathcal{G} (w,z_1) \p_w\mathcal{G} (w,z_2)  S(z_1,z_2,z_3) \non \\  &&-\int_{\S^3}\prod_{i=1}^3 d^2 z_i \mathcal{G} (z_1,z_3) \Big[\p_{z_1}\mathcal{G} (z_1,z_2) \p_w\mathcal{G} (w,z_1)R(w;z_2,z_3;\overline{z_1})\non \\ &&+ \p_{z_2}\mathcal{G} (z_1,z_2) \p_w\mathcal{G} (w,z_2)\Big(R(w;z_1,z_3;\overline{z_2})- P(z_1,z_3)(w,\overline{z_2})\Big)\Big]\non \\ &&+\frac{1}{4}\int_{\S^4}\prod_{i=1}^4 d^2 z_i \mathcal{G} (z_1,z_3) (w,\overline{z_4})S(z_1,z_2,z_3)\p_{z_4} \mathcal{G} (z_2,z_4) \p_w\mathcal{G} (w,z_4) ,\eea
where we have defined\footnote{Note that $S(z_i,z_j,z_k)$ is symmetric in its three arguments. Also\be R(z_k;z_i,z_j;\overline{z_k}) = S(z_i,z_j,z_k).\ee}
\bea S(z_i,z_j,z_k) &= &(z_i,\overline{z_j}) (z_j,\overline{z_k}) (z_k,\overline{z_i})+ (z_i,\overline{z_k}) (z_k,\overline{z_j}) (z_j,\overline{z_i}), \non \\ R(z_k;z_i,z_j;\overline{z_l}) &= & (z_k,\overline{z_i})(z_i,\overline{z_j})(z_j,\overline{z_l})+ (z_k,\overline{z_j})(z_j,\overline{z_i})(z_i,\overline{z_l})\eea
which involve various dressing factors.

\subsection{Equation involving $\mathcal{B}_6^{(2,0)}$}

Finally from \C{graph6}, we have that
\bea \label{6hol}&&\frac{1}{2}\delta_{ww} \mathcal{B}_6^{(2,0)} = -\int_{\S^4}\prod_{i=1}^4 d^2 z_i \mathcal{G} (z_2,z_3) \p_w \mathcal{G} (w,z_1) \p_w \mathcal{G} (w,z_4) (z_1,\overline{z_4}) (z_4,\overline{z_3}) (z_3,\overline{z_2}) (z_2,\overline{z_1}) \non \\&&+ \int_{\S^4}\prod_{i=1}^4 d^2 z_i   \mathcal{G} (z_2,z_3) \p_w \mathcal{G} (w,z_1) \p_{z_1} \mathcal{G} (z_1,z_4) \Big[ (w,\overline{z_3}) (z_3,\overline{z_2}) (z_2,\overline{z_4}) (z_4,\overline{z_1}) \non \eea

\bea &&+ (w,\overline{z_4}) (z_4,\overline{z_3}) (z_3,\overline{z_2}) (z_2,\overline{z_1}) - (z_4,\overline{z_2})(z_2,\overline{z_3}) (z_3,\overline{z_4}) (w,\overline{z_1})\Big].\eea

Using \C{Eigen}, we see that
\be \overline\p_w (\delta_{ww}\mathcal{B}^{(2,0)}_i) = 0\ee
for $i=4,5,6$. Thus the holomorphic variations are indeed holomorphic.

\section{Calculating the anti--holomorphic variations of the holomorphic variations}

We next calculate the various anti--holomorphic variations of the holomorphic variations given by \C{4hol}, \C{5hol} and \C{6hol}. 

\subsection{Equation involving $\mathcal{B}_4^{(2,0)}$}

From \C{4hol}, we have that
\be \frac{1}{8}\overline\delta_{uu} \delta_{ww} \mathcal{B}^{(2,0)}_4 = \sum_{\a=A}^K \Phi_{4,\a} ,\ee
where we now list the various contributions.
 
$\Phi_{4,A}$ is the $O(\p_w^2 \overline\p_u^2)$ term given by
\be \Phi_{4,A} = \int_{\S^2}\prod_{i=1}^2 d^2 z_i P(z_1,z_2) \p_w \mathcal{G} (w,z_1) \p_w\mathcal{G} (w,z_2) \overline\p_u \mathcal{G} (u,z_1) \overline\p_u \mathcal{G} (u,z_2),\ee
which has same the skeleton graph as $\Phi_{1,A}$.
$\Phi_{4,B}$ is the $O(\p_w\overline\p_u^2 \p_{z_i})$ term and its complex conjugate, and is given by
\bea \Phi_{4,B} &=& -2 \int_{\S^2}\prod_{i=1}^2 d^2 z_i \p_w \mathcal{G} (w,z_1) \overline\p_u \mathcal{G} (u,z_1)\Big[ (w,\overline{z_2})(z_2,\overline{z_1})\overline\p_u \mathcal{G} (u,z_2) \p_{z_1}\mathcal{G} (z_1,z_2)\non \\ &&+ (z_2,\overline{u}) (z_1,\overline{z_2})\p_w\mathcal{G} (w,z_2) \overline\p_{z_1} \mathcal{G} (z_1,z_2)\Big] ,\eea
which has the same skeleton graph as $\Phi_{1,B}$.
$\Phi_{4,C}$ is an $O(\p_w\overline\p_u \p_{z_i} \overline\p_{z_j})$ term given by
\be \Phi_{4,C} = 2(w,\overline{u}) \int_{\S^2}\prod_{i=1}^2 d^2 z_i (z_2,\overline{z_1})\p_w\mathcal{G} (w,z_1) \overline\p_u\mathcal{G} (u,z_2) \p_{z_1} \mathcal{G} (z_1,z_2) \overline\p_{z_2} \mathcal{G} (z_1,z_2),\ee
which has the same skeleton graph as $\Phi_{1,C}$.
$\Phi_{4,D}$ is another $O(\p_w\overline\p_u \p_{z_i} \overline\p_{z_j})$ term given by
\be \Phi_{4,D} = 2  \int_{\S^2}\prod_{i=1}^2 d^2 z_i (w,\overline{z_2}) (z_2,\overline{u})\p_w\mathcal{G} (w,z_1) \overline\p_u \mathcal{G} (u,z_1) \p_{z_1} {G} (z_1,z_2) \overline\p_{z_1} \mathcal{G} (z_1,z_2),\ee
which has the same skeleton graph as $\Phi_{1,D}$, while
$\Phi_{4,E}$ is another $O(\p_w\overline\p_u \p_{z_i} \overline\p_{z_j})$ term given by
\bea \Phi_{4,E} &=& -\frac{1}{2}  \int_{\S^3}\prod_{i=1}^3 d^2 z_i \Big[(w,\overline{z_1})(z_2,\overline{z_3})(z_3,\overline{u}) + (z_2,\overline{u})(w,\overline{z_3})(z_3,\overline{z_1})\Big] \non \\ &&\times \p_w\mathcal{G} (w,z_1) \overline\p_u\mathcal{G} (u,z_2) \p_{z_1} \mathcal{G} (z_1,z_3) \overline\p_{z_2} \mathcal{G} (z_2,z_3),\eea
which has the same skeleton graph as $\Phi_{1,E}$. Thus for all these contributions involving four derivatives, the skeleton graph of $\Phi_{4,\a}$ is the same as that of $\Phi_{1,\a}$.

Next $\Phi_{4,F}$ involves two derivatives and is given by
\bea \label{4F}&&\Phi_{4,F} = \Lambda_0 +\frac{\Lambda_1}{2} -\frac{4\pi}{{\rm det}Y} \int_{\S^2} \prod_{i=1}^2 d^2 z_i \mathcal{G} (z_1,z_2) \p_w\mathcal{G} (w,z_1) \overline\p_u\mathcal{G} (u,z_2) (z_2,\overline{z_1})\Delta (w,z_1) \overline{\Delta(u,z_2)} \non \\&& -\frac{\pi}{4{\rm det}Y} \int_{\S^3} \prod_{i=1}^3 d^2 z_i P(z_1,z_2) \mathcal{G}(z_1,z_2) \p_w\mathcal{G}(w,z_3) \overline\p_u\mathcal{G} (u,z_3) \Delta(w,z_3) \overline{\Delta(u,z_3)},\non \\ \eea
where we have that
\bea \label{def0} \Lambda_0 = \frac{\pi}{2} \int_{\S^3} \prod_{i=1}^3 d^2 z_i (z_1,\overline{z_2})(w,\overline{z_3})\mathcal{G}(z_1,z_2)\p_w \mathcal{G} (w,z_3) \overline\p_u \mathcal{G} (u,z_1) \Big[(z_2,\overline{u})(z_3,\overline{z_1}) - (z_3,\overline{u})(z_2,\overline{z_1})\Big]\non \\ +\frac{\pi}{2} \int_{\S^3} \prod_{i=1}^3 d^2 z_i (z_2,\overline{z_3})(z_1,\overline{u})\mathcal{G}(z_2,z_3)\p_w \mathcal{G} (w,z_3) \overline\p_u \mathcal{G} (u,z_1)\Big[ (w,\overline{z_2})(z_3,\overline{z_1}) - (w,\overline{z_1})(z_3,\overline{z_2})\Big], \non \\\eea
and
\be \label{defL}\Lambda_1 = -\frac{\pi}{4}  \int_{\S^4}\prod_{i=1}^4 d^2 z_i P(z_1,z_2) \mathcal{G} (z_1,z_2) (w,\overline{z_3})(z_3,\overline{z_4}) (z_4,\overline{u}) \p_w\mathcal{G} (w,z_3) \overline\p_u\mathcal{G} (u,z_4).\ee

$\Phi_{4,G}$ involves no derivatives and is given by
\be \label{4G}\Phi_{4,G} = \frac{\pi^2}{{\rm det}Y} \int_{\S^2}\prod_{i=1}^2 d^2 z_i  (w,\overline{z_1})(z_2,\overline{u})\mathcal{G}^2 (z_1,z_2) \Delta(w,z_1)\overline{\Delta(u,z_2)}.\ee

Thus for all these contributions, we see that the structure is similar to that arising from the variation of $\mathcal{B}_1^{(2,0)}$. However, there are some extra contributions which we now mention. 

$\Phi_{4,H}$ is another $O(\p_w\overline\p_u^2 \p_{z_i})$ term and its complex conjugate, and is given by
\bea \Phi_{4,H} &=& \frac{1}{2} \int_{\S^3}\prod_{i=1}^3 d^2 z_i P(z_1,z_2) \Big[  (w,\overline{z_3})\overline\p_u\mathcal{G} (u,z_1) \overline\p_u\mathcal{G} (u,z_2) \p_w\mathcal{G} (w,z_3) \p_{z_3} \mathcal{G} (z_1,z_3) \non \\ &&+(z_3,\overline{u})\p_w\mathcal{G} (w,z_1) \p_w\mathcal{G} (w,z_2)  \overline\p_u\mathcal{G} (u,z_3) \overline\p_{z_3} \mathcal{G} (z_1,z_3) \Big], \eea
which has the same skeleton graph as one of those in $\Phi_{2,B}$.
$\Phi_{4,I}$ is another $O(\p_w\overline\p_u \p_{z_i} \overline\p_{z_j})$ term given by
\bea \Phi_{4,I} &=& -\frac{1}{2}\int_{\S^3}\prod_{i=1}^3 d^2 z_i \Big[ (w,\overline{z_2})(z_2,\overline{z_1})(z_3,\overline{u})\p_w\mathcal{G} (w,z_1) \overline\p_u \mathcal{G} (u,z_3) \p_{z_1} \mathcal{G} (z_1,z_2) \overline\p_{z_3} \mathcal{G} (z_1,z_3) \non \\ &&+(z_1,\overline{z_2})(z_2,\overline{u})(w,\overline{z_3})\p_w\mathcal{G} (w,z_3) \overline\p_u \mathcal{G} (u,z_1) \p_{z_3} \mathcal{G} (z_1,z_3) \overline\p_{z_1} \mathcal{G} (z_1,z_2)\Big], \eea
which has the same skeleton graph as $\Phi_{2,C}$.
$\Phi_{4,J}$ is yet another $O(\p_w\overline\p_u \p_{z_i} \overline\p_{z_j})$ term given by
\be \Phi_{4,J} = \frac{1}{4} \int_{\S^3}\prod_{i=1}^3 d^2 z_i \m(z_1) (w,\overline{z_2})(z_3,\overline{u}) \p_w\mathcal{G} (w,z_2) \overline\p_u \mathcal{G}(u,z_3) \p_{z_2}\mathcal{G}(z_1,z_2) \overline\p_{z_3} \mathcal{G}(z_1,z_3),\ee
which has the same skeleton graph as one of the graphs in $\Phi_{2,D}$\footnote{Some graphs can have the skeleton graph of other $\Phi_{I,\a}$ as well, we simply mention one of them.}. 
Finally, $\Phi_{4,K}$ is an $O(\p_w\overline\p_u \p_{z_i} \overline\p_{z_j})$ term given by
\be \Phi_{4,K} =\frac{1}{8} \int_{\S^4}\prod_{i=1}^4 d^2 z_i P(z_2,z_3) (w,\overline{z_1})(z_4,\overline{u})\p_w\mathcal{G}(w,z_1) \overline\p_u \mathcal{G}(u,z_4) \p_{z_1}\mathcal{G}(z_1,z_2) \overline\p_{z_4} \mathcal{G} (z_3,z_4),\ee
which has the same skeleton graph as one of the graphs in $\Phi_{2,E}$.

Thus for the contributions $\Phi_{4,H}$, $\Phi_{4,I}$, $\Phi_{4,J}$ and $\Phi_{4,K}$ we  see that the structure is similar to some of the terms arising from the variation of $\mathcal{B}_2^{(2,0)}$. Thus the variation of $\mathcal{B}_4^{(2,0)}$ having the skeleton graph in figure 2 (i), leads us automatically to consider variations that arise from figure 2 (ii) as well.

We have that $\overline\delta_{uu} \delta_{ww} \mathcal{B}^{(2,0)}_4$ is hermitian, as well as holomorphic in $w$, and anti--holomorphic in $u$~\footnote{We always ignore divergent contact terms of the form $\mathcal{G}(z_1,z_2)\delta^2 (z_1-z_2)$ for reasons mentioned before.}. 

In fact in checking holomorhphy in $w$ of the variation, we are finally left with 
\bea \label{0}\frac{\pi^2}{2} \m(w) \int_{\S^2}\prod_{i=1}^2 d^2 z_i (z_1,\overline{z_2})\mathcal{G} (z_1,z_2) \overline\p_u \mathcal{G} (u,z_1) \Big[ (w,\overline{z_1})(z_2,\overline{u}) - (w,\overline{u})(z_2,\overline{z_1})\Big] \non \\ -\frac{\pi^2}{2} \m(w) \int_{\S^2}\prod_{i=1}^2 d^2 z_i (w,\overline{z_2})\mathcal{G} (z_1,z_2) \overline\p_u \mathcal{G} (u,z_1) \Big[ \m(z_1) (z_2,\overline{u} )- (z_1,\overline{u})(z_2,\overline{z_1})\Big]\eea
resulting from the second term in the right hand side of the second equation in \C{Eigen}. However \C{0} is proportional to
\be \m(w) \int_{\S^2}\prod_{i=1}^2 d^2 z_i \m(z_2)\mathcal{G} (z_1,z_2) \overline\p_u \mathcal{G} (u,z_1) \Delta(w,z_1) \overline{\Delta(u,z_1)} =0.\ee
Similar manipulations are needed to check holomorphy in $w$ of the variations of $\mathcal{B}_5^{(2,0)}$ and $\mathcal{B}_6^{(2,0)}$ in the analysis below.  

\subsection{Equation involving $\mathcal{B}_5^{(2,0)}$}

Next from \C{5hol}, we have that\be -\frac{1}{4}\overline\delta_{uu} \delta_{ww} \mathcal{B}^{(2,0)}_5 = \sum_{\a=A}^H \Phi_{5,\a},\ee
where we describe the various contributions below.

$\Phi_{5,A}$ is the $O(\p_w^2 \overline\p_u^2)$ term given by 
\be \Phi_{5,A} = -\int_{\S^3}\prod_{i=1}^3 d^2 z_i S(z_1,z_2,z_3) \p_w\mathcal{G} (w,z_1) \p_w \mathcal{G} (w,z_2) \overline\p_u \mathcal{G} (u,z_1) \overline\p_u \mathcal{G} (u,z_3),\ee
which has the same skeleton graph as $\Phi_{2,A}$.
$\Phi_{5,B}$ is an $O(\p_w\overline\p_u^2 \p_{z_i})$ term and its complex conjugate, and is given by
\bea \Phi_{5,B} &=& \int_{\S^3}\prod_{i=1}^3 d^2 z_i \overline\p_u \mathcal{G} (u,z_1) \overline\p_u\mathcal{G} (u,z_2) \Big[ R(w;z_2,z_3;\overline{z_1})\p_{z_1} \mathcal{G} (z_1,z_3) \p_w\mathcal{G} (w,z_1) \non \\ &&+R(w;z_1,z_2;\overline{z_3})\p_{z_3} \mathcal{G}(z_1,z_3) \p_w\mathcal{G}(w,z_3)\Big] \non \\ &&+  \int_{\S^3}\prod_{i=1}^3 d^2 z_i \p_w \mathcal{G} (w,z_1) \p_w\mathcal{G} (w,z_2)\Big[ R(z_1;z_2,z_3;\overline{u})\overline\p_{z_1} \mathcal{G} (z_1,z_3) \overline\p_u\mathcal{G} (u,z_1) \non \\ &&+R(z_3;z_1,z_2;\overline{u})\overline\p_{z_3} \mathcal{G}(z_1,z_3) \overline\p_u\mathcal{G}(u,z_3)\Big]- 2\Phi_{4,H} ,\eea
which has the same skeleton graph as $\Phi_{2,B}$.
$\Phi_{5,C}$ is an $O(\p_w\overline\p_u \p_{z_i} \overline\p_{z_j})$ term and is given by
\bea &&\Phi_{5,C} = -\non \int_{\S^3}\prod_{i=1}^3 d^2 z_i \Big[\p_w\mathcal{G} (w,z_1) \overline\p_u \mathcal{G} (u,z_3) \p_{z_1}\mathcal{G} (z_1,z_2)\overline\p_{z_3} \mathcal{G} (z_1,z_3) \Big((w,\overline{u})(z_3,\overline{z_2})(z_2,\overline{z_1})\non \\ &&+ (w,\overline{z_2})(z_2,\overline{u})(z_3,\overline{z_1})\Big) +\p_w\mathcal{G} (w,z_3) \overline\p_u\mathcal{G} (u,z_1) \p_{z_3}\mathcal{G}(z_1,z_3) \overline\p_{z_1} \mathcal{G}(z_1,z_2) \non \\ &&\times \Big((w,\overline{u})(z_1,\overline{z_2})(z_2,\overline{z_3})+ (w,\overline{z_2})(z_2,\overline{u})(z_1,\overline{z_3})\Big)\Big]-2\Phi_{4,I},\eea
which has the same skeleton graph as $\Phi_{2,C}$.
$\Phi_{5,D}$ is yet another $O(\p_w\overline\p_u \p_{z_i} \overline\p_{z_j})$ term and is given by
\bea &&\Phi_{5,D} = -\int_{\S^3}\prod_{i=1}^3 d^2 z_i  R(w;z_2,z_3;\overline{u})\p_w\mathcal{G} (w,z_1) \overline\p_u \mathcal{G} (u,z_1)  \p_{z_1} \mathcal{G} (z_1,z_2) \overline\p_{z_1} \mathcal{G} (z_1,z_3)\non \\ && -\int_{\S^3}\prod_{i=1}^3 d^2 z_i \p_w\mathcal{G} (w,z_2) \overline\p_u \mathcal{G} (u,z_3)\p_{z_2} \mathcal{G} (z_1,z_2) \overline\p_{z_3} \mathcal{G} (z_1,z_3)\Big[ (w,\overline{u})(z_3,\overline{z_1}) (z_1,\overline{z_2}) \non \\ &&+(w,\overline{z_1}) (z_1,\overline{u})(z_3,\overline{z_2}) -(w,\overline{z_1})(z_1,\overline{z_2})(z_3,\overline{u}) - (w,\overline{z_2})(z_3,\overline{z_1})(z_1,\overline{u})] -2\Phi_{4,J},\non \\ \eea
which has the same skeleton graph as $\Phi_{2,D}$. $\Phi_{5,E}$ is the final set of terms involving $O(\p_w\overline\p_u \p_{z_i} \overline\p_{z_j})$ and is given by 
\bea \Phi_{5,E} &=& 
\frac{1}{2} \int_{\S^4}\prod_{i=1}^4 d^2 z_i (z_4,\overline{u}) R(w;z_2,z_3;\overline{z_1})\p_{z_1}\mathcal{G} (z_1,z_2) \p_w\mathcal{G} (w,z_1) \overline\p_u \mathcal{G} (u,z_4) \overline\p_{z_4} \mathcal{G} (z_3,z_4) \non \\ &&+\frac{1}{2} \int_{\S^4}\prod_{i=1}^4 d^2 z_i (w,\overline{z_4}) R(z_1;z_2,z_3;\overline{u})\overline\p_{z_1}\mathcal{G} (z_1,z_2) \overline\p_u\mathcal{G} (u,z_1) \p_w \mathcal{G} (w,z_4) \p_{z_4} \mathcal{G} (z_3,z_4) \non \\ &&-6\Phi_{4,K},\non \\ \eea
which has the same skeleton graph as one of those in $\Phi_{2,E}$.

$\Phi_{5,F}$ involves two derivatives and is given by
\bea \label{5F}\Phi_{5,F} &=& \frac{2\pi}{{\rm det}Y} \int_{\S^3}\prod_{i=1}^3 d^2 z_i (z_2,\overline{z_3})(z_3,\overline{z_1}) \p_w\mathcal{G}(w,z_1) \overline\p_u \mathcal{G} (u,z_2) \Delta(w,z_1) \overline{\Delta(u,z_2)}\non\\&& \times \Big(\mathcal{G}(z_1,z_3)+\mathcal{G}(z_2,z_3)\Big)  -2\Lambda_0 -2\Lambda_1+\Lambda_2 ,\eea
where $\Lambda_0$ and $\Lambda_1$ are given by \C{def0} and \C{defL} respectively, and 
\bea \label{defL2}&&\Lambda_2 = -\frac{\pi}{4}\int_{\S^4}\prod_{i=1}^4 d^2 z_i \p_w\mathcal{G}(w,z_2) \overline\p_u \mathcal{G}(u,z_4) \mathcal{G}(z_1,z_3) \Big[ R(w;z_1,z_3;\overline{z_2}) (z_2,\overline{z_4})(z_4,\overline{u})\non \\&& +R(z_4;z_1,z_3;\overline{u})(w,\overline{z_2})(z_2,\overline{z_4})-S(z_1,z_2,z_3)(w,\overline{z_4})(z_4,\overline{u})-S(z_1,z_3,z_4) (w,\overline{z_2})(z_2,\overline{u})  \Big].\non \\ \eea

In obtaining $\Phi_{5,F}$ in \C{5F} we also encounter, along with its hermitian conjugate, the expression
\bea &&-\pi  \int_{\S^3}\prod_{i=1}^3 d^2 z_i (z_1,\overline{z_3}) \mathcal{G} (z_1,z_2) \p_w\mathcal{G} (w,z_3) \overline\p_u \mathcal{G} (u,z_1) \Big[ (z_2,\overline{u})\Big((w,\overline{z_1})(z_3,\overline{z_2})- (w,\overline{z_2})(z_3,\overline{z_1})\Big) \non \\ &&+ (z_2,\overline{z_1}) \Big((w,\overline{z_2})(z_3,\overline{u})- (w,\overline{u})(z_3,\overline{z_2})\Big)\Big] \non \\&& = \frac{\pi}{{\rm det}Y} \int_{\S^3}\prod_{i=1}^3 d^2 z_i \m(z_2) (z_1,\overline{z_3})\mathcal{G}(z_1,z_2) \p_w\mathcal{G}(w,z_3) \overline\p_u\mathcal{G}(u,z_1)\Delta(w,z_3) \overline{\Delta(u,z_1)} \non \\&&=0,  \eea
leading to a vanishing contribution.

$\Phi_{5,G}$ involves terms with no derivatives and is given by
\bea \label{5G}\Phi_{5,G} = -\frac{\pi^2}{{\rm det}Y}  \int_{\S^3}\prod_{i=1}^3 d^2 z_i \mathcal{G} (z_1,z_2) \mathcal{G}(z_1,z_3) \Big[(z_2,\overline{u})(w,\overline{z_1})(z_1,\overline{z_3}) \Delta(w,z_3)\overline{\Delta(u,z_2)}\non \\ + (z_2,\overline{u})(w,\overline{z_3})(z_3,\overline{z_1})\Delta(w,z_1)\overline{\Delta(u,z_2)} + (w,\overline{z_2})(z_2,\overline{z_3})(z_1,\overline{u})\Delta(w,z_3)\overline{\Delta(u,z_1)}\Big].\non \\\eea

These terms are similar in structure of those obtained in the variation of $\mathcal{B}_2^{(2,0)}$. However, there is an extra contribution given by
\bea \Phi_{5,H} = -\frac{1}{4}\int_{\S^4}\prod_{i=1}^4 d^2 z_i S(z_1,z_2,z_3)(z_4,\overline{u})  \p_w \mathcal{G} (w,z_1) \p_w\mathcal{G} (w,z_2) \overline\p_{z_4} \mathcal{G} (z_3,z_4) \overline\p_u\mathcal{G}(u,z_4)\non \\ -\frac{1}{4}\int_{\S^4}\prod_{i=1}^4 d^2 z_i S(z_1,z_2,z_3)(w,\overline{z_4})  \overline\p_u \mathcal{G} (u,z_1) \overline\p_u\mathcal{G} (u,z_2) \p_{z_4} \mathcal{G} (z_3,z_4) \p_w\mathcal{G}(w,z_4),\non \\\eea
which has the same skeleton graph as $\Phi_{3,B}$. Thus this contribution leads us to consider terms that arise in the variation of a graph with skeleton diagram given by figure 2 (iii).  

Again, we have that $\overline\delta_{uu} \delta_{ww} \mathcal{B}^{(2,0)}_5$ is hermitian, as well as holomorphic in $w$, and anti--holomorphic in $u$. 

\subsection{Equation involving $\mathcal{B}_6^{(2,0)}$}

Finally, from \C{6hol} we obtain
\be \frac{1}{2}\overline\delta_{uu} \delta_{ww} \mathcal{B}^{(2,0)}_6 = \sum_{\a=A}^G\Phi_{6,\a},\ee
yielding the terms we now list. $\Phi_{6,A}$ is the $O(\p_w^2\overline\p_u^2)$ term given by
\be \Phi_{6,A} = \int_{\S^4}\prod_{i=1}^4 d^2 z_i (z_1,\overline{z_4})(z_4,\overline{z_3})(z_3,\overline{z_2})(z_2,\overline{z_1})\p_w\mathcal{G} (w,z_1) \p_w\mathcal{G} (w,z_4) \overline\p_u\mathcal{G}(u,z_2) \overline\p_u\mathcal{G}(u,z_3),\ee
which has the same skeleton graph as $\Phi_{3,A}$. $\Phi_{6,B}$ is the $O(\p_w\overline\p_u^2 \p_{z_i})$ term and its complex conjugate, and is given by
\bea \Phi_{6,B} &=& -\frac{1}{2}\int_{\S^4}\prod_{i=1}^4 d^2 z_i \p_w\mathcal{G} (w,z_1) \p_w\mathcal{G} (w,z_2) \overline\p_{z_4} \mathcal{G}(z_3,z_4) \overline\p_u\mathcal{G} (u,z_4) \non \\ &&\times \Big[ (z_4,\overline{z_3}) R(z_3;z_1,z_2;\overline{u}) + (z_3,\overline{u})R(z_4;z_1,z_2;\overline{z_3})\Big] \non \\ &&-\frac{1}{2}\int_{\S^4}\prod_{i=1}^4 d^2 z_i \overline\p_u\mathcal{G}(u,z_1)\overline\p_u \mathcal{G} (u,z_2) \p_{z_4} \mathcal{G} (z_3,z_4) \p_w\mathcal{G} (w,z_4)\non \\ &&\times \Big[ (z_3,\overline{z_4})R(w;z_1,z_2;\overline{z_3})+ (w,\overline{z_3})R(z_3;z_1,z_2;\overline{z_4})\Big]-2\Phi_{5,H},\eea
which has the same skeleton graph as $\Phi_{3,B}$. 

$\Phi_{6,C}$, $\Phi_{6,D}$ and $\Phi_{6,E}$ are $O(\p_w\overline\p_u \p_{z_i} \overline\p_{z_j})$ terms  and are given by
\bea  \label{split}\Phi_{6,C} &=&  \int_{\S^4}\prod_{i=1}^4 d^2 z_i \p_w\mathcal{G} (w,z_1) \overline\p_u\mathcal{G}(u,z_2) \p_{z_1} \mathcal{G} (z_1,z_4) \overline\p_{z_2} \mathcal{G} (z_2,z_3)\non \\ &&\times \Big[ (w,\overline{z_3}) (z_3,\overline{u})(z_2,\overline{z_4})(z_4,\overline{z_1}) +(w,\overline{z_4}) (z_4,\overline{u})(z_2,\overline{z_3})(z_3,\overline{z_1})\Big],\non \\
\Phi_{6,D} &=& \int_{\S^4}\prod_{i=1}^4 d^2 z_i \p_w\mathcal{G} (w,z_1) \overline\p_u\mathcal{G}(u,z_2) \p_{z_1} \mathcal{G} (z_1,z_4) \overline\p_{z_2} \mathcal{G} (z_2,z_3)\non \\ &&\times \Big[ (w,\overline{z_4}) (z_4,\overline{z_3})(z_3,\overline{u})(z_2,\overline{z_1})+ (w,\overline{u}) (z_2,\overline{z_3})(z_3,\overline{z_4})(z_4,\overline{z_1}) \non \\ &&-  (z_2,\overline{z_3}) (z_3,\overline{z_4})(z_4,\overline{u})(w,\overline{z_1}) -(z_2,\overline{u})(w,\overline{z_3})(z_3,\overline{z_4})(z_4,\overline{z_1})
\Big], \non \eea
\bea
\Phi_{6,E} &=& -\int_{\S^4}\prod_{i=1}^4 d^2 z_i \p_w\mathcal{G} (w,z_1) \overline\p_u\mathcal{G}(u,z_2) \p_{z_1} \mathcal{G} (z_1,z_4) \overline\p_{z_2} \mathcal{G} (z_2,z_3)\non \\ &&\times \Big[ (z_2,\overline{u})(w,\overline{z_4})(z_4,\overline{z_3})(z_3,\overline{z_1}) + (w,\overline{z_1})(z_4,\overline{z_3}) (z_3,\overline{u})(z_2,\overline{z_4})\Big]+8\Phi_{4,K},\non \\ \eea
and they have the same skeleton graph as $\Phi_{3,C}, \Phi_{3,D}$ and $\Phi_{3,E}$. We have split the three contributions as in \C{split} as this is useful for our purposes. 

Next $\Phi_{6,F}$ contains terms with two derivatives and is given in appendix D. This can be simplified giving contributions having two derivatives, as well as without derivatives.  Following the analysis in appendix D, we get that 
\bea \label{6F}\Phi_{6,F} &=& \frac{\pi}{{\rm det}Y} \int_{\S^3}\prod_{i=1}^3 d^2 z_i \mathcal{G} (z_2,z_3) P(z_2,z_3) \p_w\mathcal{G} (w,z_1) \overline\p_u\mathcal{G}(u,z_1) \Delta (w,z_1) \overline{\Delta (u,z_1)}\non \\ && +\frac{8\pi}{{\rm det}Y} \int_{\S^2}\prod_{i=1}^2 d^2 z_i (z_2,\overline{z_1})\mathcal{G} (z_1,z_2) \p_w\mathcal{G}(w,z_1) \overline\p_u\mathcal{G} (u,z_2) \Delta(w,z_1)\overline{\Delta(u,z_2)}\non \\ && - \frac{4\pi}{{\rm det}Y} \int_{\S^3}\prod_{i=1}^3 d^2 z_i (z_2,\overline{z_3})(z_3,\overline{z_1})\p_w\mathcal{G}(w,z_1) \overline\p_u \mathcal{G}(u,z_2) \Delta(w,z_1) \overline{\Delta(u,z_2)} \non \\ &&\times \Big(\mathcal{G}(z_1,z_3)+\mathcal{G}(z_2,z_3)\Big)\non \\ &&- \frac{4\pi^2}{{\rm det}Y} \int_{\S^3}\prod_{i=1}^3 \m(z_3) (w,\overline{z_1})(z_2,\overline{u})\mathcal{G}(z_1,z_3)\mathcal{G}(z_2,z_3) \Delta(w,z_1)\overline{\Delta(u,z_2)}\non \\ && +\frac{\pi^2}{{\rm det}Y} \int_{\S^4}\prod_{i=1}^4 d^2 z_i P(z_2,z_3)(w,\overline{z_1})(z_4,\overline{u})\Delta(w,z_1) \overline{\Delta(u,z_4)}\non \\ &&\times \Big(\mathcal{G} (z_1,z_4) \mathcal{G} (z_2,z_3) +2\mathcal{G}(z_1,z_3)\mathcal{G}(z_2,z_4)\Big)+2(\Lambda_1 -\Lambda_2),\eea
where $\Lambda_1$ and $\Lambda_2$ are given in \C{defL} and \C{defL2} respectively\footnote{This is analogous to the analysis for $\Phi_{3,F}$ which has terms with two derivatives to start with given by \C{3Fmore}, but can be simplified to yield terms with two as well as no derivatives as well given by \C{fin3F}.}. 

$\Phi_{6,G}$ contains no derivatives and is given by
\bea \label{6G}\Phi_{6,G} &=& \frac{\pi^2}{{\rm det}Y} \int_{\S^4}\prod_{i=1}^4 d^2 z_i \mathcal{G} (z_1,z_4) \mathcal{G} (z_2,z_3) \Big[(w,\overline{z_3})(z_2,\overline{z_4})\Delta(w,z_4)\non \\ &&+ (w,\overline{z_4})(z_4,\overline{z_3})\Delta(w,z_2)\Big](z_1,\overline{u})(z_3,\overline{z_2})\overline{\Delta(u,z_1)}.\eea

Again, we have that $\overline\delta_{uu} \delta_{ww} \mathcal{B}^{(2,0)}_6$ is hermitian, as well as holomorphic in $w$, and anti--holomorpic in $u$.

\section{Simplifying the structure of variations of the new modular graph functions}

From the above analysis, we see that each of the graphs yields a complicated expression under the variation $\overline\delta_{uu} \delta_{ww}$. Thus we would like to relate variations among them in order to obtain simplifications, as in the previous analysis. We shall see that the structure we obtain is more intricate than what we had before. Earlier we could relate variations of $\mathcal{B}^{(2,0)}_1$ and $\mathcal{B}^{(2,0)}_2$, and separately we could relate variations of $\mathcal{B}^{(2,0)}_2$ and $\mathcal{B}^{(2,0)}_3$. Now we shall be able to relate variations of all the three modular graphs together.

\subsection{Relating variations of $\mathcal{B}_4^{(2,0)}$, $\mathcal{B}^{(2,0)}_5$ and $\mathcal{B}^{(2,0)}_6$}

To begin with, we consider
\be \label{45A}\Phi_{45,A} = \int_{\S^3} \prod_{i=1}^3 d^2 z_i (z_2,\overline{z_1}) (z_1,\overline{z_3})\p_w\mathcal{G} (w,z_1) \p_w\mathcal{G} (w,z_2) \overline\p_u \mathcal{G} (u,z_1) \overline\p_u\mathcal{G} (u,z_3) \overline\p_{z_2} \p_{z_3} \mathcal{G} (z_2,z_3)\ee
which has the same skeleton graph as $\Phi_{12,A}$, and  
\be \label{56A}\Phi_{56,A} = \int_{\S^4} \prod_{i=1}^4 d^2 z_i (z_1,\overline{z_2})(z_2,\overline{z_3})(z_3,\overline{z_4})\p_w\mathcal{G} (w,z_1) \p_w\mathcal{G}(w,z_2) \overline\p_u\mathcal{G} (u,z_3) \overline\p_u \mathcal{G}(u,z_4) \overline\p_{z_1} \p_{z_4} \mathcal{G}(z_1,z_4)\ee
which has the same skeleton graph of $\Phi_{23,A}$.

Using \C{Eigen}, we have that
\be \Phi_{45,A} - \frac{1}{2} \Phi_{56,A} = \pi\Big( 2 \Phi_{4,A} +\Phi_{5,A} +\frac{1}{2} \Phi_{6,A}\Big).\ee
Note that the contribution to $\Phi_{5,A}$ arises from both \C{45A} and \C{56A} put together. This kind of mixing between the various contributions to obtain the final expression is also true for several of the equations below.  

We next consider
\bea \label{45B}\Phi_{45,B} &=& \int_{\S^3} \prod_{i=1}^3 d^2 z_i  \overline\p_u \mathcal{G} (u,z_1) \overline\p_u \mathcal{G} (u,z_2) \Big[ \p_w \mathcal{G} (w,z_1)   \p_{z_1} \mathcal{G} (z_1,z_3)(z_3,\overline{z_1})(w,\overline{z_2})\non \\&& + \p_w \mathcal{G} (w,z_3) \p_{z_3} \mathcal{G} (z_1,z_3)(w,\overline{z_1})(z_1,\overline{z_2}) \Big] \overline\p_{z_3} \p_{z_2} \mathcal{G} (z_2,z_3)\non \\ &&+\int_{\S^3} \prod_{i=1}^3 d^2 z_i  \p_w \mathcal{G} (w,z_1)  \p_w \mathcal{G} (w,z_2) \Big[ \overline\p_u \mathcal{G} (u,z_1) \overline\p_{z_1} \mathcal{G} (z_1,z_3) (z_2,\overline{u}) (z_1, \overline{z_3}) \non \\ &&+\overline\p_u \mathcal{G} (u,z_3) \overline\p_{z_3} \mathcal{G} (z_1,z_3) (z_2, \overline{z_1})(z_1,\overline{u})  \Big]\overline\p_{z_2} \p_{z_3} \mathcal{G} (z_2,z_3),\eea
which has the same skeleton graph as $\Phi_{12,B}$, as well as
\bea  \label{56,B}\Phi_{56,B} &=& \int_{\S^4} \prod_{i=1}^4 d^2 z_i  \p_w \mathcal{G} (w,z_1) \p_{z_1} \mathcal{G} (z_1,z_4) \overline\p_u \mathcal{G} (u,z_2) \bar\p_u \mathcal{G} (u,z_3) \non \\ &&\times \Big[ (w,\overline{z_4})(z_4,\overline{z_3})(z_3,\overline{z_2})\overline\p_{z_1} \p_{z_2} \mathcal{G} (z_1,z_2) 
+(w,\overline{z_2})(z_2,\overline{z_3})(z_4,\overline{z_1}) \overline\p_{z_4} \p_{z_3} \mathcal{G} (z_3,z_4)\Big] \non \\ &&+ \int_{\S^4} \prod_{i=1}^4 d^2 z_i  \overline\p_u \mathcal{G} (u,z_1) \overline\p_{z_1} \mathcal{G} (z_1,z_4) \p_w \mathcal{G} (w,z_2) \p_w \mathcal{G} (w,z_3) \non \\ &&\times \Big[ (z_2,\overline{z_3})(z_3,\overline{z_4})(z_4,\overline{u})\overline\p_{z_2} \p_{z_1} \mathcal{G} (z_1,z_2) 
+(z_3,\overline{z_2})(z_2,\overline{u})(z_1,\overline{z_4}) \overline\p_{z_3} \p_{z_4} \mathcal{G} (z_3,z_4)\Big],\non \\ \eea
which has the same skeleton graph as $\Phi_{23,B}$.

Again using \C{Eigen}, we get that
\be \Phi_{45,B} -\frac{1}{2}\Phi_{56,B} = -\pi\Big(2\Phi_{4,B} +2\Phi_{4,H} +\Phi_{5,B} +\Phi_{5,H} +\frac{1}{2}\Phi_{6,B}\Big).\ee

We now consider
\bea \Phi_{45,C} &=&(w,\overline{u})\int_{\S^3} \prod_{i=1}^3 d^2 z_i  \Big[ \overline\p_{z_2} \mathcal{G} (z_1,z_2) \p_{z_1} \mathcal{G} (z_1,z_3) (z_3,\overline{z_1}) \overline\p_{z_3} \p_{z_2} \mathcal{G} (z_2,z_3)\non \\ &&+\overline\p_{z_2} \mathcal{G} (z_2,z_3) \p_{z_1} \mathcal{G} (z_1,z_2)  (z_2,\overline{z_3}) \overline\p_{z_1} \p_{z_3} \mathcal{G} (z_1,z_3)\Big]\p_w \mathcal{G} (w,z_1) \overline\p_u \mathcal{G} (u,z_2), \non \\ \eea
which has the same skeleton graph as $\Phi_{12,C}$, and
\bea &&\Phi_{56,C} = \int_{\S^4} \prod_{i=1}^4d^2 z_i  \p_w \mathcal{G} (w,z_1) \p_{z_1} \mathcal{G} (z_1,z_4)  \overline\p_u \mathcal{G} (u,z_2) \overline\p_{z_2} \mathcal{G} (z_2,z_3) \non \\ &&\times \Big[(w,\overline{z_4}) (z_4,\overline{u})(z_2,\overline{z_3})\overline\p_{z_1} \p_{z_3} \mathcal{G} (z_1,z_3) + (w,\overline{z_3}) (z_3,\overline{u})(z_4,\overline{z_1})\overline\p_{z_4} \p_{z_2} \mathcal{G} (z_2,z_4)\Big],\non \\\eea
which has the same skeleton graph as $\Phi_{23,C}$.

Proceeding as before, we have that
\be \Phi_{45,C} - \frac{1}{2} \Phi_{56,C} = \pi\Big(2\Phi_{4,C} + 2\Phi_{4,I} +\Phi_{5,C}+\frac{1}{2}\Phi_{6,C}\Big).\ee

Also we consider
\bea &&\Phi_{45,D} =  \int_{\S^3} \prod_{i=1}^3d^2 z_i  \Big[ (w,\overline{z_3})(z_2,\overline{u})\overline\p_u \mathcal{G} (u,z_1) \overline\p_{z_1} \mathcal{G} (z_1,z_3)  \overline\p_{z_2} \p_{z_3} \mathcal{G} (z_2,z_3)\non \\ &&+ (w,\overline{z_2})(z_2,\overline{u})\overline\p_u \mathcal{G} (u,z_3) \overline\p_{z_3} \mathcal{G} (z_2,z_3) \overline\p_{z_1} \p_{z_3} \mathcal{G} (z_1,z_3)\Big] \p_w \mathcal{G} (w,z_1) \p_{z_1} \mathcal{G} (z_1,z_2),\non \\ \eea
which has the same skeleton graph as $\Phi_{12,D}$, and
\bea \label{56D}\Phi_{56,D} = \int_{\S^4} \prod_{i=1}^4d^2 z_i \p_w \mathcal{G} (w,z_1) \p_{z_1} \mathcal{G} (z_1,z_4)\overline\p_u \mathcal{G} (u,z_2) \overline\p_{z_2} \mathcal{G} (z_2,z_3)\non \\  \times \Big[ (w,\overline{z_4})(z_4,\overline{z_3})(z_3,\overline{u}) \overline\p_{z_1} \p_{z_2} \mathcal{G} (z_1,z_2)+ \Big((w,\overline{u})(z_2,\overline{z_3})(z_4,\overline{z_1}) \non \\ - (z_4,\overline{z_1})(w,\overline{z_3})(z_2,\overline{u})- (z_2,\overline{z_3})(w,\overline{z_1})(z_4,\overline{u})\Big)\overline\p_{z_4} \p_{z_3} \mathcal{G} (z_3,z_4)\Big],\eea
which has the same skeleton graph as $\Phi_{23,D}$.

Again, we have that
\be \Phi_{45,D} - \frac{1}{2} \Phi_{56,D} = \pi\Big(2\Phi_{4,D} + 2\Phi_{4,J} +\Phi_{5,D}+\frac{1}{2}\Phi_{6,D}\Big).\ee

We finally consider\footnote{Note that from \C{56D} and \C{456E} we have that
\bea \Phi_{56,D} +\Phi_{456,E} = \int_{\S^4} \prod_{i=1}^4d^2 z_i \p_w \mathcal{G} (w,z_1) \p_{z_1} \mathcal{G} (z_1,z_4)\overline\p_u \mathcal{G} (u,z_2) \overline\p_{z_2} \mathcal{G} (z_2,z_3)\non \\  \times \Big[ (w,\overline{z_4})(z_4,\overline{z_3})(z_3,\overline{u}) \overline\p_{z_1} \p_{z_2} \mathcal{G} (z_1,z_2)+ (w,\overline{u})(z_2,\overline{z_3})(z_4,\overline{z_1}) \overline\p_{z_4} \p_{z_3} \mathcal{G} (z_3,z_4)\Big]\eea 
as certain terms cancel, which simplifies our analysis later.}
\bea \label{456E}\Phi_{456,E} &=& \int_{\S^4} \prod_{i=1}^4d^2 z_i \p_w\mathcal{G} (w,z_1) \p_{z_1}\mathcal{G} (z_1,z_3) \overline\p_u \mathcal{G} (u,z_2) \overline\p_{z_2} \mathcal{G} (z_2,z_4) \overline\p_{z_3} \p_{z_4} \mathcal{G} (z_3,z_4)\non \\ &&\times \Big[ (z_2,\overline{z_4})(w,\overline{z_1})(z_3,\overline{u})+ (z_3,\overline{z_1})(w,\overline{z_4})(z_2,\overline{u})\Big],\eea
which has the same skeleton graph as $\Phi_{12,E} = \Phi_{23,E}$. 

We have that
\be \Phi_{456,E} = -2\pi\Big(2\Phi_{4,E} + 2\Phi_{4,K} +\Phi_{5,E} +\frac{1}{2} \Phi_{6,E}\Big)\ee
proceeding along the lines of the previous analysis.

Thus adding the various contributions, we have that
\bea \label{alsomainone}&&\frac{1}{4} \overline{\delta}_{uu} \delta_{ww} \Big(\mathcal{B}^{(2,0)}_4 - \mathcal{B}^{(2,0)}_5 +\mathcal{B}^{(2,0)}_6\Big) = 2(\Phi_{4,F} +\Phi_{4,G}) + \Phi_{5,F} +\Phi_{5,G} \non \\ &&+\frac{1}{2} \Big( \Phi_{6,F} +\Phi_{6,G}\Big)+\frac{1}{2\pi} \Big[ \Big(2\Phi_{45,A} - \Phi_{56,A}\Big) - \Big(2\Phi_{45,B} - \Phi_{56,B}\Big) \non \\ &&+ \Big(2\Phi_{45,C} - \Phi_{56,C}\Big) + \Big(2\Phi_{45,D} - \Phi_{56,D}\Big)-\Phi_{456,E}\Big], \eea
which has a very similar structure to \C{mainone}. 

\subsection{Simplifying contributions to $\mathcal{B}^{(2,0)}_4 - \mathcal{B}^{(2,0)}_5 +\mathcal{B}^{(2,0)}_6$}

From \C{alsomainone}, we now simplify the contributions involving the auxiliary graphs $\Phi_{45,\a}$, $\Phi_{56,\a}$ ($\a=  A,\ldots,D$) and $\Phi_{456,E}$ using \C{Eigen}.

\subsubsection{The contributions to $\Phi_{45,A} - \Phi_{45,B} +\Phi_{45,C} +\Phi_{45,D}$}

To start with, we consider the contributions to
\be \Phi_{45,A} - \Phi_{45,B} +\Phi_{45,C} +\Phi_{45,D}.\ee
A potential contribution with a skeleton graph depicted by figure 16 (i) vanishes as in section 7.2.

Adding the various graphs, we get that
\be \label{45}\Phi_{45,A} - \Phi_{45,B} +\Phi_{45,C} +\Phi_{45,D} = \Psi_2 -\frac{\Psi_6}{2}+\sum_{i=9}^{12}\Psi_i,\ee
where
\bea \Psi_9 &=& \frac{\pi^2}{2}\int_{\S^3} \prod_{i=1}^3 d^2 z_i (w,\overline{z_1})(w,\overline{z_2}) (z_2,\overline{u}) \m(z_3) \mathcal{G} (z_1,z_2) \p_{z_1}\mathcal{G} (z_1,z_3) \overline\p_u\mathcal{G} (u,z_3)\non \\ &&+\frac{\pi^2}{2} \int_{\S^3} \prod_{i=1}^3 d^2 z_i (w,\overline{z_2})(z_2,\overline{u})(z_1,\overline{u}) \m(z_3) \mathcal{G}(z_1,z_2) \overline\p_{z_1} \mathcal{G} (z_1,z_3) \p_w\mathcal{G} (w,z_3)\non \\ \eea
with the skeleton graph depicted by figure 16 (ii). Also
\bea \Psi_{10} &=& \pi^2 \int_{\S^2} \prod_{i=1}^2 d^2 z_i \mathcal{G} (z_1,z_2) \p_w\mathcal{G} (w,z_1) \overline\p_u \mathcal{G} (u,z_2) \Big[\frac{5}{2}\m(z_1)\m(z_2)(w,\overline{u}) \non \\ && -\m(z_2) (w,\overline{z_1})(z_1,\overline{u})- \m(z_1) (w,\overline{z_2})(z_2,\overline{u}) \Big],\eea
with the skeleton graph depicted by figure 16 (iv). 

We also have that
\bea \Psi_{11} = -\frac{\pi^2}{2{\rm det}Y} \int_{\S^3} \prod_{i=1}^3 d^2 z_i \m(z_1)(z_2,\overline{z_3})\mathcal{G} (z_1,z_3) \p_w\mathcal{G}(w,z_1) \overline\p_u\mathcal{G} (u,z_2)\Delta(w,z_3) \overline{\Delta(u,z_2)}\non \\-\frac{\pi^2}{2{\rm det}Y} \int_{\S^3} \prod_{i=1}^3 d^2 z_i \m(z_2) (z_3,\overline{z_1}) \mathcal{G}(z_2,z_3) \p_w\mathcal{G}(w,z_1) \overline\p_u\mathcal{G} (u,z_2)\Delta(w,z_1) \overline{\Delta(u,z_3)}\non \\\eea
with the skeleton graph depicted by figure 16 (v). Thus all these contributions involve two derivatives.

Finally, we note that $\Psi_{12}$ which has no derivatives, is given by
\bea \Psi_{12} &=& 2\pi^3 \int_{\S^2} \prod_{i=1}^2 d^2 z_i (w,\overline{z_2})(z_2,\overline{u})(w,\overline{z_1})(z_1,\overline{u}) \mathcal{G} (z_1,z_2)^2 \non \\ &&- \pi^3 \int_{\S^3} \prod_{i=1}^3 d^2 z_i (w,\overline{z_2})(z_2,\overline{u})(w,\overline{z_1})(z_1,\overline{z_3})(z_3,\overline{u}) \mathcal{G}(z_1,z_2) \mathcal{G}(z_2,z_3) .\eea

\subsubsection{The contributions to $\Phi_{56,A} - \Phi_{56,B} +\Phi_{56,C} +\Phi_{56,D} +\Phi_{456,E}$}

We next consider contributions to
\be \label{56}\Phi_{56,A} - \Phi_{56,B} +\Phi_{56,C} +\Phi_{56,D} +\Phi_{456,E}.\ee

There is a contribution to \C{56} depicted by the skeleton graph in 21 and its hermitian conjugate. Proceeding as earlier, this contribution is equal to
\bea &&\frac{2\pi^3}{{\rm det}Y} \int_{\S^3} \prod_{i=1}^3 d^2 z_i \mathcal{G} (z_1,z_2) \mathcal{G} (z_1,z_3) \Big[ (w,\overline{z_1})(w,\overline{z_2}) (z_3,\overline{u}) \Delta(z_1,z_2) \overline{\Delta(u,z_3)}\non \\ &&+ (z_1,\overline{u})(z_2,\overline{u}) (w,\overline{z_3}) \overline{\Delta(z_1,z_2)}\Delta(w,z_3)\Big]\non \\ &&-\frac{\pi^3}{{\rm det}Y} \int_{\S^4} \prod_{i=1}^4 d^2 z_i \mathcal{G} (z_1,z_2) \mathcal{G} (z_3,z_4) \Big[ (w,\overline{z_1})(w,\overline{z_2}) (z_3,\overline{u})(z_1,\overline{z_4})\Delta(z_4,z_2) \overline{\Delta(u,z_3)}\non\\ &&+ (z_1,\overline{u})(z_2,\overline{u}) (w,\overline{z_3}) (z_4,\overline{z_1}) \overline{\Delta(z_4,z_2)}\Delta(w,z_3)\Big]\non \\ && +\frac{2\pi^2}{{\rm det}Y} \int_{\S^3} \prod_{i=1}^3 d^2 z_i  (w,\overline{z_1}) (w,\overline{z_3})\p_{z_1} \mathcal{G} (z_1,z_2) \overline\p_u \mathcal{G} (u,z_2) \mathcal{G} (z_1,z_3)  \Delta(z_2,z_3)\overline{\Delta(u,z_2)}\non \\ && +\frac{2\pi^2}{{\rm det}Y} \int_{\S^3} \prod_{i=1}^3 d^2 z_i  (z_1,\overline{u})(z_3,\overline{u})\overline\p_{z_1} \mathcal{G} (z_1,z_2) \p_w\mathcal{G} (w,z_2) \mathcal{G} (z_1,z_3) \overline{\Delta(z_2,z_3)}\Delta(w,z_2) .\non \\\eea
Note that the terms without derivatives are exactly the same as the ones in \C{sim}.

Adding the various contributions, we get that
\be \label{456}\Phi_{56,A} - \Phi_{56,B} +\Phi_{56,C} +\Phi_{56,D} +\Phi_{456,E} = 2\sum_{i=9}^{11}\Psi_i  +\Psi_{13}.\ee

Contributions involving $\Psi_7$ arise at an intermediate stage but cancel in the sum, and so does a contribution of the form
\bea &&\frac{2\pi^2}{{\rm det}Y} \int_{\S^3} \prod_{i=1}^3 d^2 z_i   \p_w\mathcal{G} (w,z_1) \overline\p_u \mathcal{G} (u,z_2) (z_2,\overline{z_3})(z_3,\overline{z_1})\non \\ &&\times \Big(\mathcal{G} (z_1,z_3)+\mathcal{G}(z_2,z_3)\Big)  \Delta(w,z_1) \overline{\Delta(u,z_2)}\eea
in the final sum. 

The only new contribution $\Psi_{13}$ has no derivatives and is given by
\bea &&\Psi_{13} = 2\pi^3 \int_{\S^3}\prod_{i=1}^3 d^2 z_i (w,\overline{z_1})(z_1,\overline{u})(w,\overline{z_2})(z_2,\overline{z_3})(z_3,\overline{u}) \mathcal{G}(z_1,z_2) \mathcal{G}(z_1,z_3)\non \\ &&-\pi^3 \int_{\S^4}\prod_{i=1}^4 d^2 z_i (w,\overline{z_1})(z_1,\overline{z_2})(w,\overline{z_4})(z_4,\overline{z_3})(z_3,\overline{u})(z_2,\overline{u})\mathcal{G}(z_1,z_4)\mathcal{G}(z_2,z_3)\non \\ &&+\frac{2\pi^3}{{\rm det}Y} \int_{\S^3}\prod_{i=1}^3 \m(z_3) (w,\overline{z_1})(z_2,\overline{u})\mathcal{G}(z_1,z_3)\mathcal{G}(z_2,z_3) \Delta(w,z_1)\overline{\Delta(u,z_2)}\non \\  &&-\frac{\pi^3}{{\rm det}Y} \int_{\S^4}\prod_{i=1}^4 d^2 z_i P(z_3,z_4)(w,\overline{z_1})(z_2,\overline{u})\mathcal{G} (z_1,z_4) \mathcal{G} (z_2,z_3)\Delta(w,z_1) \overline{\Delta(u,z_2)} \non \\ &&+ \frac{2\pi^3}{{\rm det}Y} \int_{\S^3} \prod_{i=1}^3 d^2 z_i \mathcal{G} (z_1,z_2) \mathcal{G} (z_1,z_3) \Big[ (w,\overline{z_1})(w,\overline{z_2}) (z_3,\overline{u}) \Delta(z_1,z_2) \overline{\Delta(u,z_3)}\non \\ &&+ (z_1,\overline{u})(z_2,\overline{u}) (w,\overline{z_3}) \overline{\Delta(z_1,z_2)}\Delta(w,z_3)\Big]\non \\ &&-\frac{\pi^3}{{\rm det}Y} \int_{\S^4} \prod_{i=1}^4 d^2 z_i \mathcal{G} (z_1,z_2) \mathcal{G} (z_3,z_4) \Big[ (w,\overline{z_1})(w,\overline{z_2}) (z_3,\overline{u})(z_1,\overline{z_4})\Delta(z_4,z_2) \overline{\Delta(u,z_3)}\non\\ &&+ (z_1,\overline{u})(z_2,\overline{u}) (w,\overline{z_3}) (z_4,\overline{z_1}) \overline{\Delta(z_4,z_2)}\Delta(w,z_3)\Big].\eea

\subsubsection{Summing these contributions}

From \C{45} and \C{456},we get that
\bea &&\frac{1}{2\pi} \Big[ \Big(2\Phi_{45,A} - \Phi_{56,A}\Big) - \Big(2\Phi_{45,B} - \Phi_{56,B}\Big) + \Big(2\Phi_{45,C} - \Phi_{56,C}\Big) \non \\ &&+ \Big(2\Phi_{45,D} - \Phi_{56,D}\Big) -\Phi_{456,E}\Big]   = \frac{1}{\pi}\Big(\Psi_2 - \frac{\Psi_6}{2}\Big) +\frac{1}{2\pi}\Big(2\Psi_{12}- \Psi_{13}\Big),\eea
and hence several graphs involving derivatives have cancelled.

Including the various contributions from \C{4F}, \C{4G}, \C{5F}, \C{5G}, \C{6F} and \C{6G}, we also have that
\bea 2(\Phi_{4F}+\Phi_{4G}) + \Phi_{5F} +\Phi_{5G}+\frac{1}{2} \Big(\Phi_{6F} +\Phi_{6G}\Big)= -\frac{\Psi_7}{2\pi}+\widetilde\Phi_1, \eea
where
\bea \widetilde\Phi_1 &=&\frac{2\pi^2}{{\rm det}Y} \int_{\S^2}\prod_{i=1}^2 d^2 z_i  (w,\overline{z_1})(z_2,\overline{u})\mathcal{G}^2 (z_1,z_2) \Delta(w,z_1)\overline{\Delta(u,z_2)} \non \\ &&- \frac{2\pi^2}{{\rm det}Y} \int_{\S^3}\prod_{i=1}^3 \m(z_3) (w,\overline{z_1})(z_2,\overline{u})\mathcal{G}(z_1,z_3)\mathcal{G}(z_2,z_3) \Delta(w,z_1)\overline{\Delta(u,z_2)}\non \eea
\bea &&-\frac{\pi^2}{{\rm det}Y}  \int_{\S^3}\prod_{i=1}^3 d^2 z_i \mathcal{G} (z_1,z_2) \mathcal{G}(z_1,z_3) \Big[(z_2,\overline{u})(w,\overline{z_1})(z_1,\overline{z_3}) \Delta(w,z_3)\overline{\Delta(u,z_2)}\non \\ &&+ (z_2,\overline{u})(w,\overline{z_3})(z_3,\overline{z_1})\Delta(w,z_1)\overline{\Delta(u,z_2)} + (w,\overline{z_2})(z_2,\overline{z_3})(z_1,\overline{u})\Delta(w,z_3)\overline{\Delta(u,z_1)}\Big] \non \\&& +\frac{\pi^2}{2{\rm det}Y} \int_{\S^4}\prod_{i=1}^4 d^2 z_i P(z_2,z_3)(w,\overline{z_1})(z_4,\overline{u})\Delta(w,z_1) \overline{\Delta(u,z_4)}\non \\ &&\times \Big(\mathcal{G} (z_1,z_4) \mathcal{G} (z_2,z_3) +2\mathcal{G}(z_1,z_3)\mathcal{G}(z_2,z_4)\Big)\non \\ &&+\frac{\pi^2}{2{\rm det}Y} \int_{\S^4}\prod_{i=1}^4 d^2 z_i \mathcal{G} (z_1,z_4) \mathcal{G} (z_2,z_3) \Big[(w,\overline{z_3})(z_2,\overline{z_4})\Delta(w,z_4)\non \\ &&+ (w,\overline{z_4})(z_4,\overline{z_3})\Delta(w,z_2)\Big](z_1,\overline{u})(z_3,\overline{z_2})\overline{\Delta(u,z_1)}\eea
includes contributions without derivatives.

Thus using the expressions for these various contributions, from \C{alsomainone} we get that
\bea \label{simple2}\frac{1}{4} \overline{\delta}_{uu} \delta_{ww} \Big(\mathcal{B}^{(2,0)}_4 - \mathcal{B}^{(2,0)}_5 +\mathcal{B}^{(2,0)}_6\Big) =\frac{1}{\pi} \Big[\Psi_2 - \frac{1}{2}\Big(\Psi_6 +\Psi_7\Big)\Big] +\frac{1}{2\pi}\Big(2\Psi_{12} -\Psi_{13}\Big)+\widetilde\Phi_1. \non \\ \eea
Thus strikingly, the terms involving derivatives ($\Psi_2$, $\Psi_6$ and $\Psi_7$) are exactly the same in \C{simple1} and \C{simple2}.  

\section{An eigenvalue equation for the modular graph functions}

We now proceed to obtain an eigenvalue equation involving the various modular graph functions, using the crucial input that the terms with derivatives in \C{simple1} and \C{simple2} are the same.

Thus defining
\be \mathcal{B} = \Big(\mathcal{B}^{(2,0)}_1 -\mathcal{B}^{(2,0)}_4\Big) - \Big(\mathcal{B}^{(2,0)}_2 - \mathcal{B}^{(2,0)}_5\Big) + \Big(\mathcal{B}^{(2,0)}_3 -\mathcal{B}^{(2,0)}_6\Big),\ee
and subtracting \C{simple2} from \C{simple1}, we have that
\be \label{B}\frac{1}{4} \overline\delta_{uu}\delta_{ww} \mathcal{B} = \Theta, \ee
where
\be \label{theta}\Theta= \frac{1}{2\pi} \Big[2\Big(\Psi_5 - \Psi_{12}\Big) - \Big(\Psi_8- \Psi_{13}\Big) \Big]+ \widetilde\Phi_0 -\widetilde\Phi_1\ee
which is independent of derivatives. 

We now see why the absence of derivatives on the right hand side of \C{B} is crucial to obtain a simple eigenvalue equation, hence justifying the previous analysis. 
From the Beltrami variations using \C{beltvar}, the left hand side of \C{B} is given by\footnote{See~\cite{D'Hoker:2014gfa} for a relevant discussion.}
\be \label{T1}\frac{1}{4}\overline\delta_{uu} \delta_{ww} \mathcal{B} = \pi^2  \omega_I (w) \omega_J (w) \overline{\omega_K (u)} \overline{\omega_L (u)}\p_{IJ} \overline\p_{KL}\mathcal{B},\ee
where we have used the expression for the partial derivative 
\be \p_{IJ} = \frac{1}{2} \Big(1+\delta_{IJ}\Big) \frac{\p}{\p\Omega_{IJ}}\ee
in the composite index notation. This follows from the fact that the holomorphic quadratic differential $\delta_{ww} \Phi$ for arbitrary $\Phi$ can be expanded in a basis of $\omega_I (w) \omega_J (w)$ for $I \leq J$, and similarly for the anti--holomorphic variation.

Since there are no derivatives on the right hand side of \C{B}, we can trivially pull out a factor of  $\omega_I (w) \omega_J (w) \overline{\omega_K (u)} \overline{\omega_L (u)}$ with coefficients that are independent of $w$ and $u$ which follows from the structure of the various terms. Note that the terms involving derivatives which have cancelled in the final expression have factors of $\p_w \mathcal{G} (w,z)$ or $\overline\p_u \mathcal{G} (u,z)$ and hence cannot be expressed in this form. Thus expressing $\Theta$ as
\be \label{T2}\Theta = 4\pi^2 \omega_I (w) \omega_J (w) \overline{\omega_K (u)} \overline{\omega_L (u)}\Theta_{IJ;KL},\ee  
from \C{T1} and \C{T2}, we have that
\be  \p_{IJ} \overline\p_{KL} \mathcal{B} =\Theta_{IJ;KL}+\Theta_{IJ;LK}+\Theta_{JI;KL}+\Theta_{JI;LK}\ee
on symmetrizing in $IJ$ and $KL$ separately.

Then using the expression for the Laplacian given by
\be \Delta = 2 \Big(Y_{IK} Y_{JL} + Y_{IL} Y_{JK}\Big)\p_{IJ} \overline\p_{KL},\ee
we get the equation
\be \frac{1}{8}\Delta \mathcal{B} = \Big(Y_{IK} Y_{JL} + Y_{IL} Y_{JK}\Big) \Theta_{IJ;KL}.\ee
The expression for $\Theta_{IJ;KL}$ is deduced in appendix E and is given by \C{DefTheta}. 

This yields
\bea &&\frac{1}{8} \Delta \mathcal{B} = \frac{3}{2} \int_{\S^2}\prod_{i=1}^2 d^2 z_i \mathcal{G}(z_1,z_2)^2 \Big(Q_1(z_1,z_2) - P(z_1,z_2)\Big)\non \\ &&+\frac{1}{4} \int_{\S^3}\prod_{i=1}^3 d^2 z_i \mathcal{G}(z_1,z_2)\mathcal{G}(z_1,z_3) \Big(8(z_1,\overline{z_2})(z_2,\overline{z_3})(z_3,\overline{z_1}) - 7 \m(z_1)P(z_2,z_3)\Big)\non \\ &&+\frac{1}{8} \int_{\S^4}\prod_{i=1}^4 d^2 z_i \mathcal{G} (z_1,z_4) \mathcal{G} (z_2,z_3) \Big( 4 P(z_1,z_2)P(z_3,z_4) - 4 (z_1,\overline{z_4})(z_4,\overline{z_3})(z_3,\overline{z_2})(z_2,\overline{z_1})\non \\ &&- (z_1,\overline{z_2})(z_2,\overline{z_4}) (z_4,\overline{z_3})(z_3,\overline{z_1})\Big), \eea
leading to the eigenvalue equation
\be \label{ev}\Delta \mathcal{B} = 3\Big(\mathcal{B}_1^{(2,0)} - \mathcal{B}^{(2,0)}_4\Big) -\frac{7}{2} \mathcal{B}^{(2,0)}_2 + 4\mathcal{B}^{(2,0)}_5 + 4\Big(\mathcal{B}_3^{(2,0)} - \mathcal{B}_6^{(2,0)}\Big)- \mathcal{B}_7^{(2,0)}\ee\
involving seven modular graph functions.

Interestingly it is only the combination $\mathcal{B}_1^{(2,0)} - \mathcal{B}^{(2,0)}_4$ which arises in the eigenvalue equation \C{ev}. Now we have that
\be \mathcal{B}_1^{(2,0)} - \mathcal{B}^{(2,0)}_4 = \frac{4}{{\rm det} Y} \int_{\S^2}\prod_{i=1}^2 d^2 z_i\mathcal{G} (z_1,z_2)^2 \Delta(z_1,z_2) \overline{\Delta(z_1,z_2)}\ee
which is free of short distance singularities associated with colliding vertex operators due to the presence of the holomorphic two form and its conjugate in the integrand. Hence various divergent terms we ignored actually cancel on adding the various contributions.

\begin{figure}[ht]
\begin{center}
\[
\mbox{\begin{picture}(200,330)(0,0)
\includegraphics[scale=.65]{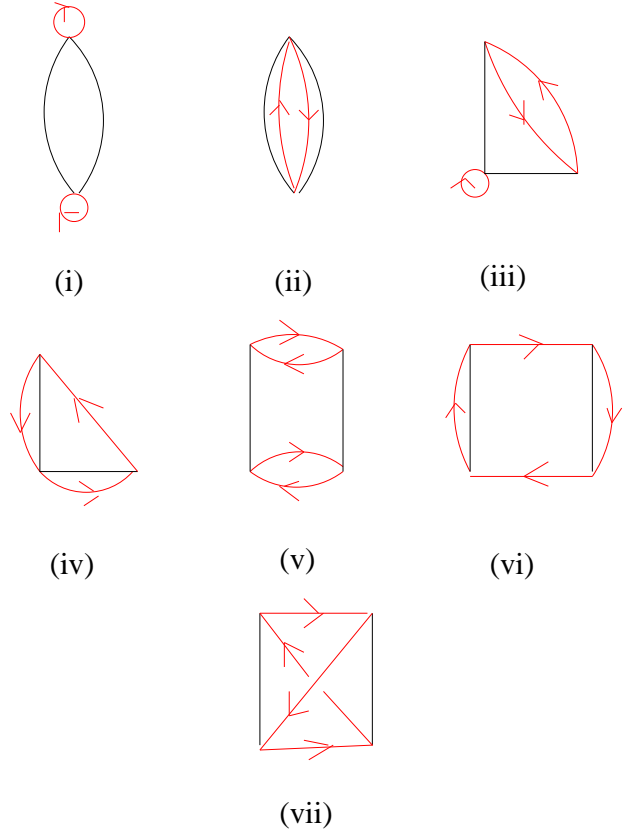}
\end{picture}}
\]
\caption{The modular graphs (i) $\mathcal{B}^{(2,0)}_1$, (ii) $\mathcal{B}^{(2,0)}_4$, (iii) $\mathcal{B}^{(2,0)}_2$, (iv) $\mathcal{B}^{(2,0)}_5$, (v) $\mathcal{B}^{(2,0)}_3$, (vi) $\mathcal{B}^{(2,0)}_6$, and (vii) $\mathcal{B}^{(2,0)}_7$} 
\end{center}
\end{figure}

Let us now denote the eigenvalue equation \C{ev} diagrammatically using modular graphs. To do, we first depict the various modular graphs by figure 26. Along with the skeleton graphs, we also denote the dressing factor $(z_i,\overline{z_j})$ by a red line from the vertex at $z_i$ to the vertex at $z_j$ with an arrow pointing from $z_i$ to $z_j$. Thus the red lines always form closed oriented loops. The eigenvalue equation is depicted by figure 27.

\begin{figure}[ht]
\begin{center}
\[
\mbox{\begin{picture}(430,260)(0,0)
\includegraphics[scale=.65]{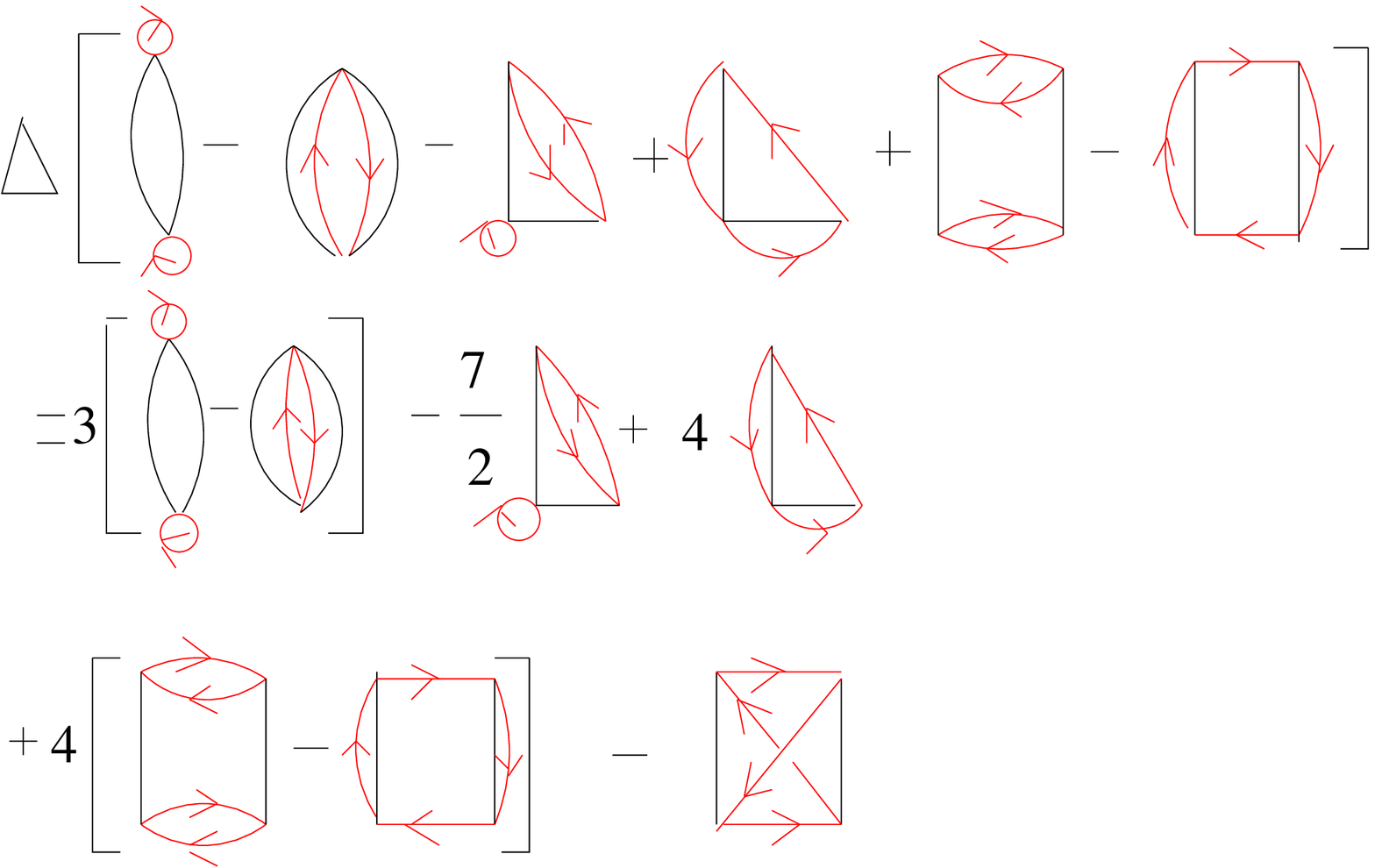}
\end{picture}}
\]
\caption{The eigenvalue equation} 
\end{center}
\end{figure}

Since the eigenvalue equation \C{ev} involves graphs other than those that arise in the integrand of the $D^8\mathcal{R}^4$ term, it is not directly useful in integrating over moduli space to obtain the coefficient of the $D^8\mathcal{R}^4$ term in the effective action. In order to do so, one must obtain more equations these graphs satisfy, which will be interesting to obtain.   

\vspace{.3cm}

{\bf{Acknowledgements}}

I am thankful to the organizers of the preparatory SERB school at the University of Hyderabad for warm hospitality where part of this work was done.

\appendix

\section{Revisiting the $D^6\mathcal{R}^4$ term}

In this appendix, we briefly revisit the eigenvalue equation satisfied by the integrand of the $D^6\mathcal{R}^4$ term. This has been obtained in~\cite{D'Hoker:2014gfa} using the string Green function in the integrand, while we perform the analysis using the Arakelov Green function. Proceeding as in the main text, from \C{D6R4}, we have that 
\bea \frac{\delta_{ww}\mathcal{B}^{(0,1)}}{16} &=& -\int_{\S^2} \prod_{i=1}^2 d^2 z_iP(z_1,z_2) \p_w \mathcal{G}(w,z_1)\p_w\mathcal{G} (w,z_2) \non \\ &&+ 2 \int_{\S^2} \prod_{i=1}^2 d^2 z_i \p_{z_1}\mathcal{G} (z_1,z_2) \p_w \mathcal{G}(w,z_1) (w,\overline{z_2}) (z_2,\overline{z_1}) ,\eea
which satisfies
\be \overline\p_w \delta_{ww} \mathcal{B}^{(0,1)}=0. \ee

This leads to
\be \label{match}\frac{1}{16}\overline{\delta}_{uu}\delta_{ww}\mathcal{B}^{(0,1)} = \Phi_{0,A} +\Phi_{0,B} +\Phi_{0,C}, \ee
where
\bea \Phi_{0,A} &=& 6\pi (Y^{-1}_{IJ} Y^{-1}_{KL} - Y^{-1}_{IL} Y^{-1}_{JK})\int_{\S} d^2 z \p_w \mathcal{G}(w,z) \overline\p_u \mathcal{G}(u,z) \omega_I (z) \overline{\omega_J (z)} \omega_K(w) \overline{\omega_L (u)}, \non \\ \Phi_{0,B} &=& -2\pi^2 (Y^{-1}_{IJ} Y^{-1}_{KL} - Y^{-1}_{IL} Y^{-1}_{JK})  Y^{-1}_{AB}Y^{-1}_{CD}\int_{\S^2} \prod_{i=1}^2 d^2 z_i \mathcal{G} (z_1,z_2) \non \\ &&\times \omega_A (z_1) \overline{\omega_B (u)}  \omega_C (w)  \overline{\omega_D (z_2)}  \omega_I (w) \overline{\omega_J (z_1)} \omega_K(z_2) \overline{\omega_L (u)}, \non \eea
\bea \Phi_{0,C} &= &-\pi P_{IJKL} Y^{-1}_{AB}\int_{\S^2} \prod_{i=1}^2 d^2 z_i \p_w \mathcal{G}(w,z_1) \overline\p_u \mathcal{G}(u,z_2) \omega_I (z_2) \overline{\omega_L (z_1)}\non \\ &&\times \Big(\omega_K (w) \omega_A (z_1) - \omega_K (z_1) \omega_A (w)\Big) \Big(\overline{\omega_J (u)} \overline{\omega_B (z_2)} - \overline{\omega_J (z_2)} \overline{\omega_B (u)}\Big) , \eea
where
\be P_{IJKL} = -Y^{-1}_{IJ} Y^{-1}_{KL} + 2 Y^{-1}_{IL} Y^{-1}_{JK}.\ee
This precisely reproduces equation (4.11) in~\cite{D'Hoker:2014gfa} with $(\Phi_{0,A},\Phi_{0,B},\Phi_{0,C}) \rightarrow -4 (\psi_A,\psi_B,\psi_C)$ and $G(w,z) \rightarrow \mathcal{G}(w,z)$.

Note that each $\Phi_{0,\alpha}$ ($\alpha= A,B,C$) is invariant under $w \leftrightarrow \overline{u}$, leading to a hermitian variation. Also 
\be \overline\p_w \Big(\overline{\delta}_{uu}\delta_{ww}\mathcal{B}^{(0,1)}\Big) =0,\ee
and hence $\p_u (\overline{\delta}_{uu}\delta_{ww}\mathcal{B}^{(0,1)}) =0$.

We express the quantities in the equations above in terms of the holomorphic bi--form\footnote{Thus $\Delta(i,j) = \Delta(z_i,z_j) dz_i \wedge dz_j $.}
\be \label{holbiform}\epsilon_{IJ} \D(z,w) =  \omega_I (z)\omega_J (w) - \omega_J (z)\omega_I (w),\ee
and hence $\Delta (z,w) =\epsilon_{IJ} \omega_I (z)\omega_J (w)$, where $\epsilon_{12} =1$. 

Apart from this analysis, we use the identity 
\be Y^{-1}_{IJ} Y^{-1}_{KL}- Y^{-1}_{IL} Y^{-1}_{JK} = \epsilon_{IK} \epsilon_{JL} ({\rm det}Y)^{-1}, \ee
as well as the identities 
\bea ({\rm det} Y)^{-1} &=& \frac{1}{2}\epsilon_{IJ} \epsilon_{KL} Y^{-1}_{IK} Y^{-1}_{JL}, \non \\ Y^{-1}_{IL} ({\rm det} Y)^{-1} &=& \epsilon_{AJ} \epsilon_{BK} Y^{-1}_{AB} Y^{-1}_{IJ} Y^{-1}_{KL}\eea
at various places in the main text.

Thus from \C{match} we have that
\be \frac{1}{16}\overline{\delta}_{uu}\delta_{ww}\mathcal{B}^{(0,1)} = \frac{5\pi^2}{{\rm det}Y} \int_{\S^2} \prod_{i=1,2} d^2 z_i \mathcal{G} (z_1,z_2) (z_1,\overline{u})(w,\overline{z_2}) \overline{\Delta(u,z_1)} \Delta (w,z_2)  ,\ee
leading to
\be \Delta \mathcal{B}^{(0,1)} = 5\mathcal{B}^{(0,1)}\ee
along the lines of discussion in the main text.

\section{An issue with holomorphy for non-conformal graphs}

In the analysis in the main text, we have considered conformally invariant graphs. As an interesting aside, let us consider the complications that arise if we consider non--conformal graphs instead. For the $D^8\mathcal{R}^4$ term, they are given by \C{O}, \C{T} and \C{Th} with $\mathcal{G}(z,w) \rightarrow G(z,w)$. Let us call these modular graphs $B^{(2,0)}_i$ rather than $\mathcal{B}^{(2,0)}_i$. 

Proceeding as in the main text, for the equation involving $B^{(2,0)}_1$, the holomorphic variation is given by
\bea \label{I}\frac{\delta_{ww} {B}^{(2,0)}_1}{8} &=&  \int_{\S^2} \prod_{i=1}^2 d^2 z_i {G}(z_1,z_2) Q_1 (z_1,z_2)\Big(\p_w G(w,z_1)^2 - \p_w {G}(w,z_1) \p_w {G}(w,z_2)\Big)\non \\ &&+ 2   \int_{\S^2}\prod_{i=1}^2 d^2 z_i {G}(z_1,z_2) \p_{z_1} {G}(z_1,z_2) \p_w {G}(w,z_1) (w,\overline{z_1}) \m(z_2) . \eea

For the equation involving $B^{(2,0)}_2$, the holomorphic variation is given by
\bea \label{I1}&&-\frac{\delta_{ww} {B}^{(2,0)}_2}{4} =  2\int_{\S^2} \prod_{i=1}^2 d^2 z_i G(z_1,z_2) \p_w G(w,z_1)^2 Q_1 (z_1,z_2) \non \\ &&  + \int_{\S^3} \prod_{i=1}^3 d^2 z_i G(z_1,z_3) Q_2 (z_1;z_2,z_3) \Big(\p_w G(w,z_2)^2 -2 \p_w G(w,z_1) \p_w G(w,z_2)\Big) \non \\ &&+2 Y^{-1}_{IJ} (Y^{-1}_{KL} Y^{-1}_{MN}-Y^{-1}_{KN}Y^{-1}_{LM}) \int_{\S^3}  \prod_{i=1}^3 d^2 z_i {G}(z_1,z_3) \Big[ \p_{z_1}{G}(z_1,z_2)  \p_w {G}(w,z_1) \omega_I (w)  \omega_K (z_2) \non \\ &&+ \p_{z_2} {G}(z_1,z_2) \p_w {G}(w,z_2) \omega_I (z_1) \omega_K (w)\Big]\overline{\omega_J (z_1)}\overline{\omega_L (z_2)} \omega_M (z_3) \overline{\omega_N (z_3)}. \eea

Finally, for the equation involving $B^{(2,0)}_3$, the holomorphic variation is given by
\bea \label{I2}&&\frac{\delta_{ww} {B}^{(2,0)}_3}{2} =  2\int_{\S^3} \prod_{i=1}^3 d^2 z_i G(z_1,z_3) \p_w G(w,z_2)^2 Q_2 (z_1;z_2,z_3) \non \\ && - \int_{\S^4} \prod_{i=1}^4 d^2 z_i {G}(z_2,z_3) Q(z_1,z_2,z_3,z_4)  \p_w {G}(w,z_1) \p_w {G}(w,z_4)\non \\ &&+ 2 (Y^{-1}_{IJ} Y^{-1}_{KL} - Y^{-1}_{IL} Y^{-1}_{JK})\int_{\S^4} \prod_{i=1}^4 d^2 z_i {G}(z_2,z_3) \p_{z_1} {G}(z_1,z_4) \p_w {G}(w,z_1)  \non \\ && \times \omega_I (w) \overline{\omega_J (z_1)} \omega_K (z_2) \overline{\omega_L (z_2)} \Big(Q_1(z_3,z_4)- P(z_3,z_4)\Big).\eea

We now calculate $\bar{\p}_w (\delta_{ww} B^{(2,0)}_i)$ for each of these variations. There is a non--vanishing contribution coming from only the terms of the form $(\p_w G(w,z))^2$ in the expressions above. This arises using
\be \bar\p_w \Big(\p_w G(w,z)\Big)^2 = -4\pi \delta^2 (w-z)\p_w G(w,z) +\ldots\ee
coming from \C{eigen}.
We further use
\be 2 \delta^2 (z-w) \p_w G(w,z) = - \p_z \delta^2 (z-w)\ee
which leads to non--vanishing contact term contributions.  
For $B^{(2,0)}_1$ from \C{I}, we get that
\be \overline\p_{w} \Big(\frac{\delta_{ww} {B}^{(2,0)}_1}{16\pi}\Big)=  -\p_w \int_{\S} d^2 z Q_1 (w,z) G(w,z),\ee
while for $B^{(2,0)}_2$ from \C{I1}, we get that
\be \overline\p_{w} \Big(\frac{\delta_{ww} {B}^{(2,0)}_2}{16\pi}\Big)=  \p_w \int_{\S} d^2 z Q_1 (w,z) G(w,z) + \frac{1}{2}\p_w \int_{\S^2} \prod_{i=1}^2 d^2 z_i G(z_1,z_2) Q_2 (z_1;w,z_2).\ee
Finally for $B^{(2,0)}_3$from \C{I2}, we get that
\be \overline\p_{w} \Big(\frac{\delta_{ww} {B}^{(2,0)}_3}{16\pi}\Big)=  - \frac{1}{2}\p_w \int_{\S^2} \prod_{i=1}^2 d^2 z_i G(z_1,z_2) Q_2 (z_1;w,z_2).\ee

Thus the violation of holomorphy is entirely due to contact terms, and the holomorphic variation is anomalous for each of the non--conformal graphs\footnote{Such an anomalous term is absent for the $D^6\mathcal{R}^4$ term if one uses the expression involving the string Green function in the integrand, as the holomorphic variation of the integrand does not have a $(\p_w G(w,z))^2$ term.}. However, we see that
\be \overline\p_{w} \delta_{ww} ({B}^{(2,0)}_1 + {B}^{(2,0)}_2 + {B}^{(2,0)}_3)=0.\ee
Hence though each of these graphs that contribute does not have a holomorphic variation, their sum does. This is simply because the $Q_1(z_1,z_2) G(z_1,z_2)(\p_w G(w,z_1))^2$ term in the integrand cancels between $\delta_{ww}B_1^{(2,0)}$ and $\delta_{ww}B_2^{(2,0)}$, while the $Q_2 (z_1,z_2,z_3) G(z_1,z_3)(\p_wG(w,z_2))^2$ term in the integrand cancels between $\delta_{ww}B_2^{(2,0)}$ and $\delta_{ww}B_3^{(2,0)}$. This is to be contrasted with the variations of $\mathcal{B}^{(2,0)}_i$ in section 5, involving the Arakelov Green function, where each variation is holomorphic.

\section{Simplifying the structure of $\Phi_{1,F}$, $\Phi_{2,F}$, $\Phi_{2,G}$, $\Phi_{3,F}$ and $\Phi_{3,G}$ }

Let us express $\Phi_{1,F}$, $\Phi_{2,F}$, $\Phi_{2,G}$, $\Phi_{3,F}$ and $\Phi_{3,G}$ given in \C{1F}, \C{2F}, \C{2G}, \C{3F} and \C{3G} respectively in a different way which will be very useful for our purposes.

Now $\Phi_{2,G}$ and $\Phi_{3,G}$ do not contain any derivatives. 
From \C{2G}, we get that
\bea \Phi_{2,G} = -\frac{\pi^2}{{\rm det} Y}\int_{\S^3} \prod_{i=1,2,3} d^2 z_i \mathcal{G} (z_1,z_2) \mathcal{G} (z_1,z_3) \m(z_1) (w,\overline{z_3}) (z_2,\overline{u}) \Delta (w,z_3) \overline{\Delta (u,z_2)} ,\eea
while from \C{3G}, we obtain
\bea \Phi_{3,G} = \frac{2\pi^2}{{\rm det}Y}  \int_{\S^4} \prod_{i=1,2,3,4} d^2 z_i \mathcal{G} (z_1,z_4) \mathcal{G} (z_2,z_3) P(z_3,z_4) (w,\overline{z_2}) (z_1,\overline{u}) \Delta (w,z_2) \overline{\Delta (u,z_1)}.\non \\ \eea

Each of the remaining expressions contain two derivatives, on $\p_w$ and one $\overline\p_u$. First from \C{1F}, we have that
\be \Phi_{1,F} = -\frac{2\pi}{{\rm det}Y} \int_{\S^2} \prod_{i=1,2} d^2 z_i \mathcal{G} (z_1,z_2) \p_w\mathcal{G} (w,z_1) \overline\p_u\mathcal{G} (u,z_2) (z_2,\overline{z_1})\Delta (w,z_1) \overline{\Delta(u,z_2)} ,\ee
while \C{2F} yields
\bea \label{2Fmore}\Phi_{2,F} &=& \frac{\pi}{{\rm det}Y} \int_{\S^3} \prod_{i=1,2,3} d^2 z_i \p_w\mathcal{G} (w,z_1) \overline\p_u\mathcal{G} (u,z_2) \Delta (w,z_1) \overline{\Delta(u,z_2)} (z_3,\overline{z_1}) (z_2,\overline{z_3}) \non \\ &&\times \Big(\mathcal{G} (z_1,z_3) +\mathcal{G} (z_2,z_3)\Big).  \eea
The first term in \C{2F} actually vanishes hence leading to the single term in \C{2Fmore}. 

Finally \C{3F} gives us
\bea \label{3Fmore}\Phi_{3,F} &=& -\frac{2\pi}{({\rm det}Y)^2} \int_{\S^3} \prod_{i=1,2,3} d^2 z_i \mathcal{G} (z_2,z_3) \p_w\mathcal{G} (w,z_1) \overline\p_u \mathcal{G} (u,z_1) \non \\ && \times \Delta (z_2,z_3) \overline{\Delta (z_2,z_3)}  \Delta (w,z_1) \overline{\Delta (u,z_1)} \non \\ &&- \frac{2\pi}{{\rm det}Y} \int_{\S^4} \prod_{i=1,2,3,4} d^2 z_i \mathcal{G} (z_2,z_3) \p_w \mathcal{G} (w,z_1) \overline\p_u \mathcal{G} (u,z_4) \non \\ &&\times (z_4,\overline{z_2}) (z_2,\overline{z_3}) (z_3,\overline{z_1}) \Delta (w,z_1) \overline{\Delta (u,z_4)}.\eea
We now express the two terms in \C{3Fmore} differently such that they have similar structure as the other terms, so that it is useful when we add the various contributions with appropriate coefficients. For the first term, we use
\be\label{special} ({\rm det}Y)^{-1} \Delta (z_2,z_3) \overline{\Delta (z_2,z_3)} = \m(z_2) \m(z_3) - P(z_2,z_3).\ee 
In fact, \C{special} is a special case of the general result
\be ({\rm det} Y)^{-1} \Delta (z_i,z_j) \overline{\Delta (z_k,z_l)} = (z_i,\overline{z_k})(z_j,\overline{z_l}) - (z_i,\overline{z_l})(z_j,\overline{z_k}).\ee
which we often use in making such simplifications. 
  
The second term has $\p_w$ and $\overline\p_u$ and is integrated over four positions. We use \C{repeat} judiciously to express it in terms of integrals where terms involving $\p_w$ and $\overline\p_u$ are not integrated over more than three positions. This gives us that
\bea \label{fin3F}\Phi_{3,F} &=&  -\frac{4\pi^2}{{\rm det}Y} \int_{\S^3} \prod_{i=1,2,3} d^2 z_i \mathcal{G} (z_1,z_2) \mathcal{G} (z_2,z_3) (w,\overline{z_1}) (z_3,\overline{u}) \m(z_2) \Delta (w,z_1) \overline{\Delta (u,z_3)} \non \\ &&+\frac{2\pi^2}{{\rm det}Y}  \int_{\S^4} \prod_{i=1,2,3,4} d^2 z_i \mathcal{G} (z_1,z_3) \mathcal{G} (z_2,z_4) P(z_2,z_3) (w,\overline{z_1}) (z_4,\overline{u})  \Delta (w,z_1) \overline{\Delta (u,z_4)} \non \\ &&+\frac{2\pi}{({\rm det}Y)} \int_{\S^3} \prod_{i=1,2,3} d^2 z_i \p_w\mathcal{G} (w,z_1) \overline\p_u \mathcal{G} (u,z_1)     \Delta (w,z_1) \overline{\Delta (u,z_1)} P(z_2,z_3)\mathcal{G} (z_2,z_3) \non \\ &&+\frac{8\pi}{{\rm det}Y}\int_{\S^2} \prod_{i=1,2} d^2 z_i \mathcal{G}(z_1,z_2) \p_w \mathcal{G} (w,z_1) \overline\p_u \mathcal{G} (u,z_2) (z_2,\overline{z_1}) \Delta (w,z_1) \overline{\Delta (u,z_2)} \non \\ &&-\frac{4\pi}{{\rm det}Y} \int_{\S^3} \prod_{i=1,2,3} d^2 z_i \p_w \mathcal{G} (w,z_1) \overline\p_u \mathcal{G} (u,z_2) (z_2,\overline{z_3})  (z_3,\overline{z_1}) \Delta (w,z_1) \overline{\Delta (u,z_2)} \non \\ &&\times  \Big(\mathcal{G} (z_1,z_3) +\mathcal{G} (z_2,z_3)\Big).\eea
Thus $\Phi_{3,F}$ also contains some contributions without derivatives. 

Adding the various contributions given above, we get that
\bea  \label{addmore}&&2(\Phi_{1,F}  + \Phi_{2,F} +\Phi_{2,G}) +\frac{1}{2} \Big( \Phi_{3,F} +\Phi_{3,G}\Big) \non \\&& = \frac{\pi}{({\rm det}Y)} \int_{\S^3} \prod_{i=1}^3 d^2 z_i \p_w\mathcal{G} (w,z_1) \overline\p_u \mathcal{G} (u,z_1)     \Delta (w,z_1) \overline{\Delta (u,z_1)} P(z_2,z_3)\mathcal{G} (z_2,z_3)+\widetilde{\Phi}_0, \non \\ \eea
where
\bea \label{tilde}\widetilde{\Phi}_0 &= &-\frac{4\pi^2}{{\rm det} Y}\int_{\S^3} \prod_{i=1}^3 d^2 z_i \mathcal{G} (z_1,z_2) \mathcal{G} (z_1,z_3) \m(z_1) (w,\overline{z_3}) (z_2,\overline{u}) \Delta (w,z_3) \overline{\Delta (u,z_2)}\non \\ &&+\frac{2\pi^2}{{\rm det}Y}  \int_{\S^4} \prod_{i=1}^4 d^2 z_i \mathcal{G} (z_1,z_4) \mathcal{G} (z_2,z_3)P(z_3,z_4) (w,\overline{z_2}) (z_1,\overline{u}) \Delta (w,z_2) \overline{\Delta (u,z_1)}.\non \\ \eea
Hence there is only one contribution having derivatives in \C{addmore}, which has the skeleton graph depicted by figure 16 (iii). 

\section{Simplifying the structure of $\Phi_{6,F}$}

In section 10.3, we have considered the variation $\overline\delta_{uu} \delta_{ww} \mathcal{B}^{(2,0)}_6/2$. The terms involving two derivatives are denoted by $\Phi_{6,F}$. We have that 

\bea &&\Phi_{6,F} = \frac{3\pi}{{\rm det}Y} \int_{\S^3}\prod_{i=1}^3 d^2 z_i \mathcal{G} (z_2,z_3) P(z_2,z_3) \p_w\mathcal{G} (w,z_1) \overline\p_u\mathcal{G}(u,z_1) \Delta (w,z_1) \overline{\Delta (u,z_1)}\non \\ &&+\pi  \int_{\S^4}\prod_{i=1}^4 d^2 z_i\mathcal{G} (z_2,z_3) \p_w\mathcal{G} (w,z_1) \overline\p_u\mathcal{G} (u,z_4)\Big[ \frac{1}{2} (w,\overline{z_4})(z_4,\overline{z_1}) R(z_1;z_2,z_3;\overline{u}) \non \\ &&+\frac{1}{2}(z_4,\overline{z_1})(z_1,\overline{u}) R(w;z_2,z_3;\overline{z_4})-\frac{1}{2} (w,\overline{u})(z_4,\overline{z_1}) R(z_1;z_2,z_3;\overline{z_4})  \non \\ &&- \frac{1}{2} P(z_1,z_4) R(w;z_2,z_3;\overline{u}) +\Big((w,\overline{z_4})(z_1,\overline{u})- (w,\overline{u})(z_1,\overline{z_4})\Big)R(z_4;z_2,z_3;\overline{z_1})\Big] \non \\ &&+2(\Lambda_1 -\Lambda_2),\eea
where $\Lambda_1$ and $\Lambda_2$ are given by \C{defL} and \C{defL2} respectively. 
We can now simplify this expression to express $\Phi_{6,F}$ in terms of graphs having two as well as no derivatives.

To do so, we obtain the relation
\bea &&\frac{\pi}{2}  \int_{\S^4}\prod_{i=1}^4 d^2 z_i(z_4,\overline{z_1})\mathcal{G} (z_2,z_3) \p_w\mathcal{G} (w,z_1) \overline\p_u\mathcal{G} (u,z_4)\Big[ (w,\overline{z_4}) R(z_1;z_2,z_3;\overline{u}) \non \\ &&+(z_1,\overline{u}) R(w;z_2,z_3;\overline{z_4})- (w,\overline{u}) R(z_1;z_2,z_3;\overline{z_4})  -  (z_1,\overline{z_4}) R(w;z_2,z_3;\overline{u}) \Big] \non \\ &&= -\frac{\pi}{{\rm det}Y}  \int_{\S^4}\prod_{i=1}^4 d^2 z_i(z_4,\overline{z_1})P(z_2,z_3)\mathcal{G} (z_2,z_3) \p_w\mathcal{G} (w,z_1) \overline\p_u\mathcal{G} (u,z_4)\Delta(w,z_1)\overline{\Delta(u,z_4)}\non \\ && = -\frac{2\pi}{{\rm det}Y} \int_{\S^3}\prod_{i=1}^3 d^2 z_iP(z_2,z_3)\mathcal{G} (z_2,z_3) \p_w\mathcal{G} (w,z_1) \overline\p_u\mathcal{G} (u,z_1)\Delta(w,z_1)\overline{\Delta(u,z_1)} \non \\ && +\frac{\pi^2}{{\rm det}Y} \int_{\S^4}\prod_{i=1}^4 d^2 z_i P(z_2,z_3)(w,\overline{z_1})(z_4,\overline{u})\mathcal{G} (z_1,z_4) \mathcal{G} (z_2,z_3)\Delta(w,z_1) \overline{\Delta(u,z_4)}, \eea
where we have used \C{repeat} for $(z_4,\overline{z_1})$ to obtain the last equation from the previous one.

We also obtain the relation
\bea &&\pi  \int_{\S^4}\prod_{i=1}^4 d^2 z_i\mathcal{G} (z_2,z_3) \p_w\mathcal{G} (w,z_1) \overline\p_u\mathcal{G} (u,z_4) \Big((w,\overline{z_4})(z_1,\overline{u})- (w,\overline{u})(z_1,\overline{z_4})\Big)R(z_4;z_2,z_3;\overline{z_1})\non \\ &&= \frac{8\pi}{{\rm det}Y} \int_{\S^2}\prod_{i=1}^2 d^2 z_i (z_2,\overline{z_1})\mathcal{G} (z_1,z_2) \p_w\mathcal{G}(w,z_1) \overline\p_u\mathcal{G} (u,z_2) \Delta(w,z_1)\overline{\Delta(u,z_2)}\non \\ &&- \frac{4\pi}{{\rm det}Y} \int_{\S^3}\prod_{i=1}^3 d^2 z_i (z_2,\overline{z_3})(z_3,\overline{z_1})\p_w\mathcal{G}(w,z_1) \overline\p_u \mathcal{G}(u,z_2) \Delta(w,z_1) \overline{\Delta(u,z_2)} \non \\ &&\times \Big(\mathcal{G}(z_1,z_3)+\mathcal{G}(z_2,z_3)\Big)\non \\ &&- \frac{4\pi^2}{{\rm det}Y} \int_{\S^3}\prod_{i=1}^3 \m(z_3) (w,\overline{z_1})(z_2,\overline{u})\mathcal{G}(z_1,z_3)\mathcal{G}(z_2,z_3) \Delta(w,z_1)\overline{\Delta(u,z_2)}\non \\  &&+\frac{2\pi^2}{{\rm det}Y} \int_{\S^4}\prod_{i=1}^4 d^2 z_i P(z_2,z_3)(w,\overline{z_1})(z_4,\overline{u})\Delta(w,z_1) \overline{\Delta(u,z_4)}\mathcal{G}(z_1,z_3)\mathcal{G}(z_2,z_4),\non \\ \eea
where we have used the relation \C{repeat} for $(z_4,\overline{z_2})$ and $(z_3,\overline{z_1})$ in the first line, noting that $R(z_4;z_2,z_3;\overline{z_1}) \rightarrow 2 (z_4,\overline{z_2})(z_2,\overline{z_3})(z_3,\overline{z_1})$ in the integral.

Hence putting the various contributions together, we get \C{6F}.

\section{Obtaining the expression for $\Theta_{IJ;KL}$}

We obtain the expression for $\Theta_{IJ;KL}$ which is defined using \C{theta} and \C{T2}. Substituting the expressions for the terms on the right hand side of \C{theta}, we define
\be \Theta =\widetilde\Theta_2 + \widetilde\Theta_3 +\widetilde\Theta_4,\ee
where $\widetilde\Theta_i$ is an integral over $i$ insertion points. Thus we have that
\bea \widetilde\Theta_2 = \frac{2\pi^2}{{\rm det}Y} \int_{\S^2}\prod_{i=1}^2 d^2 z_i \mathcal{G}(z_1,z_2)^2(w,\overline{z_1})\Big[(z_1,\overline{u})\Delta(w,z_2) - (z_2,\overline{u})\Delta(w,z_1)\Big]\overline{\Delta(u,z_2)},\non \\ \eea
involving two integrals,
\bea &&\widetilde\Theta_3 = \frac{\pi^2}{{\rm det}Y}\int_{\S^3}\prod_{i=1}^3 d^2 z_i \mathcal{G} (z_1,z_2)\mathcal{G}(z_1,z_3)\Big[ -(w,\overline{z_2})(z_2,\overline{z_3})(z_3,\overline{u})\Delta(w,z_1)\overline{\Delta(u,z_1)}\non \\ &&-\Big(\m(z_1)(w,\overline{z_3})(z_2,\overline{u})+ (w,\overline{z_1})(z_1,\overline{u})(z_2,\overline{z_3}) - (w,\overline{z_1})(z_1,\overline{z_3})(z_2,\overline{u})\Big)\Delta(w,z_3)\overline{\Delta(u,z_2)}\non \\ &&+(z_2,\overline{u})(w,\overline{z_3})(z_3,\overline{z_1})\Delta(w,z_1)\overline{\Delta(u,z_2)}+ (z_1,\overline{u})(w,\overline{z_3})(z_3,\overline{z_2}) \Delta(w,z_2) \overline{\Delta(u,z_1)}\Big],\non \\ \eea
involving three integrals, and 
\bea \widetilde\Theta_4 &=&  \frac{\pi^2}{2{\rm det}Y}\int_{\S^4}\prod_{i=1}^4 d^2 z_i  \mathcal{G}(z_1,z_4)\mathcal{G} (z_2,z_3)\Big[ (z_4,\overline{z_3})\Big((z_3,\overline{z_4})(w,\overline{z_2})(z_1,\overline{u}) \non \\ &&+ (w,\overline{z_4})(z_3,\overline{u})(z_1,\overline{z_2})- (w,\overline{z_4})(z_3,\overline{z_2})(z_1,\overline{u})\Big)\Delta(w,z_2)\overline{\Delta(u,z_1)} \non \\ &&- (w,\overline{z_3})(z_3,\overline{z_2})(z_2,\overline{z_4})(z_1,\overline{u})\Delta(w,z_4)\overline{\Delta(u,z_1)}\Big]\eea 
involving four integrals. Note that each $\widetilde\Theta_i$ is hermitian. 
 
We next define
\bea \widetilde\Theta_i = 4\pi^2 \omega_I (w) \omega_J (w) \overline{\omega_K (u)} \overline{\omega_L (u)}\Theta_{(i)IJ;KL}.\eea
This leads to the expressions
\bea \Theta_{(2)IJ;KL} &=& \frac{1}{2} \int_{\S^2}\prod_{i=1}^2 d^2 z_i \mathcal{G}(z_1,z_2)^2 Y^{-1}_{IP} Y^{-1}_{KQ} \overline{\omega_P (z_1)} \non \\ &&\times \Big[\omega_Q (z_1) \Big(Y^{-1}_{JL} \m(z_2) - Y^{-1}_{JM} Y^{-1}_{LN} \overline{\omega_M (z_2)} \omega_N (z_2)\Big)  \non \\ &&- \omega_Q (z_2) \Big(Y^{-1}_{JL} (z_1,\overline{z_2}) -  Y^{-1}_{JM} Y^{-1}_{LN} \overline{\omega_M (z_2)} \omega_N (z_1)\Big)\Big]\eea
from $\widetilde{\Theta}_2$,
\bea \Theta_{(3)IJ;KL} &=& \frac{1}{4} \int_{\S^3}\prod_{i=1}^3 d^2 z_i \mathcal{G}(z_1,z_2)\mathcal{G}(z_1,z_3) \Big[ -Y^{-1}_{IP} Y^{-1}_{QK}(z_2,\overline{z_3}) \overline{\omega_P (z_2)} \omega_Q (z_3)\non \\ &&\times\Big(Y^{-1}_{JL} \m(z_1) - Y^{-1}_{JM} Y^{-1}_{LN} \overline{\omega_M (z_1)} \omega_N (z_1) \Big) \non \\ 
&&- Y^{-1}_{JP} Y^{-1}_{LQ}\Big(Y^{-1}_{IK} (z_3,\overline{z_2}) - Y^{-1}_{IM} Y^{-1}_{KN} \overline{\omega_M (z_2)} \omega_N (z_3)\Big)\Big(\m(z_1)  \overline{\omega_P (z_3)} \omega_Q (z_2)\non \\ &&+ (z_2,\overline{z_3}) \overline{\omega_P (z_1)} \omega_Q (z_1) - (z_1,\overline{z_3}) \overline{\omega_P (z_1)}\omega_Q (z_2)\Big)\non \\ &&+ Y^{-1}_{IM} Y^{-1}_{KN} (z_3,\overline{z_1}) \overline{\omega_M (z_3)} \omega_N (z_2) \Big(Y^{-1}_{JL} (z_1,\overline{z_2}) - Y^{-1}_{JP} Y^{-1}_{LQ} \overline{\omega_P (z_2)} \omega_Q (z_1)\Big)\non \\ &&+ Y^{-1}_{IM} Y^{-1}_{KN} (z_3,\overline{z_2}) \overline{\omega_M (z_3)} \omega_N (z_1) \Big(Y^{-1}_{JL} (z_2,\overline{z_1}) - Y^{-1}_{JP} Y^{-1}_{LQ} \overline{\omega_P (z_1)} \omega_Q (z_2)\Big) \Big]\non \\ \eea
from $\widetilde{\Theta}_3$, and finally
\bea \Theta_{(4)IJ;KL} = \frac{1}{8} \int_{\S^4}\prod_{i=1}^4 d^2 z_i \mathcal{G}(z_1,z_4) \mathcal{G}(z_2,z_3) \Big[(z_4,\overline{z_3}) \Big(Y^{-1}_{IK} (z_2,\overline{z_1}) - Y^{-1}_{IM} Y^{-1}_{KN} \overline{\omega_M (z_1)} \omega_N (z_2)\Big)\non \\ \times Y^{-1}_{JP} Y^{-1}_{LQ}\Big((z_3,\overline{z_4})  \overline{\omega_P (z_2)} \omega_Q (z_1) + (z_1,\overline{z_2}) \overline{\omega_P (z_4)} \omega_Q (z_3) -(z_3,\overline{z_2}) \overline{\omega_P (z_4)} \omega_Q (z_1)\Big)\non \\ -(z_3,\overline{z_2})(z_2,\overline{z_4}) Y^{-1}_{IM} Y^{-1}_{KN} \overline{\omega_M (z_3)} \omega_N (z_1) \Big(Y^{-1}_{JL} (z_4,\overline{z_1}) - Y^{-1}_{JP} Y^{-1}_{LQ} \overline{\omega_P (z_1)}\omega_Q (z_4)\Big)\Big]\non \\ \eea
from $\widetilde\Theta_4$.

Thus we have that
\be \label{DefTheta}\Theta_{IJ;KL} = \sum_{i=2}^4 \Theta_{(i)IJ;KL}.\ee


\providecommand{\href}[2]{#2}\begingroup\raggedright\endgroup

\end{document}